\documentclass[12pt]{article}
\pdfoutput=1
\usepackage{jheppub}
\usepackage{amssymb,amsmath,amstext,amsfonts}
\usepackage{bbold,ulem,bm}
\usepackage{graphics}
\usepackage{mathtools}
\usepackage{hyperref}
\usepackage{color}

\usepackage{graphicx}
\usepackage{subcaption}
\usepackage{mwe}
\usepackage{mathrsfs}
\usepackage{multirow}

\newcommand{\beq}{\begin{equation}}
\newcommand{\eeq}{\end{equation}}
\newcommand{\bal}{\begin{aligned}}
\newcommand{\eal}{\end{aligned}}

\def\vphi{\varphi}
\def\Mp{M_{{\rm Pl}}}
\def\etaperp{\eta_{\perp}}
\def\Tdot#1{{{#1}^{\hbox{.}}}}
\def\s{\sigma}
\def\msh{m^{2}_{s\, ({\rm eff})}}
\def\c{|c_s|}
\def\m{\star}
\def\eq{{\rm eq}}
\def\orth{{\rm orth}}

\renewcommand{\arraystretch}{2}

\begin{document}

\title{Primordial fluctuations and non-Gaussianities in sidetracked inflation} 

\author[a]{Sebastian Garcia-Saenz,}
\affiliation[a]{Institut d'Astrophysique de Paris, GReCO, UMR 7095 du CNRS et de Sorbonne Universit\'e, 98bis boulevard Arago, 75014 Paris, France}
\author[a]{S\'ebastien Renaux-Petel}
\author[b]{and John Ronayne}
\affiliation[b]{School of Physics and Astronomy, Queen Mary University of London, Mile End Road, London, E1 4NS, UK}

  \emailAdd{sebastian.garcia-saenz@iap.fr}
\emailAdd{renaux@iap.fr}
\emailAdd{j.ronayne@qmul.ac.uk }

\abstract{Heavy scalar fields can undergo an instability during inflation as a result of their kinetic couplings with the inflaton. This is known as the geometrical destabilization of inflation, as it relies on the effect of the negative curvature of the field-space manifold overcoming the stabilizing force of the potential. This instability can drive the system away from its original path in field space into a new inflationary attractor, a scenario that we dub {\it sidetracked inflation}. We study this second phase and its observable consequences in several classes of two-field models. We show that cosmological fluctuations exhibit varied behaviours depending on the potential and the field space geometry, and that they can be captured by single-field effective theories with either a modified dispersion relation, a reduced speed of sound, or an imaginary one --- the latter case describing a transient tachyonic growth of the fluctuations. We also numerically calculate the bispectrum with the transport approach, finding large non-Gaussianities of equilateral and orthogonal shapes. In the hyperbolic geometry the potentials of our models present a pole at the boundary of the Poincar\'e disk and we discuss their relationships with $\alpha$-attractors.}

\maketitle


\section{Introduction} \label{sec:intro}

Multi-field inflation provides an extension of the minimal single-field inflationary para\-digm that is most natural from a theoretical point of view. Multiple scalars are generically present in most top-down scenarios of the very early universe, including constructions in the contexts of string theory \cite{Baumann:2014nda}, supergravity \cite{Yamaguchi:2011kg}, and other theories beyond the Standard Model \cite{Linde:2005ht}. Nevertheless, in view of the spectacular agreement of the predictions of slow-roll single-field inflation with experimental data \cite{Ade:2015lrj}, it is commonly argued that the additional fields must be very heavy, with masses parametrically larger than the inflationary Hubble scale $H$, and should therefore play no important role in the cosmological dynamics. A more precise statement is that these ``spectator'' fields can be integrated out to yield an effectively single-field description that is valid throughout the epoch of inflation \cite{Tolley:2009fg,Cremonini:2010ua,Achucarro:2010da}, and hence one may expect the heavy scalars to affect inflation only indirectly through the renormalization of operators controlling the dynamics of the inflaton.

Recently, however, it has been shown that heavy scalar fields with bare masses $m_h^2\gg H^2$ can undergo, under very general conditions, a tachyonic-like instability induced by kinetic couplings with the inflaton, as one generically has in nonlinear sigma models. This has been named {\it the geometrical destabilization} of inflation \cite{Renaux-Petel:2015mga}. It is akin to the instability that arises in models of hybrid inflation \cite{Linde:1993cn} where the heavy scalars become tachyonic as a result of their coupling to the inflaton at the level of the potential. The geometrical destabilization on the other hand is triggered by the rolling of the inflaton in a negatively curved internal field space, and may take place well before any potential-driven ``waterfalls'' along the inflationary trajectory.

If the geometrical destabilization does occur, its outcome is quite uncertain. Standard perturbation theory breaks down at the onset of the instability and the vacuum state that describes inflation can no longer be trusted. Nevertheless, on physical grounds we may expect either of two things to happen depending on the interactions and the scales involved. The first possibility is that the universe becomes dominated by inhomogeneities and that inflation ends prematurely, that is at a time much before the end of the slow-roll phase as it would have happened in the absence of any instability. The consequence is that cosmological modes that are observable through cosmic microwave background (CMB) and large scale structure data probe a different part of the inflaton potential, leading to modifications to their correlation functions and the corresponding predictions for the cosmological parameters of interest. This scenario was analyzed in reference \cite{Renaux-Petel:2017dia} for a large class of inflationary models using Bayesian techniques, and it was quantified how such a premature end of inflation results in sizable changes to the constraints on theoretical models from experimental data.

\begin{figure*}[t]
\centering
\includegraphics[width=0.6\textwidth]{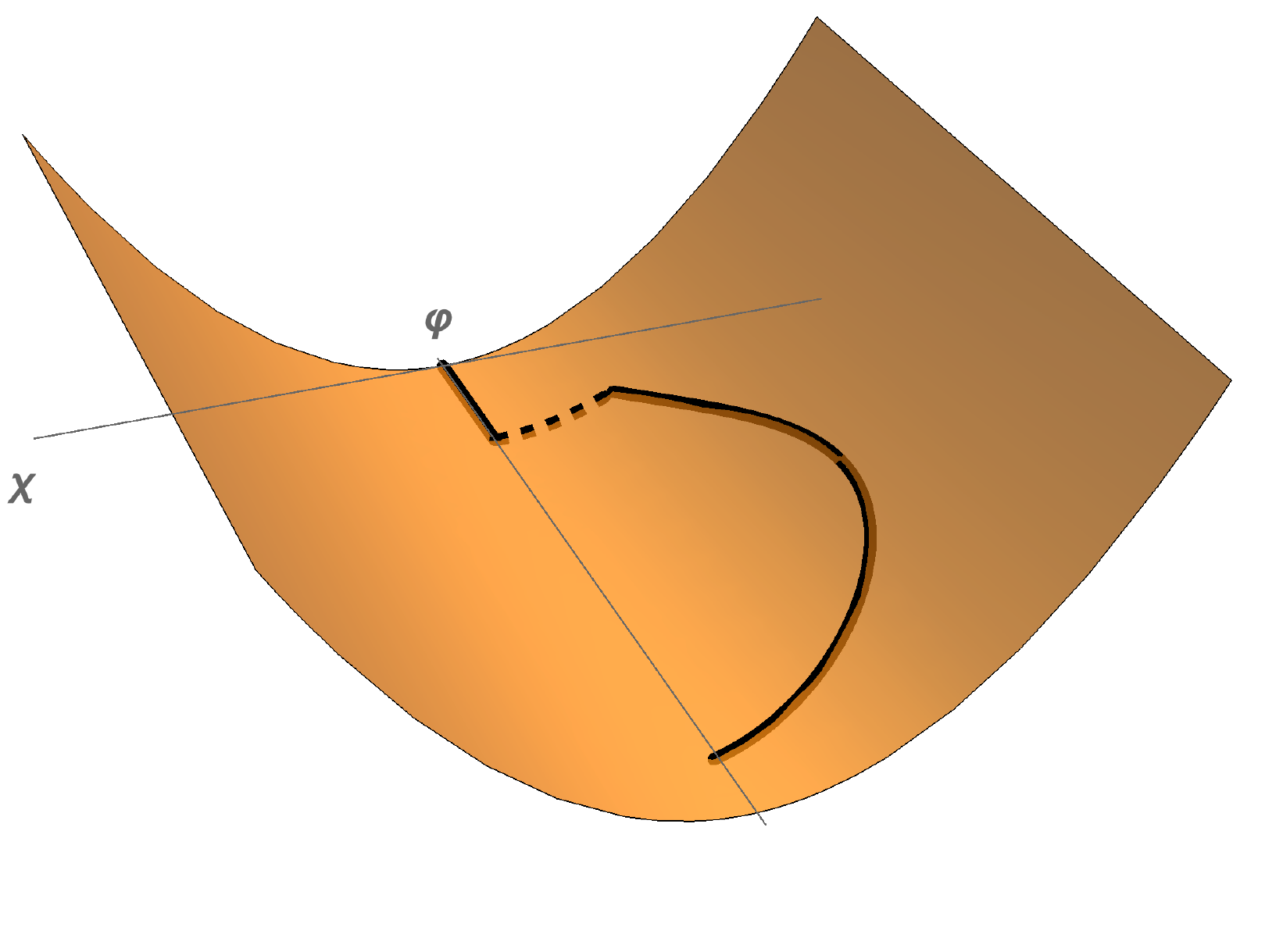}
\caption{Schematic plot of the field-space trajectory in sidetracked inflation. The field $\varphi$ corresponds to the inflaton and $\chi$ is the heavy field that becomes unstable. The dashed line represents the instability phase where the classical field picture is lost. The orange surface represents the potential. The information on the field space geometry is not represented.}
\label{fig:schematic}
\end{figure*}

The second possible outcome is that the exponential growth of the unstable fields drives the system to a new inflationary vacuum. Thus, in this set-up, the universe undergoes a second phase of inflation in which one (or more) of the heavy scalars evolves along a path away from its ground state. We dub this scenario {\it sidetracked inflation}, owing to the way the geometric destabilization causes the field-space trajectory to divert from its original, effectively single-field path (see fig.\ \ref{fig:schematic}). It is the objective of the present paper to perform an analysis of the dynamics and properties of sidetracked inflation, with a focus on its peculiar multi-field effects on the power spectrum and primordial non-Gaussianities.

This picture obviously glosses over the details related to the physics of the instability, which as explained lie beyond the reach of perturbative field theory. It is motivated however by the fact that, at least in a large number of cases, the equations of motion indeed admit a nontrivial time-dependent attractor away from the inflationary valley of the potential, as we have investigated with a broad class of two-field models.
Thus, provided inflation does not abruptly end as described above in the first scenario, it is natural to expect that the unstable system will eventually settle into this attractor and give rise to another phase of inflation. We will see that this second phase features some very interesting dynamics: it can last extremely long, as the increase in the value of the non-canonical kinetic term of the inflaton translates into an effective flattening of the potential; its path in the internal field space can deviate very strongly from a geodesic; and multifield effects are very important.

In particular, we show that the dynamics of linear cosmological fluctuations can be effectively described by a single-field effective theory, that is characterized, depending on the field space manifold and the potential that is considered, by a modified dispersion relation, a reduced speed of sound, or an imaginary one, describing a transient tachyonic instability. The bispectrum is generically large in these models, with shapes that can be of equilateral but also of orthogonal type, in particular in models with hyperbolic field spaces and that feature an effective imaginary speed of sound.

Hyperbolic field space geometry is an essential aspect of so-called cosmological attractors---inflationary models whose predictions are insensitive to the form of the inflaton potential \cite{Kallosh:2013hoa,Kallosh:2013yoa,Kallosh:2013tua}. It has been explained that this universality stems from the presence of a pole in the kinetic term of the inflaton \cite{Galante:2014ifa,Kallosh:2015zsa}, which translates into an exponential flattening of the potential upon canonical normalization. It is thus natural to ask whether the geometrical destabilization and sidetracked inflation could be relevant for such theories and possibly hinder some of their successful features. We will show however that, on closer inspection, our models present a subtle but crucial difference with cosmological attractors, to wit the fact that in our case it is both the kinetic term {\it and} the potential that exhibit a pole (in a suitable parametrization). This possibility appears to have been overlooked perhaps because it doesn't arise in the single-field context where a singular potential would be unnatural, but we will argue that it can be perfectly generic within multi-field scenarios.

In the next section we explain our implementation of the sidetracked inflation scenario, including the way we reach this attractor phase from a heuristic modeling of the geometrical destabilization that precedes the second inflationary phase. We also review the necessary formalism and list the two-field models that we scrutinize. We study in section \ref{sec:background} the background dynamics of the sidetracked phase, before devoting section \ref{sec:powerspectrum} to a detailed study of the properties of linear cosmological fluctuations. We show how single-field effective theories for the fluctuations can reproduce with a good accuracy the exact results from numerical computations in the full two-field system, and give results for the tensor-to-scalar ratio $r$ and spectral index $n_s$ of the curvature power spectrum computed for each model. Similarly in section \ref{sec:bispectrum} we give numerical results from the full two-field picture concerning the primordial bispectrum. We discuss the relationship between the class of models that we analyze and the models of $\alpha$-attractors in section \ref{sec:alpha}, before concluding in section \ref{sec:conclusion} with a summary and discussion of our results.

\vspace{1em}

\noindent {\it Conventions:} We adopt the mostly-plus signature for the spacetime metric. Derivatives with respect to the cosmic time $t$ will be denoted by a dot, whereas derivatives with respect to the number of e-folds $N$ will be denoted by a prime. We employ natural units but we will sometimes re-instore explicitly the Planck mass when it is illuminating.


\section{Formalism and setup} \label{sec:models}

\subsection{Nonlinear sigma models of multi-field inflation}

We consider nonlinear sigma models for a set of scalar fields $\phi^I$ minimally coupled to gravity,
\beq \label{eq:nonlinsigma}
S=\int d^4x\sqrt{-g}\bigg[\frac{1}{2}\,R(g)-\frac{1}{2}\,G_{IJ}(\phi)\nabla^{\mu}\phi^I\nabla_{\mu}\phi^J-V(\phi)\bigg]\,,
\eeq
with $G_{IJ}$ the metric of the internal field space manifold. We will further denote by $\Gamma^I_{JK}$ and ${\cal R}^I_{\phantom{I}JKL}$ the corresponding Levi-Civita connection and Riemann tensor, respectively. \\

\noindent
On spatially flat Friedman-Lema\^itre-Robertson Walker spacetimes of line element $ds^2=-dt^2+a^2(t)d{\vec x}^2$, where $a(t)$ is the scale factor, the Einstein and Klein-Gordon equations of motion for homogeneous scalar fields read
\begin{eqnarray}
 3H^2&=&\frac12 \dot \sigma^2+V\,,  \\
 \dot{H}&=&-\frac12\dot \sigma^2\,, \\
{\cal D}_t \dot \phi^I  +3H  \dot \phi^I &+&G^{IJ} V_{,J}=0\,, \label{eom-scalars}
\end{eqnarray}
where dots denote derivatives with respect to $t$, $H \equiv \dot a/a$ is the Hubble parameter, $\frac{1}{2}\dot \sigma^2 \equiv \frac{1}{2}G_{IJ} \dot \phi^I \dot \phi^J$ is the kinetic energy of the fields, and, here and in the following, ${\cal D}_t A^I \equiv \dot{A^I} + \Gamma^I_{JK} \dot \phi^J A^K$ for a field space vector $A^I$. As is customary, we introduce the `deceleration' parameter $\epsilon$, which is less then one by definition during inflation, such that
\beq
\epsilon \equiv -\frac{\dot H}{H^2}=\frac{\dot \sigma^2}{2 H^2}\,.
\eeq
It will be useful, both at the level of the background as well as the one of the perturbations, to split the equations of motion into an instantaneous adiabatic part, corresponding to quantities projected along the fields' velocities, and an entropic part, corresponding to quantities projected orthogonally to it \cite{Gordon:2000hv,Bartolo:2001rt} (note that as we consider two-field models, the entropic sector is one-dimensional). For this purpose, we introduce the adiabatic unit vector $e_{\sigma}^I \equiv \dot \phi^I /{\dot \sigma}$ and the entropic unit vector $e_s^I$, orthogonal to $e_\sigma^I$ and such that the orthonormal field space basis $(e_\sigma^I,e_s^I)$ is right-handed. The projection of the equations of motion \eqref{eom-scalars} along $e_\sigma^I$ reads
\beq
\ddot \sigma+3H \dot \sigma+V_{,\sigma}=0
\label{sigmaddot}
\eeq
where $V_{,\sigma} \equiv e_\sigma^I V_{,I}$, and similarly for analogous quantities, whereas its projection along $e_s^I$ gives the rate of change of the adiabatic basis vector in terms of the entropic basis vector $e_s^I$, as
\beq
{\cal D}_t e_\sigma^I= H \etaperp e_s^I\,, \quad \etaperp \equiv - V_s/(H \dot \sigma)\,.
\label{etaperp}
\eeq
Here, we introduced the important parameter $\etaperp$ \cite{GrootNibbelink:2000vx,GrootNibbelink:2001qt}. By definition, it quantities the acceleration of the scalar fields perpendicular to their velocities, so it is a dimensionless measure of the deviation of the background trajectory from a geodesic: the larger $\etaperp$, the more curved the trajectory in field space. This perpendicular acceleration parameter also plays an important role in the physics of the fluctuations about the background, to which we now turn.

Cosmological perturbation theory in non linear sigma models of inflation has been extensively studied (see \textit{e.g.} \cite{Gong:2016qmq} for a recent review), and we will make a detailed analysis of the cosmological fluctuations in sidetracked inflation in sections \ref{sec:powerspectrum} and \ref{sec:bispectrum}. Here it is enough to briefly summarize formal aspects of linear perturbation theory about the above background. 
In particular, the dynamics of scalar linear fluctuations is dictated by the second-order action \cite{Sasaki:1995aw,GrootNibbelink:2001qt,Langlois:2008mn}
 \begin{eqnarray}
S_{(2)}= \int  dt\, d^3x \,a^3\left(G_{IJ}\mathcal{D}_tQ^I\mathcal{D}_tQ^J-\frac{1}{a^2}G_{IJ}\partial_i Q^I \partial^i Q^J-M_{IJ}Q^IQ^J\right)\,,
\label{S2}
\end{eqnarray}
where the $Q^I$ are the field fluctuations in the spatially flat gauge and the mass (squared) matrix is given by
\begin{eqnarray} \label{masssquared}
M_{IJ} &=& V_{; IJ} - \mathcal{R}_{IKLJ}\dot \phi^K \dot \phi^L -\frac{1}{a^3}\mathcal{D}_t\left[\frac{a^3}{H} \dot \phi_
I \dot \phi_J\right]\,.
\end{eqnarray}
Here $ V_{;IJ} \equiv V_{,IJ}-\Gamma_{IJ}^K V_{,K}$, and field space indices are raised and lowered using
$G_{IJ}$. One can easily deduce from Eq.~\eqref{S2} the equations of motion for the linear fluctuations (in Fourier space):
\begin{eqnarray}
{\cal D}_t {\cal D}_t Q^I  +3H {\cal D}_t Q^I +\frac{k^2}{a^2} Q^I +M^I_{\,J} Q^J=0\,.
\label{pert}
\end{eqnarray}

\subsection{Geometrical destabilization and sidetracked inflation}

We now review the mechanism behind the geometrical destabilization uncovered in \cite{Renaux-Petel:2015mga}, and our implementation of the sidetracked inflation scenario that we described qualitatively in section \ref{sec:intro}. We specify to the case of a two-dimensional field space, although the analysis can in principle be generalized to arbitrarily many fields. As mentioned above, we introduce the adiabatic and entropic fluctuations, respectively defined by $Q_\sigma \equiv e_{\sigma I} Q^I$ and $Q_s \equiv e_{s I} Q^I$. On super-Hubble scales, one can deduce from Eq.~\eqref{pert} (see section \ref{fluctuations-adiabatic-entropic} for details) that the latter satisfies the following equation of motion:
\beq
\ddot Q_s+3 H \dot Q_s +\msh Q_s \approx 0\,,
\label{Qs-super-Hubble}
\eeq
where $\msh$ is a super-Hubble effective mass such that (reintroducing the Planck mass for the sake of clarity)
\beq \label{eq:entropicmass}
\frac{\msh}{H^2}\equiv\frac{V_{;ss}}{H^2}+3\eta_{\perp}^2+\epsilon R_{\rm fs} \Mp^2\,.
\eeq
Here $V_{;ss}\equiv e_s^I e_s^J V_{;IJ}$ is the projection of the Hessian matrix along the entropic direction, $\eta_{\perp}$ has been defined in Eq.~\eqref{etaperp}, and $R_{\rm fs}$ is the scalar curvature of the internal manifold. The crucial observation is that $\msh$ can become negative, for realistic values of the scales involved, whenever the field space manifold is negatively curved.

This can be made more explicit by considering the following scenario, which is the one we will focus on in the remainder of the paper. Assume a model with two scalar fields: an inflaton $\varphi$ that initially drives inflation in a standard slow-roll fashion, and a spectator field $\chi$ with a large bare mass $m_h$ sitting at the bottom of the potential valley at $\chi=0$, corresponding to a field space geodesic. Thus, at this stage, we have $V_{;ss}=m_h^2$ and $\eta_{\perp}=0$. We define the curvature scale $M$ of the field space manifold in such a way that $R_{\rm fs}=-4/M^2$ when restricted to the $\chi=0$ line (or exactly if the space has constant curvature).\footnote{This assumes, as we will do, that the internal metric depends only on $\chi$. We will further comment on this point later.} During this ``primary'' inflationary phase Eq.\ \eqref{eq:entropicmass} becomes 
\beq
\frac{\msh}{H^2}=\frac{m_h^2}{H^2}-4\epsilon\,\frac{\Mp^2}{M^2}\,,
\eeq
which implies that the super-Hubble entropic perturbation $Q_s$ becomes tachyonic, and therefore the instability of the background, at a critical time $N=N_c$ when the slow-roll parameter $\epsilon$ reaches the value
\beq \label{eq:epscritical}
\epsilon_c=\frac{M^2}{4\Mp^2}\,\frac{m_h^2}{H_c^2}\,,
\eeq
and with $H_c=H(N_c)$. Notice that even though $H$ decreases during inflation, $\epsilon$ typically grows at a faster rate during a slow-roll regime,\footnote{More precisely, a necessary condition for the geometrical destabilization to occur in this setup is that the quantity $\epsilon H^2$ be an increasing function of time. This translates into the inequality 
$\epsilon'/\epsilon>2\epsilon$,
which holds for concave potentials and even some convex ones \cite{Renaux-Petel:2015mga}.} and therefore the instability can be quite generic for reasonable values of $m_h$ and $M$. Taking for instance $m_h\sim 10H\sim 10H_c$ and $M\sim M_{\rm GUT}\sim 10^{-3}\,M_P$ one has $\epsilon_c\ll1$, and hence the instability can take place well before the end of inflation as it would occur in a single-field context.

As explained in \cite{Renaux-Petel:2015mga} and above in section \ref{sec:intro}, what happens after the geometrical destabilization is so far highly uncertain, and very likely model-dependent anyway. It therefore makes sense, as a first step towards a more thorough understanding of the physics involved, to adopt a specific outcome as a working assumption and study its consequences for a broad and generic class of models. This is what we do in this paper for the situation where inflation doesn't end as a result of the instability (the case studied in \cite{Renaux-Petel:2017dia}), but instead continues along a ``sidetracked'' trajectory away from the bottom of the potential valley at $\chi=0$.

Our modeling of the sidetracked inflation scenario will be as a two-step process (see fig.\ \ref{fig:schematic}). The system is assumed to start in the standard inflationary vacuum, with $\chi=0$ and a slowly rolling inflaton $\varphi$. At the critical time of the instability, defined by eq.\ \eqref{eq:epscritical}, we displace the heavy field by an amount $\chi_c\equiv H_c/2\pi$, which is a typical value for the amplitude of quantum fluctuations in a massless field. Together with the inflaton field's amplitude $\varphi_c$ (and its derivative) at the time $N_c$, this provides the initial conditions for the second phase of inflation. The latter then ends in a standard manner through slow-roll violation defined by the condition $\epsilon=1$.

This is admittedly a very blurry picture of the dynamics involved, but it is motivated by the fact that the second inflationary trajectory corresponds to an attractor of the equations of motion, at least in the models we have analyzed. Indeed, we have checked numerically that varying the initial conditions described in the previous paragraph, even by a large amount, doesn't affect any of the conclusions, as the system is inevitably driven towards the sidetracked attractor where the heavy field slowly evolves with a typical amplitude $\chi\sim M$, as might be expected on dimensional grounds\footnote{This heuristic picture will be refined in section \ref{sec:background}.}. The evolution then ends as both $\chi$ and the inflaton $\varphi$ fall into the stable minimum of the potential (or in any case by slow-roll violation when the potential chosen to model the inflationary phase solely does not admit a stable minimum).

The existence of this attractor solution may be heuristically understood as arising from an interplay between the repulsive force of the negatively curved field manifold and the stabilizing force of the $\chi$ potential, so that one can expect a regime where the two effects compensate each other allowing for a stable inflationary phase --- this intuitive picture will be confirmed analytically and numerically in section \refeq{sec:background}. It is worth emphasizing that, as we will see below, the field trajectory in this set-up is typically very different from a geodesic, which is a central feature of the sidetracked scenario that we are putting forth.

\subsection{Geometries and potentials}
\label{two-field}

We will study two classes of nonlinear sigma models with scalar fields $\phi^I=(\varphi,\chi)$, characterized by different field space metrics $G_{IJ}$. For each class, we will consider several choices of the potential $V$ that are theoretically motivated and extensively studied in the context of single-field inflation \cite{Martin:2013tda}.

The first internal metric we scrutinize is
\beq \label{eq:minimalmetric}
G_{IJ}d\phi^Id\phi^J=\left(1+\frac{2\chi^2}{M^2}\right)d\varphi^2+d\chi^2\,.
\eeq
We will refer to this as the ``minimal'' model, as it amounts to the addition of a single dimension-6 operator to the standard scalar field action; it is also the minimal sample realization of the geometrical destabilization used in reference \cite{Renaux-Petel:2015mga}. The corresponding scalar curvature  is
\beq
R_{\rm fs}=-\frac{4}{M^2(1+2\chi^2/M^2)^2}\,,
\label{R-minimal}
\eeq
and so indeed $R_{\rm fs}\simeq-4/M^2$ before the time of the geometrical destabilization when $\chi\simeq0$; see the previous subsection.

The second case is the metric of the hyperbolic plane,
\beq \label{eq:hyperbolicmetric}
G_{IJ}d\phi^Id\phi^J=\left(1+\frac{2\chi^2}{M^2}\right)d\varphi^2+\frac{2\sqrt{2}\,\chi}{M}\,d\varphi d\chi+d\chi^2\,,
\eeq
which has a {\it constant} scalar curvature $R_{\rm fs}=-4/M^2$. We have chosen this particular parametrization for the reason that it gives a seemingly ``small'' deformation of the above minimal model (although in fact the extra operator is less irrelevant as it is dimension-5), and hence may allow us to better understand the physical effects due to changes in the field space manifold.
We remark that \eqref{eq:hyperbolicmetric} can be obtained from the dilatonic-type metric $d\varphi^2+e^{-2\sqrt{2}\,\varphi/M}d\psi^2$ upon letting $e^{-\sqrt{2}\,\varphi/M}\psi\equiv\chi$. The two theories are of course inequivalent, however, as a field redefinition will have the effect of changing the form of the potential. We will further elaborate on this point in section \ref{sec:alpha} where we comment on the relation between our set-up and the models of cosmological attractors.\\

The potentials we will consider are all of the form
\beq \label{eq:potential1}
V=\Lambda^4{\mathcal{V}}(\varphi)+\frac{1}{2}\,m_h^2\chi^2\,,
\eeq
where ${\mathcal{V}}(\varphi)$ is a dimensionless function of $\varphi$. Similarly to \cite{Renaux-Petel:2015mga,Renaux-Petel:2017dia}, we choose $m_h=10H_c$, so that, according to eq.\ \eqref{eq:epscritical}, one has $\epsilon_c=25(M/M_P)^2$. As usual the energy scale $\Lambda$ will determine the overall scale of the power spectrum and can therefore be fixed a posteriori to match the observed amplitude of the curvature power spectrum. 

The four specific models we have studied are shown in table \ref{tab:models}. We refer the reader to \cite{Renaux-Petel:2017dia,Martin:2013tda} for details about these models and their relevant parameter spaces, but in the following we give a brief rationale for our choices.

The first case we investigate is the Starobinsky potential (SI) \cite{Starobinsky:1980te,Bezrukov:2007ep}, a prototypical example of plateau models. It is an interesting case study because it has no free parameter and, in its single-field realization, is in excellent agreement with experimental constraints. Then we consider three characteristic hilltop models: natural inflation (NI) \cite{Freese:1990rb}, quadratic small-field inflation ${\mathrm{SFI}}_2$, and quartic small-field inflation ${\mathrm{SFI}}_4$. The case of ${\mathrm{SFI}}_2$ can be regarded as truncation of NI if we take the scale $\mu=2f$, and hence the comparison between the two models gives a way to study the effects of the nonlinearities of the potential. We use $f={1,10,100}$, which are the orders of values commonly assumed in order to have agreement with data. On the other hand ${\mathrm{SFI}}_4$ has a vanishing mass at the hilltop, $V''(\varphi=0)=0$, and is therefore a priori in a different class. To enable comparison, we choose for it the same values of the scale $\mu$ as in ${\mathrm{SFI}}_2$. 

Eventually, to study the influence of the results on the curvature scale $M$, we consider the three values $M=\left(10^{-2},10^{-2.5},10^{-3}\right)\Mp$, although for NI, ${\rm SFI}_2$ and ${\rm SFI}_4$, we did that only for the central values of the parameters $f=10$ and $\mu=20$.

Although we studied all these models (36 with the various parameters' choice), and we will indeed give results for the observables for each of them, when we display detailed results and comparison with analytical formulae in the central part of the paper, we decided to use two representative examples: Starobinsky inflation and Natural Inflation with $f=10$, each with $M=10^{-3}$ (which is the value by default, if not otherwise specified), as they exhibit both characteristic features and varied properties. 

\setlength{\tabcolsep}{8pt}
\renewcommand{\arraystretch}{2}
\begin{table}[h]
\centering
\begin{tabular}{cccc}
\hline
Model & Acronym & Inflaton potential ${\mathcal{V}}(\varphi)$ & Parameter values \\
\hline\hline
Starobinsky inflation & SI & $\left(1-e^{-\sqrt{2/3}\,\varphi}\right)^2$ & --- \\
Natural inflation & NI & $1+\cos\left(\frac{\varphi}{f}\right)$ & $f=\{1,10,100\}$ \\
Quadratic small field & ${\mathrm{SFI}}_2$ & $1-\left(\frac{\varphi}{\mu}\right)^2$ & $\mu=\{2,20,200\}$ \\
Quartic small field & ${\mathrm{SFI}}_4$ & $1-\left(\frac{\varphi}{\mu}\right)^4$ & $\mu=\{2,20,200\}$ \\
\hline
\end{tabular}
\caption{List of inflationary models and values considered for the free parameters. The dimensionless potential function ${\mathcal{V}}(\varphi)$ is introduced in eq.\ \eqref{eq:potential1}.}
\label{tab:models}
\end{table}


\section{Background dynamics of sidetracked inflation} \label{sec:background}

In this section, we describe in more details the physics of sidetracked inflation at the level of the background. In a conventional phase of inflation driven by multiple scalar fields all slowly rolling down the potential, the acceleration term ${\cal D}_t \phi^I$ in \eqref{eom-scalars} is negligible compared to the Hubble friction and to the effect of the potential, so that all fields approximately follow $\dot \phi^I \simeq -V^{,I}/(3H)$. The sidetracked phase is markedly different, as we will see, as a central ingredient of it is that some acceleration terms are large. This is not in conflict with the existence of a phase of inflation, as the latter only requires $\epsilon \equiv -\dot H/H^2$ as well as its time derivative $\eta \equiv \dot \epsilon/(H \epsilon)$ to be small, for inflation to occur and to last long enough. Given that $\epsilon=\frac12 \dot \sigma^2/H^2$, this readily implies that $\ddot \sigma \ll 3 H \dot \sigma$, in other words that the acceleration of the fields tangential to the background trajectory be small. The perpendicular acceleration, quantified by the parameter $\etaperp$, need not be small in general, and indeed it will be large in sidetracked inflation.

\subsection{Background trajectories and qualitative understanding}

\begin{figure*}
        \centering
        \begin{subfigure}[b]{0.475\textwidth}
            \centering
            \includegraphics[width=\textwidth]{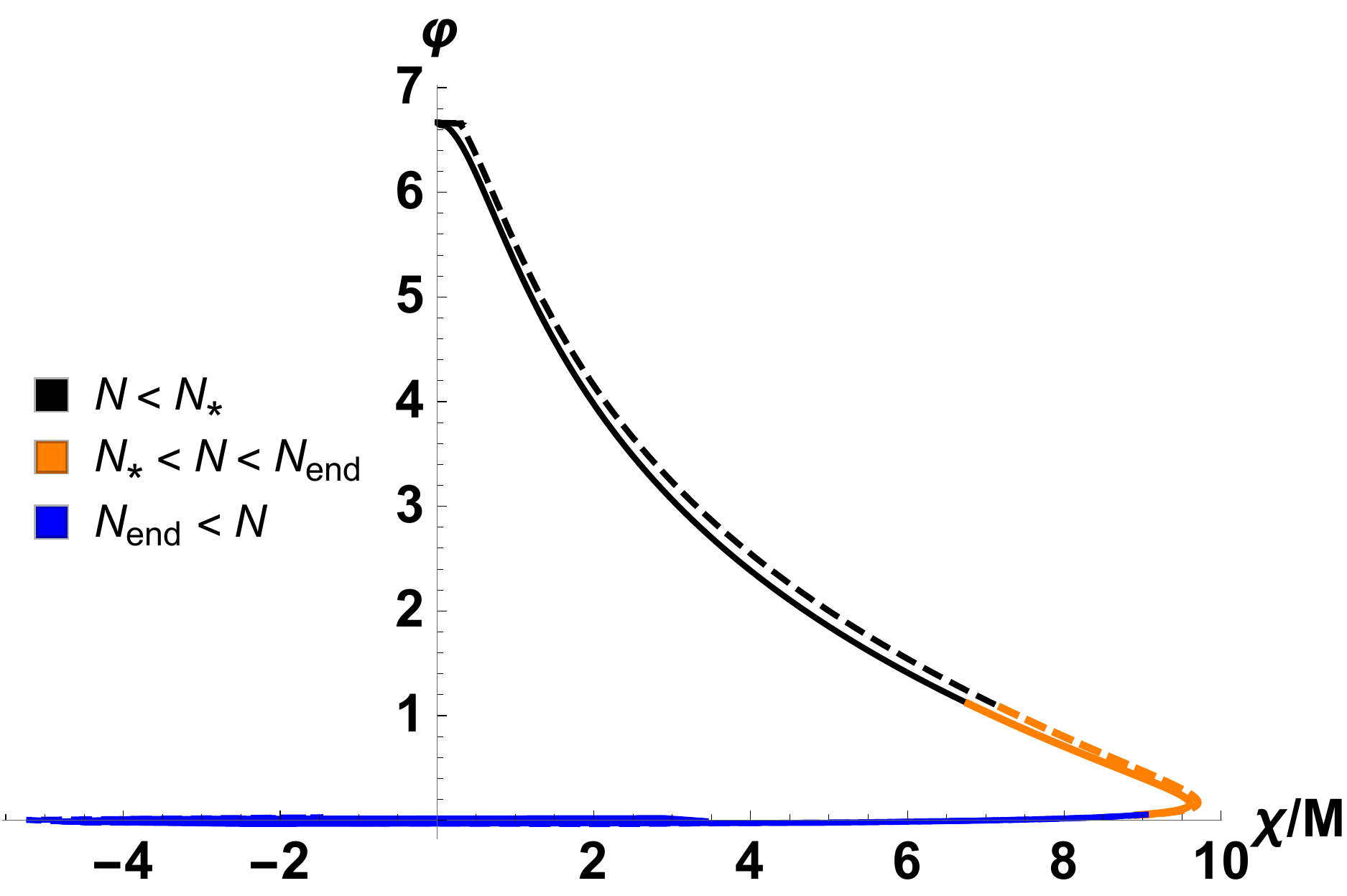}
            \caption{{\small Starobinsky inflation}}    
            \label{fig:trajectory SI}
        \end{subfigure}
        \hfill
        \begin{subfigure}[b]{0.475\textwidth}  
            \centering 
            \includegraphics[width=\textwidth]{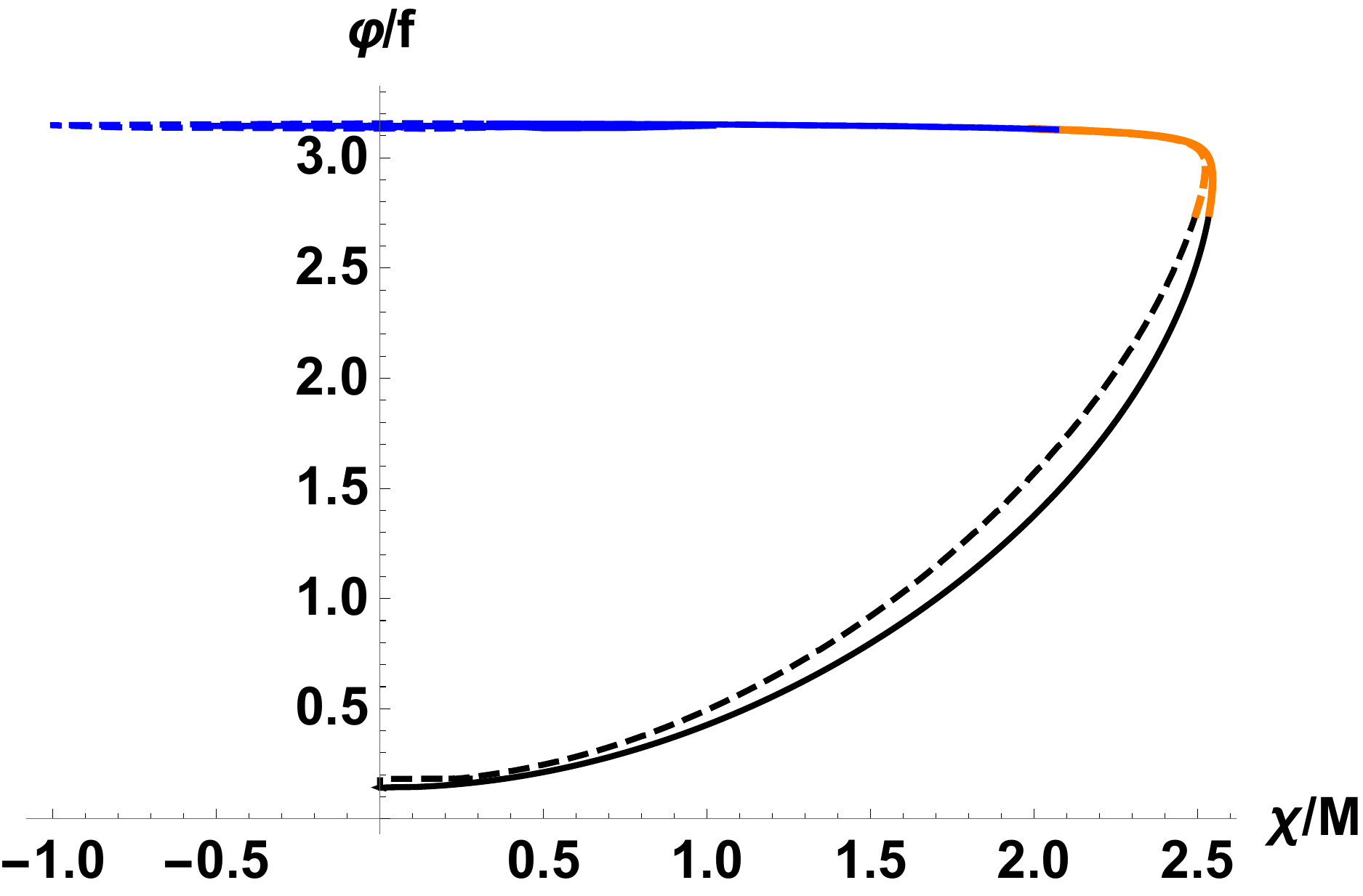}
            \caption{{\small Natural inflation ($f=10$)}}    
            \label{fig:trajectory NI}
        \end{subfigure}
        \caption{Field space trajectories for the SI and NI potentials (with $f=10$), for both the minimal (solid line) and hyperbolic (dashed line) field space geometries. The three portions of the curve indicated in the legend correspond to the phase from the time of the instability to the Hubble crossing at $N=N_{*}$ of the CMB pivot scale; the phase of inflation from $N_{*}$ to the time at which inflation ends at $N=N_{\rm end}$; and the phase after this instant obtained by continuing the integration for a few more e-folds, ignoring any other physical effects. We use the representative value $\Delta N_{\rm pivot}\equiv N_{\rm end}-N_{*}=55$.}
        \label{fig:trajectories}
    \end{figure*}

To gain a qualitative understanding of the second phase of inflation following the geometrical destabilization, we begin by displaying some representative field space trajectories. We show these for the SI and NI potentials in fig.\ \ref{fig:trajectories} (taking the NI scale $f=10$). We have displayed each curve as divided into three portions: (1) the first part starts at the time of the instability and ends at time $N=N_{*}$ when perturbations of the CMB pivot scale size exit the Hubble radius; (2) the second is the phase of inflation that goes from $N_{*}$ to the time at which inflation ends at $N=N_{\rm end}$ by slow-roll violation (i.e.\ when $\epsilon=1$), and corresponds roughly to the range of field values, and hence the part of the potential, that can be probed via cosmological and astrophysical observations; (3) the third phase shows how the curve continues for a few more e-folds after the end of inflation. The curve of phase (1) is of course uncertain in its initial part as a field trajectory cannot be defined immediately following the geometrical destabilization. Similarly phase (3) is simply a qualitative representation of how the fields settle down into the stable minimum of the potential, as we expect other physical effects to become important after inflation ends.

\begin{figure*}
        \centering
        \begin{subfigure}[b]{0.475\textwidth}
            \centering
            \includegraphics[width=\textwidth]{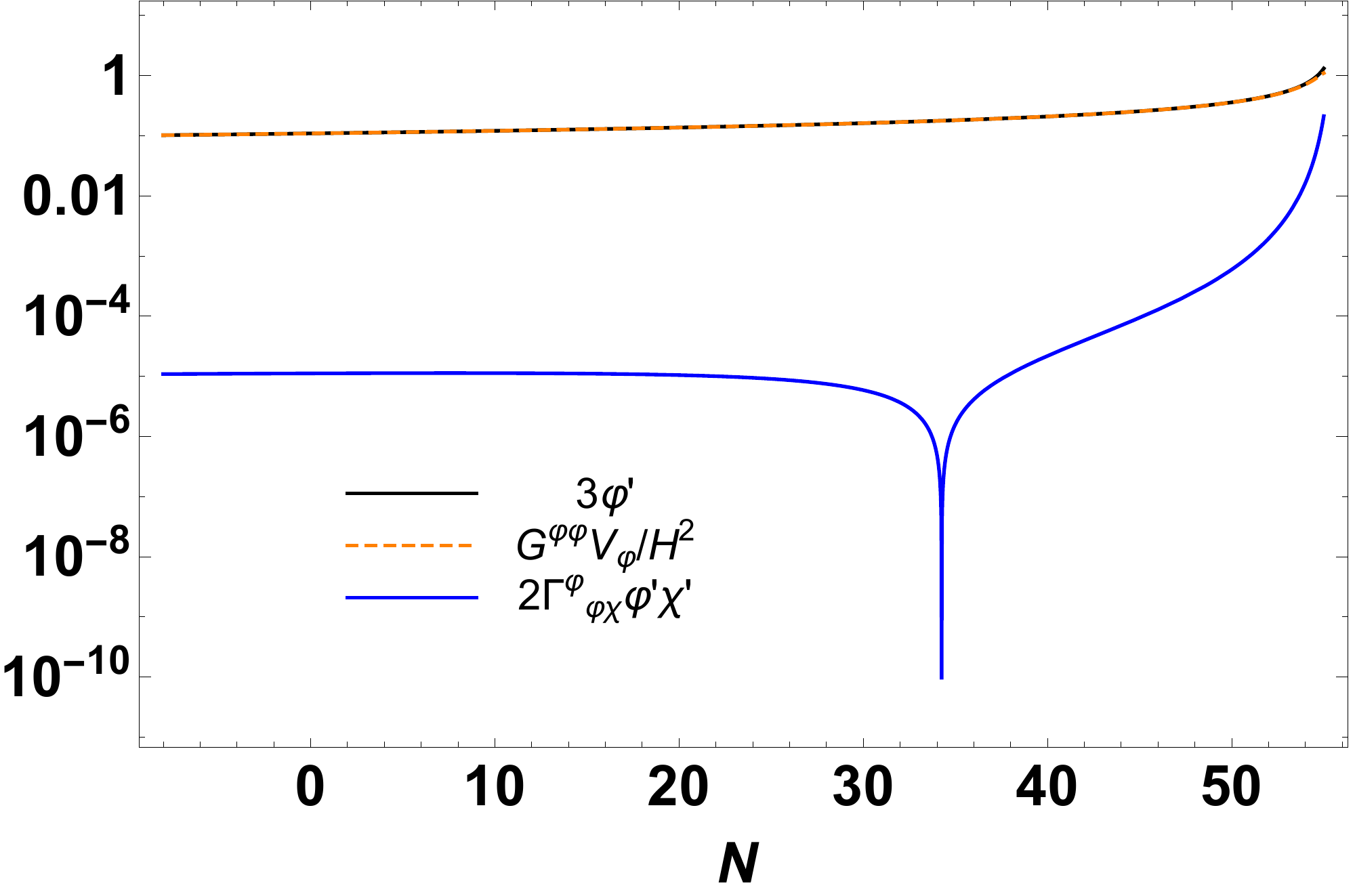}
            \caption{{\small Terms in the equation of motion for $\vphi$ in the minimal geometry.}}    
            \label{figa}
        \end{subfigure}
        \hfill
        \begin{subfigure}[b]{0.475\textwidth}  
            \centering 
            \includegraphics[width=\textwidth]{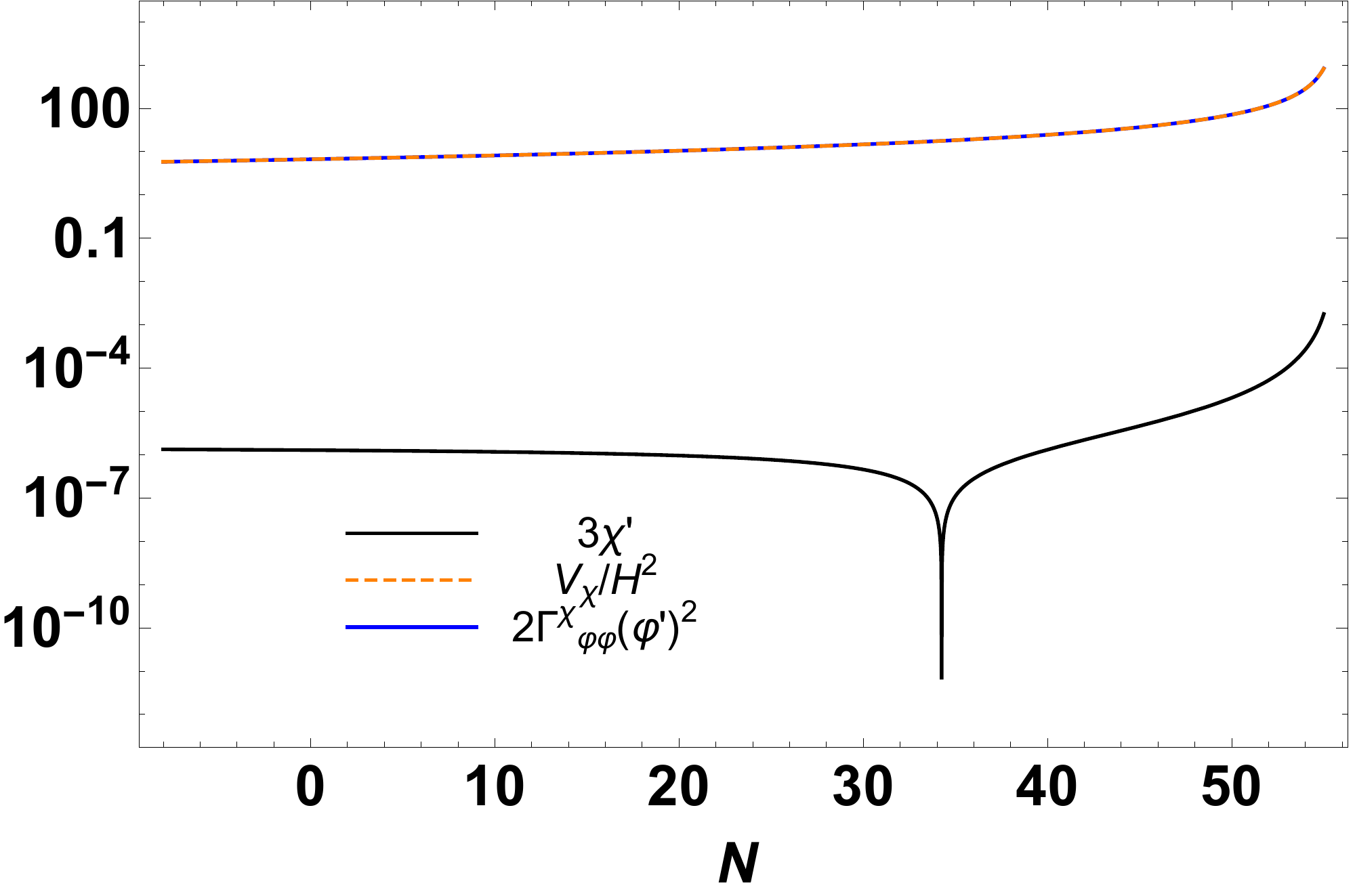}
            \caption{{\small Terms in the equation of motion for $\chi$ in the minimal geometry.}}    
            \label{figb}
        \end{subfigure}
        \vskip\baselineskip
        \begin{subfigure}[b]{0.475\textwidth}   
            \centering 
            \includegraphics[width=\textwidth]{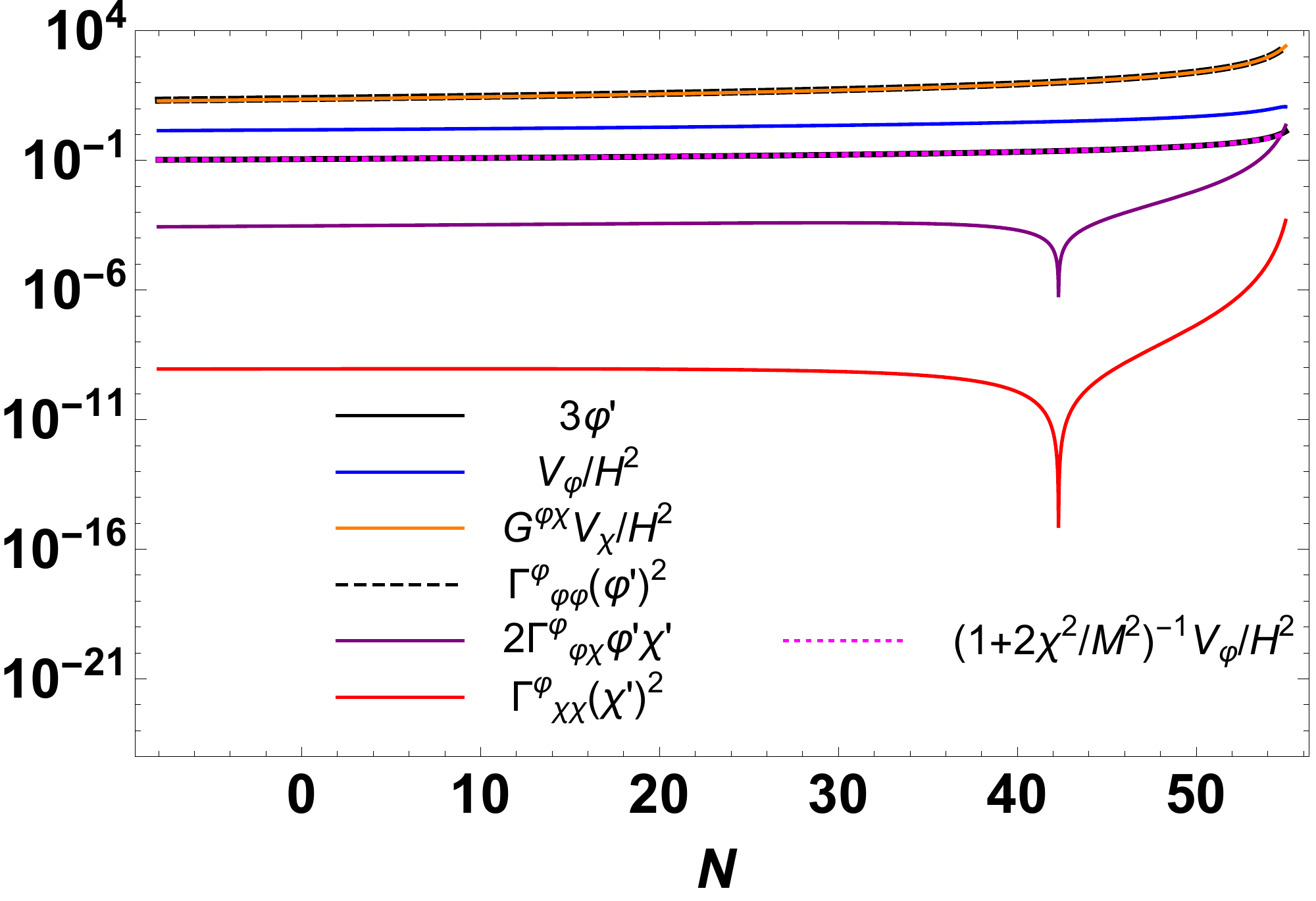}
            \caption{{\small Terms in the equation of motion for $\vphi$ in the hyperbolic geometry.}}    
            \label{figc}
        \end{subfigure}
        \quad
        \begin{subfigure}[b]{0.475\textwidth}   
            \centering 
            \includegraphics[width=\textwidth]{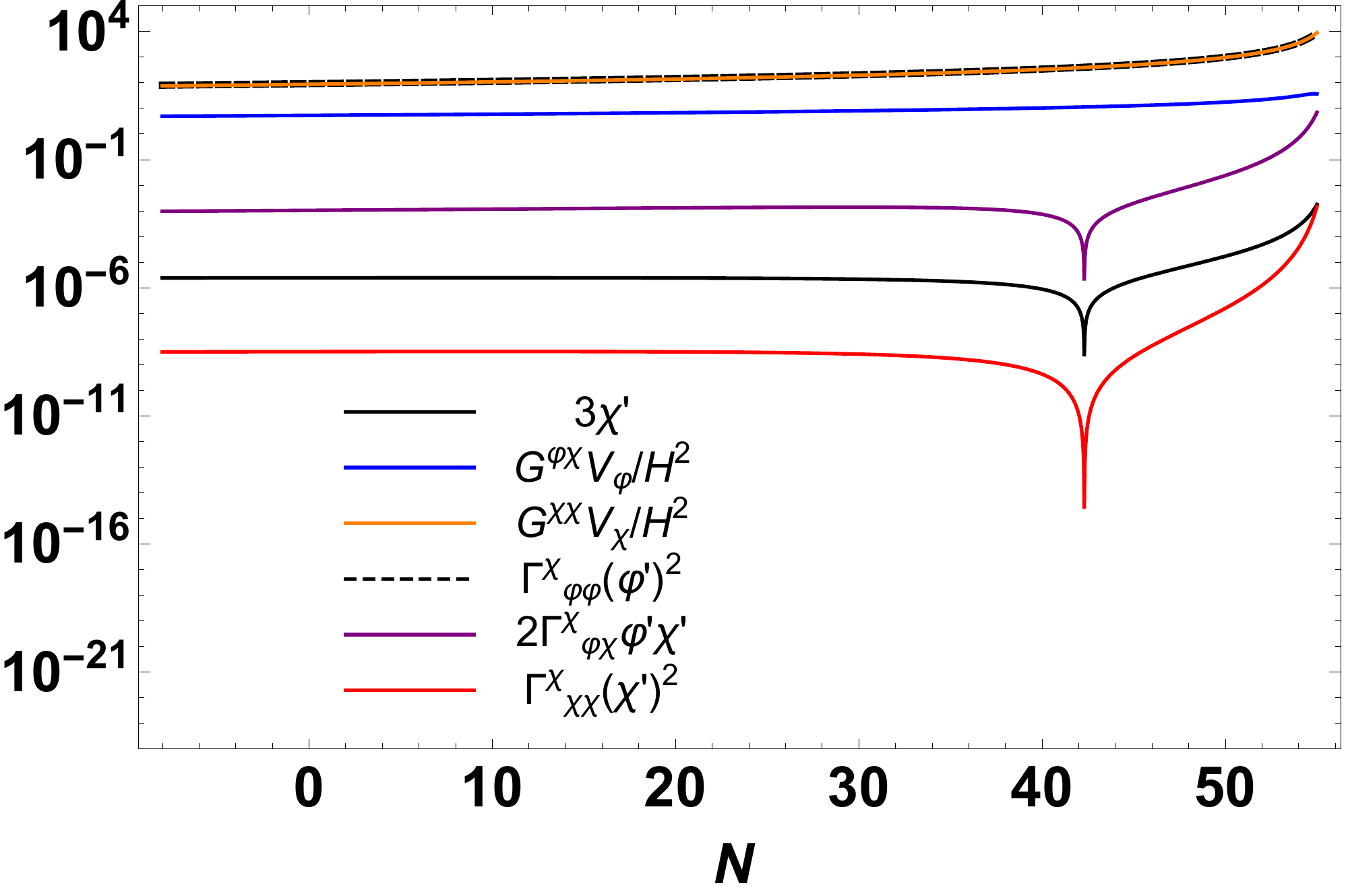}
            \caption{{\small Terms in the equation of motion for $\chi$ in the hyperbolic geometry.}}    
            \label{figd}
        \end{subfigure}
        \caption{Relative contributions of the absolute values of the different terms in the equations of motion for the scalar fields $\vphi$ (left) and $\chi$ (right), for the minimal geometry (top) and the hyperbolic geometry (bottom). The model is Natural Inflation with $f=10$ and $M=10^{-3}$, and the plots show the last 63 e-folds of inflation. One can explicitly check that the terms dominating the dynamics are the ones described in the main text. We made use of derivatives with respect to the number of e-folds, denoted by a prime.} 
        \label{fig:understanding-background}
    \end{figure*}

One important observation is that, for each inflaton potential, the trajectories obtained with the minimal and hyperbolic field spaces are very nearly the same. Although we only display two cases, we have checked that the same conclusion applies for all the models we have studied, and we will prove this feature analytically below. Another important feature, not visible in fig.\ \ref{fig:trajectories}, is that the sidetracked phase of inflation last in general very long, comparatively much longer than along the (unstable) single-field trajectory lying at $
\chi=0$: taking the minimal model for concreteness, while there are 170 e-folds of inflation (respectively 531) left along $\chi=0$ starting from the critical point in SI (respectively in NI), the corresponding sidetracked phase lasts about 770 e-folds (respectively about 2630). 
Eventually, the velocity of $\chi$ is negligible compared to the one of $\phi$ (see fig.\ \ref{fig:understanding-background}), and with $\chi$ of order $M$, it is straightforward to see that $\chi$ gives a negligible contribution to the total potential energy. The simple intuitive picture that emerges from these observations is that the sidetracked phase of inflation is supported by a slowly-varying inflaton field $\vphi$, slowed down on its potential due its non-canonical normalization provided by the almost constant non-zero value of $\chi$.

\subsection{Minimal geometry}

Let us determine the conditions under which this can be realized, considering first the minimal model. With the field space metric \eqref{eq:minimalmetric}, the scalar fields' equations of motion \eqref{eom-scalars} then take the form:
\begin{eqnarray}
\ddot{\vphi}+3H \dot{\vphi}+\frac{4 \chi}{M^2 \left(1+\frac{2\chi^2}{M^2} \right)} \dot{\chi} \dot{\vphi}+\frac{V_{,\vphi}}{1+ \frac{2\chi^2}{M^2}}=0\,, \label{eq-phi-minimal} \\
\ddot{\chi}+3 H \dot{\chi}-2 \frac{\dot{\vphi}^2}{M^2} \chi+V_{,\chi}=0\,, \label{eq-chi-minimal}
\end{eqnarray}
where remember that we make use of the potential \eqref{eq:potential1}, so that $V_{,\chi}=m_h^2 \chi$. A non-zero approximately constant $\chi$ can only be a solution of Eq.~\eqref{eq-chi-minimal} provided that 
\beq
2\frac{\dot \vphi^2}{M^2} \simeq m_h^2,
\label{velocity-phi-1}
\eeq
expressing the almost cancellation between the repulsive force originating from the field space geometry and the one from the potential. The fact that this relation holds, and that the last two terms in \eqref{eq-chi-minimal} completely dominate the equation of motion of $\chi$, can be seen in Fig.\ \ref{figb} for the representative example of NI with $f=10$.  As we previously said, this is in sharp contrast with a field slowly rolling down its potential, for which the dominating terms would be the Hubble friction term and the gradient of the potential. This standard situation is at play however for the inflaton $\vphi$, as can be checked in Fig.\ \ref{figa}, where the third term in \eqref{eq-phi-minimal}, originating from the non-standard field space metric, and suppressed by the velocity of $\chi$, is shown to be negligible in the dynamics. The inflaton field therefore approximately verifies
\beq
3H \dot{\vphi} \simeq -\frac{V_{,\vphi}}{1+ \frac{2\chi^2}{M^2}}\,,
\label{velocity-phi-2}
\eeq
showing clearly how the non-standard normalization of $\vphi$ generates an effective flattened potential compared to the single-field case. Obviously, the agreement between the two expressions \eqref{velocity-phi-1} and \eqref{velocity-phi-2} determines the yet unknown value of the field $\chi$ that enables to support the sidetracked phase, such that:
\beq
1+ \frac{2\chi^2}{M^2} \simeq \sqrt{\frac23} \frac{ \Mp |V_{,\vphi}|}{m_h M\sqrt{V(\vphi)} }\,.\label{chi-phi}
\eeq
Here, we used that $3 H^2 \Mp^2 \simeq \Lambda^4{\mathcal{V}}(\varphi) \equiv V(\vphi)$ to explicitly express that Eq.~\eqref{chi-phi} fixes $\chi$ as a function of $\vphi$. As all the approximate relations given in this section, one can check that the above relation is verified to a very good accuracy, determining $\chi$ to a few $0.1\%$ in NI with $f=10$ for instance. To have a better understanding of the order of magnitude of $\chi$ along sidetracked inflation, one can rewrite Eq.~\eqref{chi-phi} as 
\beq
1+ \frac{2\chi^2}{M^2} \simeq \sqrt{2}  \left(\Mp\frac{ |V_{,\vphi}| }{V(\vphi)} \right)\left(\frac{1}{m_h/H}\right)\left( \frac{\Mp}{M}\right)\,,
\label{chi-phi-2}
\eeq
where the first two terms in parentheses are small, due to the flatness of the inflaton potential and the heavy bare mass of $\chi$, while the last term is enhanced by the hierarchy between the curvature and the Planck scale. It is hard therefore to conclude in general about the amplitude of $\chi$. As a very rough estimate though, one can assume that the first two terms have a similar order of magnitude than at the critical time, despite the very long duration of the sidetracked phase\footnote{This holds for the potentials we have studied, but it would not necessarily be true for potentials whose shape is vastly different in the sidetracked phase and around the critical time. The values at which $\chi$ is stabilized could then differ from $M$ by a large amount, but our analysis and our analytical estimates would still apply in that case.}. Together with Eq.~\eqref{eq:epscritical}, and the slow-roll expression $\epsilon_c \simeq \Mp^2/2 \left(V_{,\vphi}/V \right)^2_c $, one therefore concludes that the left hand side of Eq.~\eqref{chi-phi} is of order one, \textit{i.e.} that $\chi$ is stabilized in the sidetracked phase at $\chi$ of order $M$, as announced. Now that $\chi$ is known as a function of $\vphi$, one can of course check the consistency of the approximations that we have performed. In particular, one can determine the velocity of $\chi$ as
\beq
\frac{\dot \chi}{\dot \vphi}\simeq \frac{{\rm sign}(V_{,\vphi})}{2 \sqrt{2}} \frac{H}{m_h} \frac{M}{\chi} \left[\Mp^2 \frac{V_{,\vphi \vphi}}{V}-\frac{\Mp^2}{2} \left(\frac{V_{,\vphi}}{V} \right)^2 \right]\,,
\label{velocity-chi}
\eeq
which shows that it is indeed suppressed compared to the one of $\vphi$, by $H/m_h \ll 1$, and by the flatness of the inflaton potential.\\

\subsection{Hyperbolic geometry}

We now turn to the hyperbolic field space, whose scalar fields' equations of motion can be out in the form:
\begin{eqnarray}
\hspace{-1.0cm}&&\ddot{\vphi}+3H \dot{\vphi}
+ \frac{4 \chi}{M^2} \dot{\chi} \dot{\vphi}+
\frac{\sqrt{2}}{M} \dot{\chi^2}
+\sqrt{2} \frac{\chi}{M} \left[2 \frac{\dot{\vphi}^2}{M^2} \chi-V_{,\chi}  \right]
+V_{,\vphi}=0 \label{eq-phi-HYP} \\
\hspace{-1.0cm}&&\ddot{\chi}+3 H \dot{\chi}
-\frac{2 \chi}{M^2}\dot{\chi}^2-4\sqrt{2} \frac{\chi^2}{M^3}\dot{\vphi} \dot{\chi}
+\left(1+2\frac{\chi^2}{M^2} \right) \left[-2  \frac{\dot{\vphi}^2}{M^2} \chi +V_{,\chi} \right]-\sqrt{2} \frac{\chi}{M} V_{,\vphi}=0 \,.\label{eq-chi-HYP}
\end{eqnarray}
As the field space metric \eqref{eq:hyperbolicmetric} is non-diagonal, equations \eqref{eq-phi-HYP}-\eqref{eq-chi-HYP} are more complicated than their minimal counterparts \eqref{eq-phi-minimal}-\eqref{eq-chi-minimal}. However, we will show that all the approximate relations we have derived above for the minimal model still hold in this seemingly more intricate case, and that the two dynamics are similar, something we have already noted by looking at the field space trajectories in fig.~\ref{fig:trajectories}. 

We start again by looking for an approximately constant $\chi$ providing a non-trivial solution of Eq.~\eqref{eq-chi-HYP}. It can exist provided now that
\beq
2\frac{\dot \vphi^2}{M^2} \simeq m_h^2 -\sqrt{2} V_{,\vphi}/M \left(1+\frac{2\chi^2}{M^2} \right)^{-1}\,,
\label{velocity-phi-1-HYP}
\eeq
where the last term is new compared to the minimal case. One can check that it is subdominant compared to the first term on the right hand side, although not always negligible. In NI with $f=10$ for instance, its value diminishes from about 15 \% of the first term in the bulk of the sidetracked phase to a few percent of it in the last 60 e-folds. The fact that it is subdominant can be understood using the same back of the enveloppe estimates as we have used below Eq.~\eqref{chi-phi}, showing that it is smaller than the first term by at least the ratio $H/m_h$. In fig.~\ref{figd}, we display the relative contributions of the various terms in Eq.~\eqref{eq-chi-HYP}, showing that the latter is indeed dominated by the cancellation between the two terms in brackets in \eqref{eq-chi-HYP}, so that Eq.~\eqref{velocity-phi-1} approximately holds, like in the minimal case. For simplicity, we do not display the even more precise cancellation between the last three `forces', which gives the relation \eqref{velocity-phi-1-HYP}. However, this refined estimate is important to understand the dynamics of the inflaton. Indeed, by using it in the equation of motion for $\vphi$, one can see that the last three terms in \eqref{eq-phi-HYP} sum up to $\simeq V_{,\vphi}/\left(1+\frac{2\chi^2}{M^2}\right)$. As the third and fourth terms, involving the velocity of $\chi$, are consistently negligible (see fig.~\ref{figc}), one deduces that the inflaton field verifies, like in the minimal case, the simple equation \eqref{velocity-phi-2}, as shown in fig.~\ref{figc} by the superposition of the black and dotted magenta lines.

Like in the minimal case, the agreement between the two expressions \eqref{velocity-phi-1-HYP} and \eqref{velocity-phi-2} of $\dot \vphi$ determine $\chi$ as a function of $\vphi$. Because of the subdominant second term in \eqref{velocity-phi-1-HYP}, $1+ \frac{2\chi^2}{M^2}$ now verifies a quadratic equation, whose solution is straightforward to write down, but that we will not need in the following, and that is not particularly illuminating. At leading order, one can thus simply employ expressions \eqref{chi-phi}-\eqref{chi-phi-2}.

\subsection{Summary and effective single-field theory for the background}
\label{single-field}

Let us summarize the main findings above. Despite small and understood differences, the background dynamics of sidetracked inflation in the minimal and in the hyperbolic field spaces are similar, and can be summarized at leading order by the simple equations \eqref{velocity-phi-2} and \eqref{chi-phi}, expressing: 1) the rolling of the inflaton $\vphi$ on its potential, further slowed-down through its interactions with the accompanying scalar field $\chi$, giving it more inertia.  2) the fact that the dynamics of the latter is being completely fixed by the inflaton, as a result of the competition between the force originating from the field space geometry and the one from the quadratic potential of $\chi$. The background dynamics can therefore be reformulated in terms of $\vphi$ uniquely, which sheds an interesting light on sidetracked inflation. In table \ref{tab:comparison}, we collect a number of useful relations that derive easily from the equations above, that we compare to their counterparts in standard single-field slow-roll inflation. 
\setlength{\tabcolsep}{8pt}
\renewcommand{\arraystretch}{2}
\begin{table}[h!]
\centering
\begin{tabular}{|l|l|l|l|l|l|}
\hline
Functions & Sidetracked inflation & Slow-roll inflation  \\
\hline
\hline
$\frac{{\rm d} \vphi}{ {\rm d} N} \simeq$&$
\left\{
\begin{array}{ll}
-\sqrt{\frac32} \Mp M m_h/\sqrt{V} \,{\rm sign}(V_{,\vphi})\\
-\frac{M}{\sqrt{2}} \, \frac{m_h}{H(\vphi)} \,{\rm sign}(V_{,\vphi}) &
\end{array}\right.$ & $ - \Mp^2 \frac{V_{,\vphi}}{V}$   \\
\hline
$\epsilon \simeq$&$
\left\{
\begin{array}{ll}
\frac12 \sqrt{\frac32} \Mp M m_h |V_{,\vphi}|/V^{3/2}\\
\frac{1}{2 \sqrt{2}}  M \frac{|V_{,\vphi}|}{V}\,\frac{m_h}{H(\vphi)}&
\end{array}\right.$ & $ \frac{\Mp^2}{2} \left( \frac{V_{,\vphi}}{V} \right)^2$  \\
\hline
$\eta=\frac{{\rm d}\,{\rm ln} \,\epsilon}{ {\rm d} N} \simeq$&$
\left\{
\begin{array}{ll}
 3 \epsilon-\sqrt{\frac{3}{2}}\Mp M m_h \frac{V_{,\vphi \vphi}}{|V_{, \vphi}| V^{1/2}}\\
 3 \epsilon-\frac{M}{\sqrt{2}} \frac{V_{,\vphi \vphi}}{|V_{, \vphi}|}\,\frac{m_h}{H(\vphi)}&
\end{array}\right.$
 & $4 \epsilon-2 \Mp^2 \frac{V_{,\vphi \vphi}}{V} $ \\
 \hline
$N-N_{{\rm ini}} \simeq$&$
\left\{
\begin{array}{ll}
-\sqrt{\frac23} \frac{1}{\Mp M m_h}  \int_{\vphi_{{\rm ini}}}^{\vphi} {\rm sign}(V_{,\vphi'}) \sqrt{V(\vphi')}    {\rm d} \vphi' \\
 -\frac{\sqrt{2}}{M}  \int_{\vphi_{{\rm ini}}}^{\vphi} {\rm sign}(V_{,\vphi'}) \frac{H(\vphi')}{m_h}   {\rm d} \vphi'    &
\end{array}\right.$
 &$-\frac{1}{\Mp^2} \int_{\vphi_{{\rm ini}}}^{\vphi} \frac{V(\vphi')}{V_{,\vphi'}} {\rm d} \vphi'$\\
\hline
\end{tabular}
\caption{Comparison between sidetracked inflation and slow-roll inflation.}
\label{tab:comparison}
\end{table}

One can see that the dynamics and the functional dependences of the various inflationary parameters on the shape of the potential are very different between standard slow-roll inflation and sidetracked inflation. For the latter, we give each time two equivalent expressions, the first one in terms of the various mass scales $\Mp, M, m_h$ and the potential, and the second one that make appear the ratio $m_h/H(\vphi)$ by using $3 H(\vphi)^2 \Mp^2 \simeq V(\vphi)$. Strictly speaking, one could envisage situations in which $m_h$ has no relationship with the Hubble scale. However, this is not the case in realistic situations, and in our approach and our numerical examples, we took $m_h=10 H_c$. The ratio $m_h/H(\vphi)$ is a dynamical quantity and is larger in sidetracked inflation than at the critical time because of the decrease of the Hubble scale, however, for qualitative estimates, one can think of it as an ${\cal O}(10)$ quantity. The second expressions of $\epsilon$ and $\eta$ make it clear therefore that the overall scale $\Lambda$ of the potential (see Eq.~\eqref{eq:potential1}) is irrelevant for the dynamics, and that only its shape ${\mathcal{V}}(\varphi)$ matters, like in slow-roll inflation. In the latter case, the expressions of $\epsilon$ and $\eta$ indicate the well known fact that the (log) potential should be flat in Planck units. In sidetracked inflation, the corresponding expressions rather indicate that the potential should be flat with respect to the curvature scale $M$, with the requirements:
\beq
M \frac{V_{,\vphi}}{V} \ll 1 \,, \quad M \frac{V_{,\vphi \vphi}}{V_{, \vphi}} \ll 1\,.
\label{conditions-sidetracked}
\eeq
Concretely, this implies that one can have a prolonged phase of inflation supported by potentials that would be too steep to allow standard slow-roll inflation. This is clearly visible in fig.\ \ref{fig:trajectories} for instance, where inflation arises on the Starobinsky potential with $\vphi \sim \Mp$, and on the natural inflation potential with $\vphi \sim f$.\footnote{Note that in the last $55$ e-folds in NI, inflation does not arise near the top of the hill, but rather near the minimum of the potential, so that it is approximately quadratic.} Additionally, the second criteria in Eq.~\eqref{conditions-sidetracked} does not involve $\Mp^2 V_{,\vphi \vphi}/V$ in the standard way, but rather $M V_{,\vphi \vphi}/V_{, \vphi}$. In NI near the top of the hill for instance, this translates into $M \ll \vphi \ll f$ rather than the standard requirement that $f \gg \Mp$. More generally, it is interesting that inflation can occur in the presence of steep potentials, both for the inflaton $\vphi$ and its parter $\chi$. This amusing feature of sidetracked inflation thus offers an interesting playground for future work, notably with respect to the eta problem. Note however that it comes at the expense of the appearance of the sub-Planckian curvature scale $M$, and is likely that a proper treatment of naturalness issues in quantum field theory renders the situation more intricate.


\section{Cosmological fluctuations and power spectrum} \label{sec:powerspectrum}

\subsection{Numerical methods}

To determine the properties of the linear cosmological fluctuations generated in sidetracked inflation, we solve the coupled equations \eqref{pert} in the natural coordinate basis $(\vphi,\chi)$, choosing Bunch-Davies initial conditions. We follow a by now standard procedure to implement numerically the quantization of the system (see \textit{e.g.} references \cite{Salopek:1988qh,vanTent:2002az,GrootNibbelink:2001qt,Tsujikawa:2002qx,Weinberg:2008zzc,Achucarro:2010da,Huston:2011fr,McAllister:2012am,Amin:2014eta}): we identify two variables (as we are dealing with two fields) that are independent deep inside the Hubble radius, each corresponding to an independent set of creation and annihilation operators whose effects add incoherently, and solve the system of equations \eqref{pert} two times, each time imposing the Bunch-Davies initial conditions for only one of the independent variables, while setting the other variables to zero initially. One then extracts power spectra by summing the relevant quantities over the two runs. Deep inside the Hubble radius, one can neglect the mass matrix in the action \eqref{S2}, so that identifying a set of independent variables is equivalent to identifying a set of vielbeins for the field space metric $G_{IJ}$, which is straightforward. In practice, we impose initial conditions eight e-folds before Hubble crossing. This is larger than what is sufficient in more conventional circumstances but, as we will see, the strong bending of the trajectory entails a  non-trivial evolution of the fluctuations inside the Hubble radius, and starting the evolution at a later time would give inaccurate results.

As we encountered a highly non-trivial behaviour of the fluctuations, we also used the completely independent so-called transport approach to determine their properties, finding excellent agreement between the two methods. In this powerful approach, the computation of the mode functions themselves is bypassed, and Hamilton's equations are used to determine evolution equations for the correlation functions of the cosmological fluctuations, which are the direct quantities of interest. We refer the interested reader to references \cite{Mulryne:2009kh,Mulryne:2010rp,Dias:2011xy,Seery:2012vj,Anderson:2012em,Mulryne:2013uka} for early theoretical works and \cite{Dias:2015rca,Dias:2016rjq,Mulryne:2016mzv,Seery:2016lko,Ronayne:2017qzn,Butchers:2018hds} for recent improvements and developments of automated tools. In this work, we make use of the publicly available code \texttt{PyTransport 2.0} \cite{Ronayne:2017qzn}, which numerically implements the transport approach to calculate the power spectrum and bispectrum in generic nonlinear sigma models of inflation.

Contrary to the study of the premature end of inflation possibly triggered by the geometrical destabilization \cite{Renaux-Petel:2017dia}, in which it is was important to take into account the uncertainties of the reheating phase, and given the exploratory nature of our study here, we do not attempt to model reheating and simply assume throughout the representative value $\Delta N_{\rm pivot}=55$ for the number of e-folds between Hubble crossing of the CMB pivot scale and the end of inflation, the latter defined by the instant at which $\epsilon=1$.

\subsection{Adiabatic and entropic fluctuations}
\label{fluctuations-adiabatic-entropic}

With the two approaches described above, it is relatively straightforward to determine the power spectra of the curvature and entropic perturbations for any scale, and therefore the scalar spectral index $n_s$ as well as the tensor-to-scalar ratio $r$\footnote{As usual, the tensor fluctuations are decoupled from the scalar sector, and the standard result ${\cal P}_{{\rm t}}(k)=2/\pi^2 H^2_{k=aH}/\Mp^2$ holds.}. However, to gain more insight into the physics of the fluctuations, it is instructive to formulate it in terms of the instantaneous adiabatic/entropic splitting. In particular, from Eq.~\eqref{pert}, one can deduce the coupled adiabatic and entropic equations of motion as
\begin{eqnarray}
\label{Qsigma}
 \ddot{Q}_{\sigma}+3H
 \dot{Q}_{\sigma}+\left(\frac{k^2}{a^2}+m_{\sigma}^2\right)  Q_{\sigma} = \Tdot{\left(2 H \eta_\perp Q_{s}\right)}
-\left(\frac{\dot H}{H}
+\frac{V_{,\sigma}}{\dot \sigma }\right)  2 H \eta_\perp\, Q_{s}\,,
\end{eqnarray}
\begin{eqnarray}
 \ddot{Q}_{s}+ 3H \dot{Q}_{s}+\left(\frac{k^2}{a^2}+m_{s}^2\right)Q_{s}=-2 \dot \sigma \eta_\perp \dot \zeta\,,
\label{delta_s_all_scales}
\end{eqnarray}
where
the adiabatic mass (squared)
$m_{\sigma}^2$ is given by
\beq
\frac{m_\sigma^2}{H^2} \equiv -\frac{3}{2}\eta-\frac14\eta^2-\frac12\epsilon \eta -\frac12\dot \eta/H\,,
\label{msigma}
\eeq
the entropic mass (squared) $m_s^2$ is given by
\begin{equation}  \label{delta_s_all_scales2}
\frac{m_{s}^2}{H^2}\equiv \frac{V_{;s s}}{H^2}+\epsilon R_{\rm fs} \Mp^2 - \eta_{\perp}^2\,,
\end{equation}
and in Eq.~\eqref{delta_s_all_scales}, we employed in the right hand side the comoving curvature perturbation $\zeta$, directly proportional to the adiabatic fluctuation, such that 
\begin{eqnarray}
\zeta =\frac{H}{{\dot \sigma}} Q_{\sigma}\,.
\label{zeta-Qsigma}
\end{eqnarray}
In a symmetric way, note that it is also useful to introduce the rescaled entropic perturbation
\beq
{\cal S}=\frac{H}{{\dot \sigma}} Q_{s}\,.
\label{S-def}
\eeq

As is well known on general grounds, on super-Hubble scales such that $k \ll a H$, there exists a first integral for $Q_{\sigma}$, which can be conveniently rewritten in terms of $\zeta$ and ${\cal S}$ as
\beq
\dot{\zeta} \approx  2 H \etaperp \, {\cal S} \,
\label{zeta-large-scales}
\eeq
(one can check indeed that the large-scale limit of (\ref{Qsigma}) is a consequence of \eqref{zeta-large-scales}). Inserting the latter result
into Eq.~\eqref{delta_s_all_scales}, one finds that on super-Hubble scales, 
\beq
\ddot{Q}_s+3H \dot{Q}_s+\left(m_s^2+4 H^2 \etaperp^2 \right) Q_s \approx 0\,,
\label{Q2-large-scales}
\eeq
which is in agreement with the already given result \eqref{eq:entropicmass}, with $\msh=m_s^2+4 H^2 \etaperp^2$. One can see that the two notions of entropic masses coincide in the case of a geodesic motion with $\etaperp=0$, but that they differ in general, a feature that plays a central role in sidetracked inflation, as we will see.

Complementary to the super-Hubble limit discussed above, it is useful to recast the equations of motion \eqref{Qsigma}-\eqref{delta_s_all_scales} in a form that is more adapted to understand the physics on sub-Hubble scales. By introducing the canonically normalized fields in conformal time $\tau$ (such that ${\rm d}t=a \,{\rm d} \tau$), $v_{\s}=a\, Q_{\s}$ and $v_{s}=a\, Q_s$, for the adiabatic and entropic fluctuations respectively, the equations can be put in the compact form, derived in reference \cite{Langlois:2008mn} in a more general context:
\begin{eqnarray}
v_{\s}''-\xi v_{s}'+\left( k^2-\frac{z''}{z}\right) v_{\s} -\frac{(z \xi)'}{z}v_{s}&=&0\,,
\label{eq_v_sigma}
\\
v_{s}''+\xi  v_{\s}'+\left(k^2- \frac{a''}{a}+a^2 m_s^2\right) v_{s} - \frac{z'}{z} \xi v_{\s}&=&0\,, \quad '={\rm d}/ {\rm d} \tau\,.
\label{eq_v_s}
\end{eqnarray}
In these two equations only, in order not to clutter the text with two many notations, we used primes to denote derivatives with respect to conformal time, whereas other instances in the rest of the text do denote derivatives with respect to the number of e-folds. These equations render it clear that in addition to the scale factor, the only other background quantities affecting the dynamics of fluctuations are 
\beq
z \equiv \frac{a \dot \s}{H}=a \sqrt{2 \epsilon}\,,
\eeq
such that $v_\sigma=z \zeta$, the entropic mass \eqref{delta_s_all_scales2}, and the time-dependent coupling between the adiabatic and entropic fluctuation
\beq
\xi \equiv 2 a H\etaperp\,.
\label{xi}
\eeq
Although detailed predictions of the cosmological fluctuations, even in a single-field context, depend on the precise time evolution of $z$, when the background evolution is close to de Sitter, with $\epsilon, \eta$ and $\dot \eta / ( H \eta)$ all much smaller than unity, one can consider at leading order that time derivatives of $z$ are dominated by the variation of the scale factor, \textit{i.e.} $\frac{1}{z}\frac{{\rm d} z}{{\rm d} \tau} \simeq  \frac{1}{a}\frac{{\rm d} a}{{\rm d} \tau} \simeq -\frac{1}{\tau}$ and $\frac{1}{z}\frac{{\rm d}^2 z}{{\rm d} \tau^2} \simeq  \frac{1}{a}\frac{{\rm d} a}{{\rm d} \tau^2} \simeq \frac{2}{\tau^2}$. This corresponds to situations in which the adiabatic mass \eqref{msigma} is negligible compared to the Hubble scale, which applies in all the cases we have considered. Were the effect of the bending trajectory, \textit{i.e} of the coupling $\xi$, negligible in the dynamics of the fluctuations, this would lead to the well known slow-roll single-field like result $v_{\s\, k}\simeq  \frac{1}{\sqrt{2k}}e^{-ik \tau }\left(1-{i\over k \tau}\right)$, and hence to the standard result 
${\cal P}_{\zeta_k}=\left(H^2/(8 \pi^2 \epsilon)\right)_{k=aH}$
 for the dimensionless power spectrum $ {\cal P}_{\zeta_k}= k^3/(2 \pi^2) P_{\zeta_k}$, where $\langle \zeta_{\boldsymbol{k}_1} \zeta_{\boldsymbol{k}_2} \rangle \equiv (2\pi)^3 \delta(\boldsymbol{k}_1+\boldsymbol{k}_2) {P}_{\zeta_k}$ with $k=|\boldsymbol{k}_1|=|\boldsymbol{k}_2|$. On the contrary, as we have mentioned in section \ref{sec:background}, the trajectory of sidetracked inflation differs strongly from a geodesic, so that it is important to take into account the coupled dynamics between the adiabatic and entropic degrees of freedom. For this, we need to understand the behaviour of the various (related) mass scales $m_s^2, H^2 \etaperp^2$ and $\msh$ compared to the Hubble scale. This is what we do in the following, building on our understanding of the background dynamics in section \ref{sec:background}.

\subsection{Analytical understanding of relevant mass scales}
\label{understanding-parameters}

We have explained in section \ref{sec:background} that $\dot \chi \ll \dot \vphi$ in the sidetracked phase of inflation (see Eq.~\eqref{velocity-chi} for instance). To a very good approximation, one can then consider that the adiabatic vector points in the direction of $\vphi$ only, \textit{i.e.} that $e_\sigma^I \propto (\dot \vphi,0)$ in the natural coordinate basis. With the expressions \eqref{eq:minimalmetric} and \eqref{eq:hyperbolicmetric} of the field space metrics, this leads to
\beq
e_\s^I=\left(\left(1+\frac{2\chi^2}{M^2} \right)^{-1/2},0\right)\,.
\label{esigma-result}
\eeq 
From this, it is straightforward to deduce the form of the entropic unit vector as
\beq
e_s^I =\left\{
\begin{array}{ll}
(0,1) \qquad \qquad \qquad \qquad \qquad \qquad \qquad \qquad \quad  {\rm minimal}\\
\left(-\sqrt{2}\, {\chi \over M} \left(1+\frac{2\chi^2}{M^2} \right)^{-1/2},\left(1+\frac{2\chi^2}{M^2} \right)^{1/2}  \right) \qquad {\rm hyperbolic}&
\end{array}\right.
\label{es-result}
\eeq
where here the two results differ in the minimal and in the hyperbolic model. Contrary to the background properties, the estimation of the various parameters, and the resulting dynamics of the fluctuations, will be rather different for the two geometries, so we treat them separately in what follows. Before that, let us just note the common formal expression of $\etaperp$ that we will use. From its definition in Eq.~\eqref{etaperp}, and the fact that the adiabatic acceleration $\ddot \sigma$ in \eqref{sigmaddot} is negligible during inflation, we deduce that
\beq
\etaperp \simeq 3 \frac{ V_{,s}}{V_{,\s}}
\label{etaperp-potential}
\eeq
to a very good approximation. 

\subsubsection{Minimal geometry}
\label{understanding-minimal}

\begin{figure*}
        \centering
        \begin{subfigure}[b]{0.475\textwidth}
            \centering
            \includegraphics[width=\textwidth]{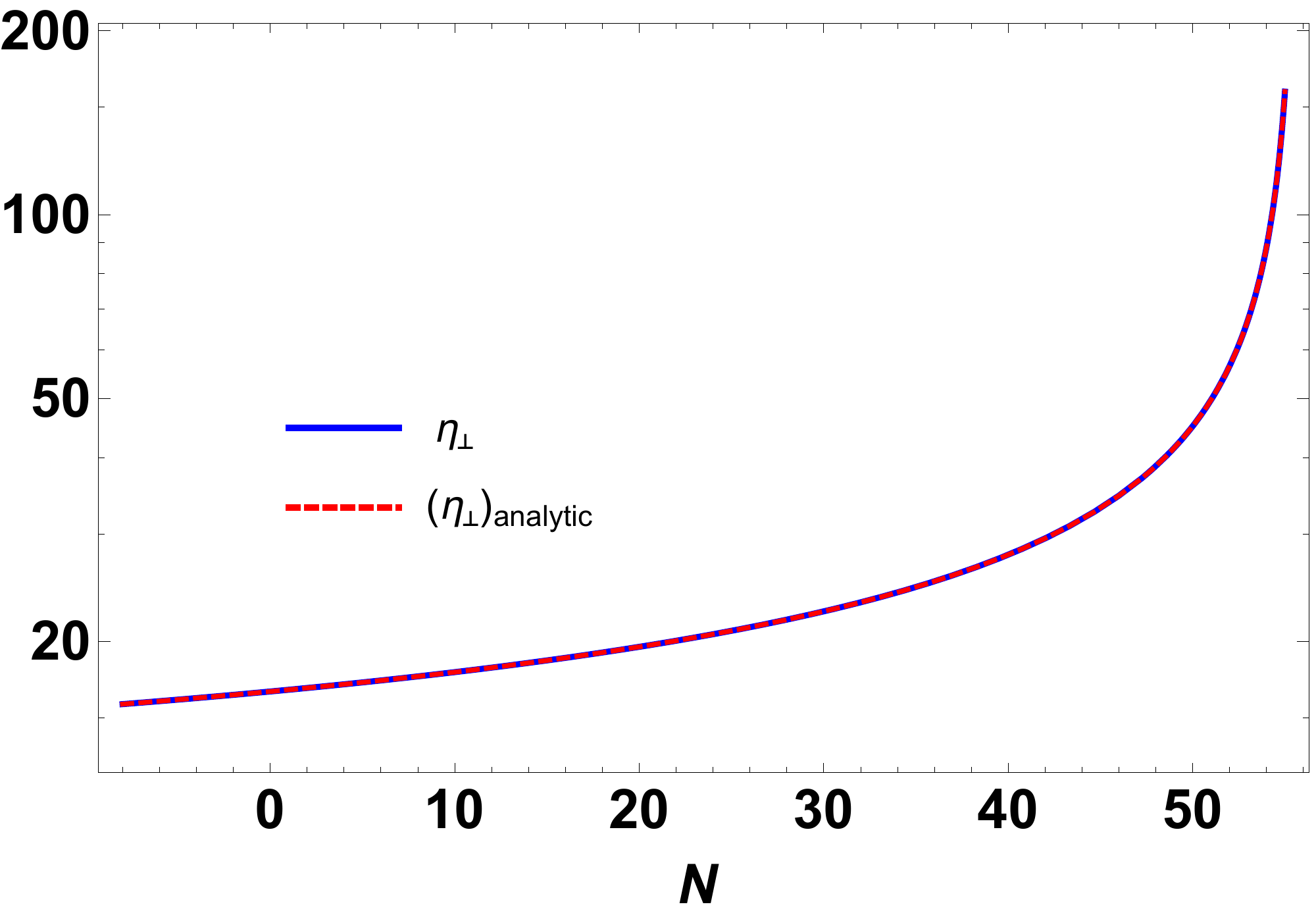}
            \caption{{\small Starobinsky inflation}}    
            \label{fig:etaperp-minimal-SI}
        \end{subfigure}
        \hfill
        \begin{subfigure}[b]{0.475\textwidth}  
            \centering 
            \includegraphics[width=\textwidth]{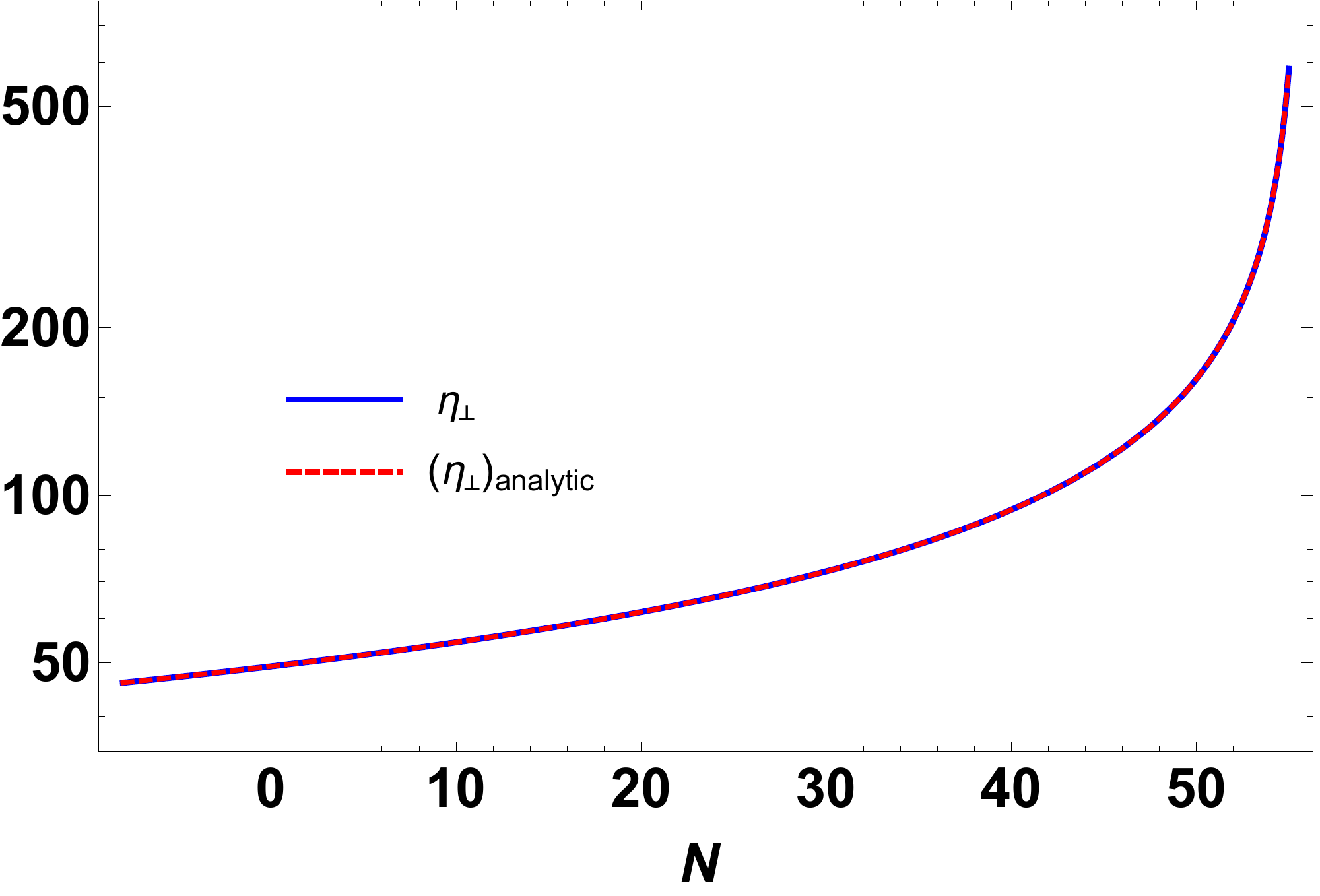}
            \caption{{\small Natural inflation ($f=10$)}}    
            \label{fig:etaperp-minimal-NI}
        \end{subfigure}
        \caption{Fully numerical and analytical result \eqref{etaperp-minimal} for the absolute value of $\etaperp$ in the last 63 e-folds of sidetracked inflation in the minimal geometry, for the potentials of SI (left) and NI, $f=10$ (right).}
        \label{fig:etaperp}
    \end{figure*}

From the expression \eqref{etaperp-potential} of $\etaperp$, together with \eqref{esigma-result}-\eqref{es-result}, one finds $\etaperp \simeq 3 V_{,\chi}/V_{,\vphi} \times \left(1+\frac{2\chi^2}{M^2} \right)^{1/2} $. Using Eq.~\eqref{velocity-phi-2} to express $V_{,\vphi}$ in terms of $\dot \vphi$, and the simple form \eqref{velocity-phi-2} of the latter, one then finds
\beq
\etaperp \simeq \frac{m_h}{H} \frac{ {\sqrt{2}\,\chi \over M}}{\left(1+\frac{2 \chi^2}{M^2} \right)^{1/2}}\times {\rm sign}(V_{,\vphi})\,.
\label{etaperp-minimal}
\eeq
One can of course express this result in terms of $\vphi$ only, by using Eq.~\eqref{chi-phi}. However, Eq.~\eqref{etaperp-minimal} is more instructive: as $\chi={\cal O}(M)$, the second multiplicative factor in \eqref{etaperp-minimal} is of order one. Let us also recall that, as a consequence of the decrease of the Hubble scale, $m_h/H > m_h/H_c=10$. One therefore reaches the conclusion that, as announced, the bending of the background trajectory, as measured by $\etaperp$, is large in sidetracked inflation in the minimal geometry. Intuitively, one can understand this result: as the sidetracked phase stems from the competition (and neutralization) of the effects from the geometry and from the potential, it is not surprising that 1) the resulting trajectory deviates from a geodesic, and 2) that it does so by an amount related to how massive the field $\chi$ is. The fully numerical result for $\etaperp$, as well as the analytical estimate \eqref{etaperp-minimal}, are shown in Fig.\ \ref{fig:etaperp} for the two examples of SI and NI with $f=10$.  Note that in each case the agreement is excellent, with a relative accuracy of order $10^{-5}$ and $10^{-7}$ (not visible in the figures).

We now determine expressions for the two entropic masses. With \eqref{es-result}, it is straightforward to find that $V_{;ss} \simeq m_h^2$. For the geometrical contribution, we use Eq.~\eqref{velocity-phi-1} to find that $\epsilon \simeq \left(1+\frac{2 \chi^2}{M^2} \right) \frac{m_h^2}{H^2} \frac{M^2}{4 \Mp^2}$, and hence, with the expression \eqref{R-minimal} of the field space curvature, that $\epsilon R_{\rm fs} \Mp^2 \simeq - \frac{m_h^2}{H^2} \left(1+\frac{2 \chi^2}{M^2} \right)^{-1}$. The three contributions --- the Hessian, the bending, and the geometrical ones --- to 
$\msh/H^2$
 (Eq.~\eqref{eq:entropicmass}) and $m_s^2/H^2$ (Eq.~\eqref{delta_s_all_scales2}) are therefore individually large, each of order of the large mass $m_h^2/H^2$. However, by summing them, we find
\begin{eqnarray}
\frac{\msh}{H^2} &\simeq&  \frac{4 m_h^2}{H^2} \frac{ {2\chi^2 \over M^2}}{\left(1+\frac{2 \chi^2}{M^2} \right)} \gg 1\,\\  
\frac{m_s^2}{H^2} &\simeq &0\,.
\end{eqnarray}
\begin{figure*}
        \centering
        \begin{subfigure}[b]{0.475\textwidth}
            \centering
            \includegraphics[width=\textwidth]{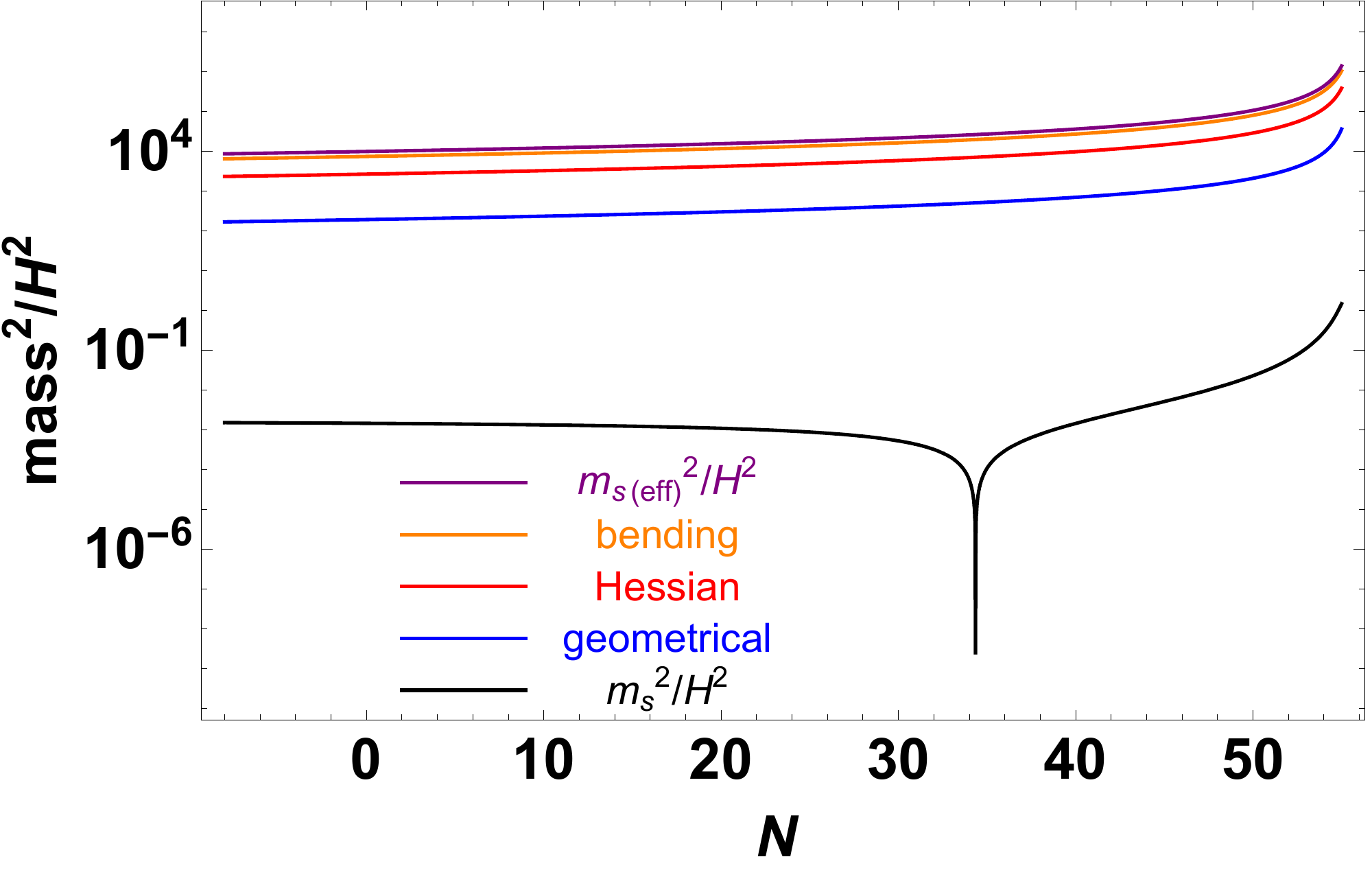}
            \caption{{\small  Natural inflation ($f=10$)}}    
            \label{fig:masses-NI}
        \end{subfigure}
        \hfill
        \begin{subfigure}[b]{0.475\textwidth}  
            \centering 
            \includegraphics[width=\textwidth]{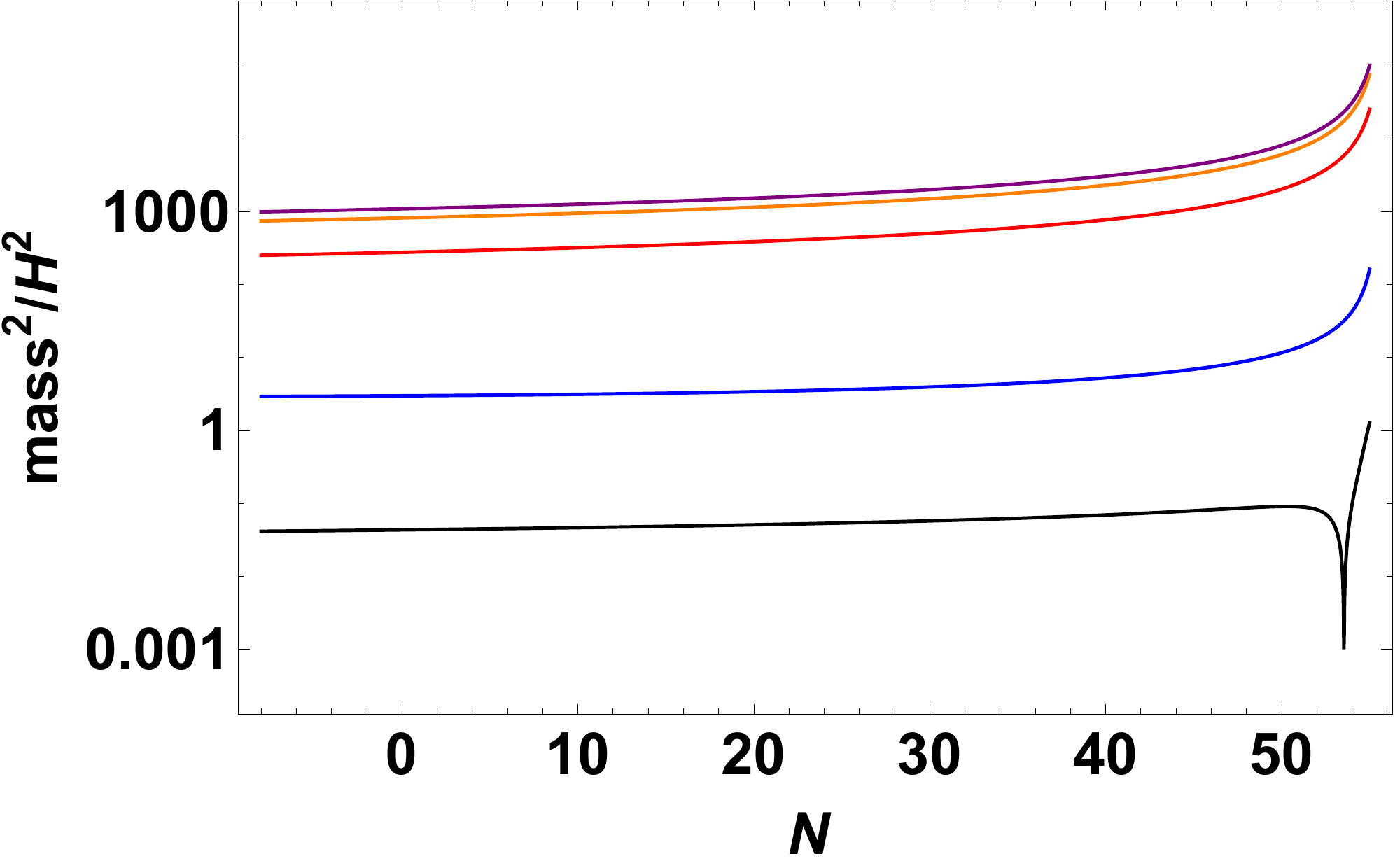}
            \caption{{\small Starobinsky inflation }}    
            \label{fig:masses-SI}
        \end{subfigure}
        \caption{Absolute values of $\msh/H^2$, its three contributions (see Eq.~\eqref{eq:entropicmass}), as well as $m_s^2/H^2$, for 
        our two representative examples of SI and NI with $f=10$, in the minimal geometry.  The plots show the last 63 e-folds of inflation. One can check that $m_s^2/H^2 \ll 1$, and one actually has $m_s^2/H^2<0$ around Hubble crossing.}
        \label{fig:masses}
    \end{figure*}
In other words, while the effective mass $\msh$, which dictates the evolution of the entropic fluctuations on super-Hubble scales according to Eq.~\eqref{Qs-super-Hubble}, is much larger than the Hubble rate, the various large contributions to the entropic mass $m_s^2$, which is important for the sub-Hubble dynamics (see Eq.~\eqref{eq_v_s}), cancel, at least in our analytical treatment. This is confirmed numerically, as we can see in Fig.\ \ref{fig:masses}, where we plot (the absolute values of) $\msh/H^2$, its three contributions, as well as $m_s^2/H^2$, for the two representative examples of SI and NI with $f=10$. In both cases, we indeed find that $m_s^2 \ll H^2$, and we also observe that $m_s^2$ is negative. Naturally, we could keep track of subleading terms in our analytical treatment, beginning with the correction to $e_\sigma^I$ induced by the non-zero velocity of $\chi$, that is suppressed by $10^{-5}$ compared to the one of $\chi$ in these two examples (see Fig.\ \ref{fig:understanding-background} and Eq.~\eqref{velocity-chi}). Although we did not attempt it, we expect it would reproduce the small value of $m_s^2$, that is indeed suppressed by $10^{-5}$ compared to $\msh$.

\subsubsection{Hyperbolic geometry}
\label{understanding-hyper}
 
As for the hyperbolic geometry, we can follow the same steps as in the minimal one, which used the estimate \eqref{velocity-phi-1} for $\dot \vphi$ in particular, finding again large individual contributions to $\msh/H^2$, of order $m_h^2/H^2$, and a vanishing $m_s^2/H^2$. While these results are indeed quantitatively correct for each of the various contributions, and for $\msh$, this result is misleading for $m_s^2$, the reason being that the subleading correction to $\dot \vphi$ in the refined expression \eqref{velocity-phi-1-HYP} has to be taken into account when the leading order result vanishes. In what follows, we give both the leading-order expressions of the various quantities involved, making use of \eqref{velocity-phi-1}, as well as refined ones, making use of \eqref{velocity-phi-1-HYP}-\eqref{velocity-phi-2}. We do this in particular because the magnitude of these parameters is most easily understood with the leading-order estimates.

For $\etaperp$, starting from \eqref{etaperp-potential}, one then finds\footnote{The complete leading-order result has an additional contribution $-3 \sqrt{2} \chi/M$, which however exceeds the accuracy of this calculation.}
\beq
\etaperp \simeq \left\{
\begin{array}{ll}
 {\sqrt{2}\,\chi \over M}\,  \frac{m_h}{H} \,{\rm sign}(V_{,\vphi}) \qquad   {\rm leading-order}\\
\frac{2 {\chi \over M}}{1+\frac{2\chi^2}{M^2} } \frac{V_{,\vphi}}{3 H^2 M} \qquad \qquad \,\,  {\rm refined}&
\end{array}\right.
\label{etaperp-hyper}
\eeq
where the leading-order (respectively the refined) estimate is accurate to the level $10^{-2}$ (respectively $10^{-4}$) for the NI potential with $f=10$ for instance. As announced, one finds a large bending, like in the minimal model, and the same remarks as in that case apply regarding an intuitive picture of its origin.

\begin{figure*}
        \centering
        \begin{subfigure}[b]{0.475\textwidth}
            \centering
            \includegraphics[width=\textwidth]{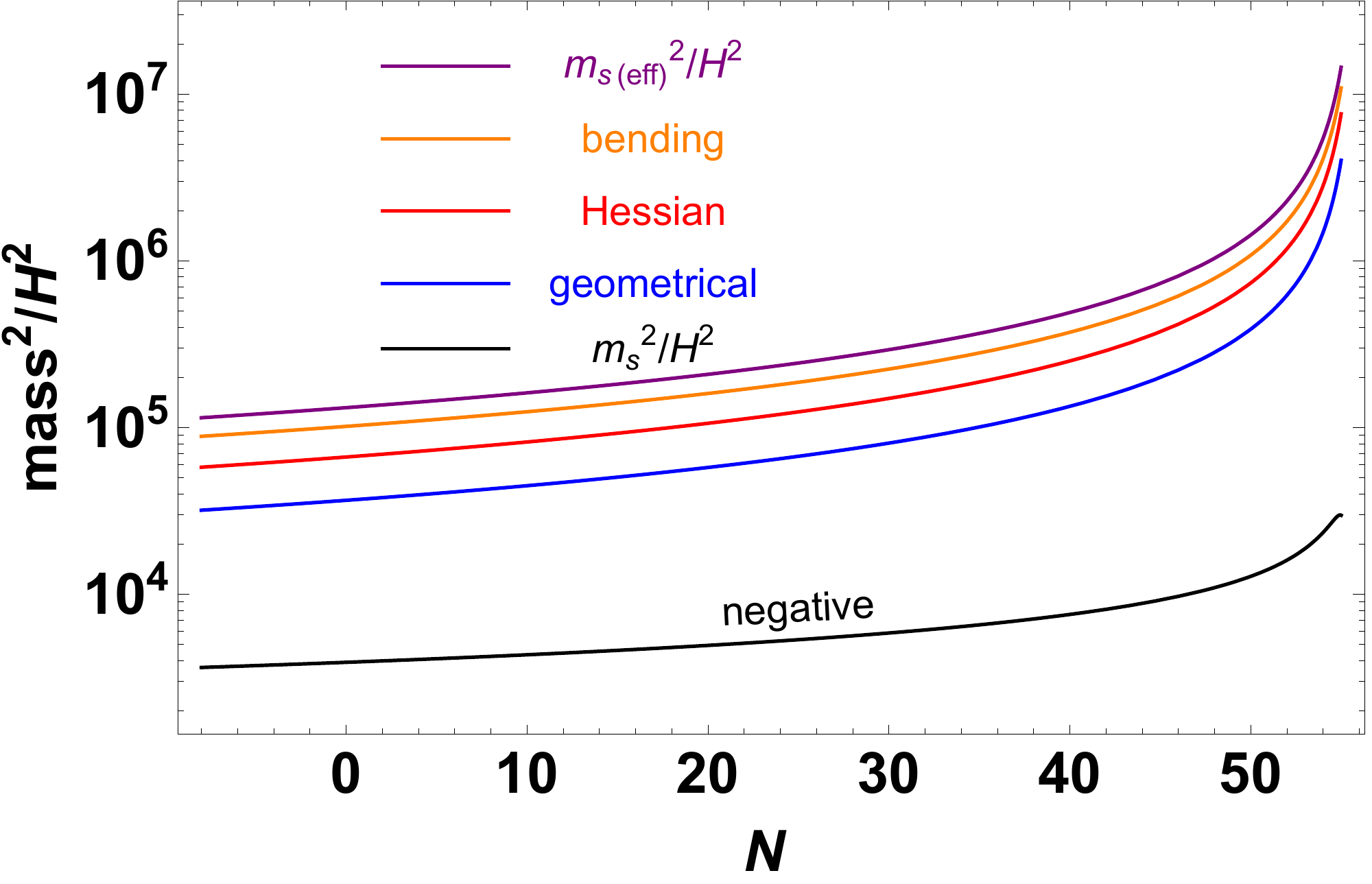}
            \caption{{\small  Natural inflation ($f=10$)}}    
            \label{fig:masses-NI-hyper}
        \end{subfigure}
        \hfill
        \begin{subfigure}[b]{0.475\textwidth}  
            \centering 
            \includegraphics[width=\textwidth]{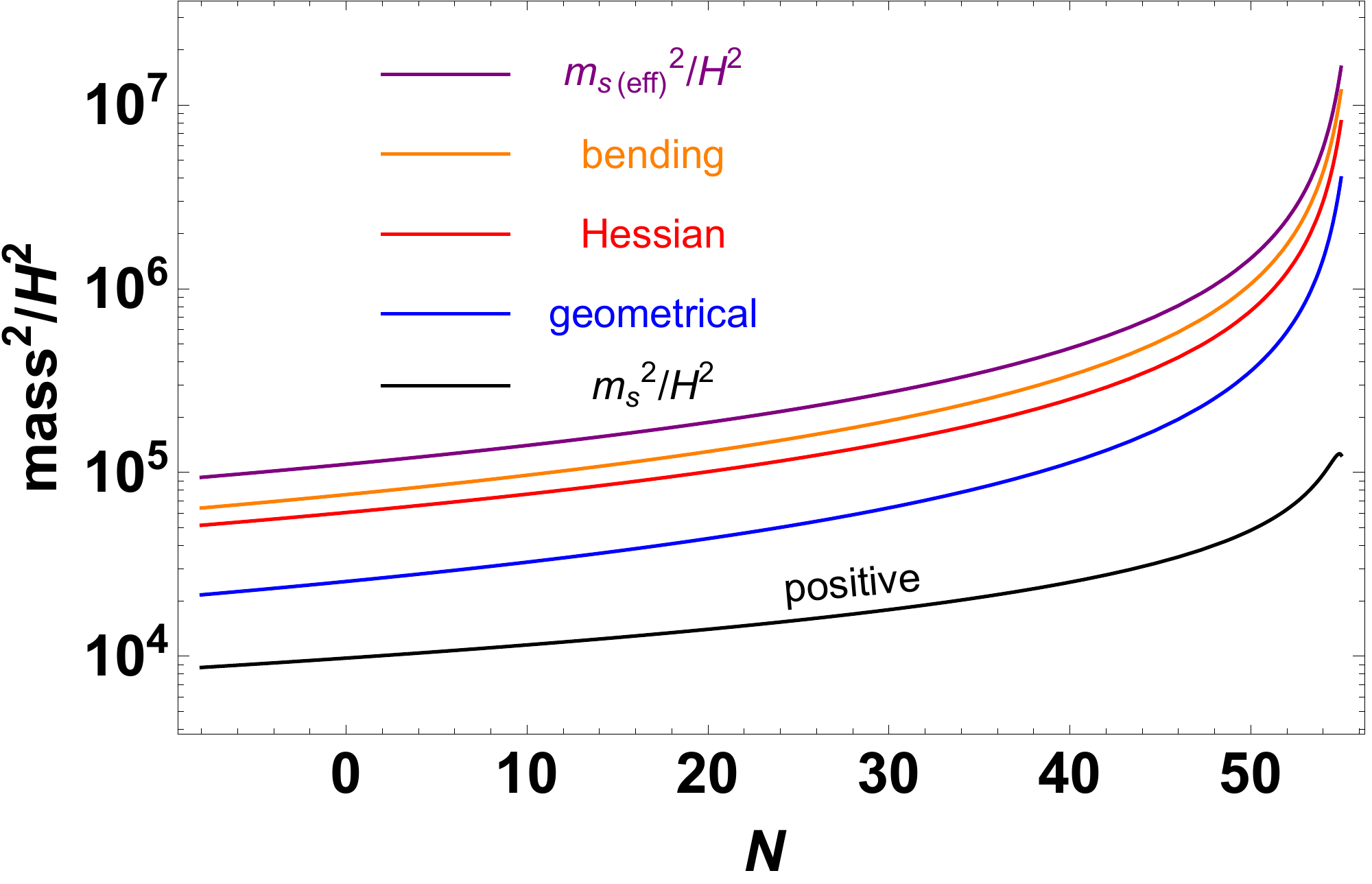}
            \caption{{\small Starobinsky inflation }}    
            \label{fig:masses-SI-hyper}
        \end{subfigure}
        \caption{Absolute values of $\msh/H^2$, its three contributions (see Eq.~\eqref{eq:entropicmass}), as well as $m_s^2/H^2$, for 
        our two representative examples of SI and NI with $f=10$, in the hyperbolic geometry. The plots show the last 63 e-folds of inflation.        
        Contrary to the minimal geometry, one has $|m_s^2|/H^2 \gg 1$ in that case. In addition, note that $m_s^2$ is positive in SI, and negative in NI.}
        \label{fig:masses-hyper}
    \end{figure*}

By using \eqref{es-result}, one can straightforwardly compute $V_{;ss}$, finding 
\beq
V_{;ss} \simeq \left(1+\frac{4\chi^2}{M^2} \right) \, m_h^2 -\sqrt{2} \frac{V_{,\vphi}}{M} \left( 1+\frac{2\chi^2}{M^2} \right)^{-1}+ \frac{ {2\chi^2 \over M^2}}{\left(1+\frac{2 \chi^2}{M^2} \right)} V_{,\vphi \vphi}\,,
\eeq
where one would keep the first term at leading-order, the second term in the refined estimate, and the last term can always be neglected for practical purposes. Eventually, using $\epsilon \simeq \left(1+\frac{2 \chi^2}{M^2} \right) \dot \vphi^2/(2 H^2 \Mp^2)$, together with $R_{\rm fs}=-4/M^2$, one deduces that
\beq
\epsilon R_{\rm fs} \Mp^2 \simeq  \left\{
\begin{array}{ll}
- \left(1+\frac{2 \chi^2}{M^2} \right) \frac{m_h^2}{H^2}  \qquad \qquad \,\, \,\, {\rm leading-order}\\
-\frac{2}{1+\frac{2\chi^2}{M^2} } \left(\frac{V_{,\vphi}}{3 H^2 M}\right)^2 \qquad \qquad   {\rm refined}\,.&
\end{array}\right.
\eeq
Summing these contributions, one deduces that
\begin{eqnarray}
\frac{\msh}{H^2} &\simeq&  8\frac{m_h^2}{H^2} {\chi^2 \over M^2} \gg 1\,\qquad \qquad \quad {\rm leading-order}\\
\frac{m_s^2}{H^2} &\simeq & 2 \sqrt{2}\frac{ {2\chi^2 \over M^2}}{\left(1+\frac{2 \chi^2}{M^2} \right)}  \frac{V_{,\vphi}}{H^2 M}\qquad {\rm refined} \,, \label{ms2-hyper-1}
\end{eqnarray}
where it is sufficient to give the leading-order form of the super-Hubble entropic mass for our purpose. Now that a non-zero result for $m_s^2$ is found, one can use Eq.~\eqref{velocity-phi-2} and the simple leading-order estimate \eqref{velocity-phi-1} to find the simple, more intuitive form 
\beq
\frac{m_s^2}{H^2} \simeq 12 \frac{m_h}{H}  \frac{\chi^2}{M^2}\,{\rm sign}(V_{,\vphi})\,.\label{ms2-hyper-2}
\eeq
This shows that, contrary to the minimal case, $|m_s^2/H^2| \gg 1$, and that it is smaller than $\msh/H^2$ only by a factor $m_h/H$. In addition to its amplitude, a crucial feature of the result \eqref{ms2-hyper-1}-\eqref{ms2-hyper-2} is that the sign of the entropic mass squared $m_s^2$ can be positive or negative --- with important observational consequences --- depending on whether the slope of the potential is positive or negative respectively. It is rather unusual in inflationary models to find a physical quantity that depends on the sign of the slope of the potential. In standard single-field inflationary models in particular, one can arbitrarily change the definition of $\vphi$ into $-\vphi$, and hence the sign of $V_{,\vphi}$, without physical consequences. One can do so because the standard kinetic term $(\partial \vphi)^2$ is trivially $\mathbb{Z}_2$ symmetric. However, while this feature is still true for the minimal field space metric \eqref{eq:minimalmetric}, this is not the case for the hyperbolic metric \eqref{eq:hyperbolicmetric}. Hence, it is not surprising that a physical quantity can depend on the choice of labelling the field $\vphi$ or $-\vphi$, and hence on the sign of $V_{,\vphi}$, simply because the starting point Lagrangian does depend on this choice in our hyperbolic model.

Similarly to the minimal model, all our analytical estimates above have been checked to agree with very high accuracy with the fully numerical results. For instance, the non-trivial result \eqref{ms2-hyper-1} for $m_s^2$ is accurate to the level $10^{-5}$ for our two representative examples of SI and NI with $f=10$. We show in Fig.\ \ref{fig:masses-hyper} (the absolute values of) $\msh/H^2$, its three contributions, as well as $m_s^2/H^2$, for these two examples.

\subsubsection{Summary}

Let us summarize the important features for the dynamics of the fluctuations that we learned in sections \ref{understanding-minimal}-\ref{understanding-hyper}, and describe their consequences.

\textbullet \,\,The deviation of sidetracked inflation's trajectory from a geodesic, and hence the coupling between the adiabatic and entropic fluctuations, is very large, as measured by the parameter $\etaperp \gg 1$.

\textbullet \,\,The super-Hubble effective mass $\msh$, governing the dynamics of entropic fluctuations once they exit the Hubble radius (see \eqref{Qs-super-Hubble}), is positive and much larger than the Hubble rate. Hence entropic fluctuations decay extremely fast outside the Hubble radius and the curvature perturbation is expected to be conserved on super-Hubble scales. An adiabatic limit is therefore reached by the end of inflation, rendering the multifield scenario of sidetracked inflation predictive without the need to describe the reheating stage.

\textbullet \,\,The entropic mass squared $m_s^2$, which dictates the evolution of the entropic fluctuations inside the Hubble radius --- together with its coupling to the adiabatic degree of freedom --- has different behaviours in the two field space geometries that we consider: it is much smaller than the Hubble rate in the minimal model (and negative), and much larger in the hyperbolic one. In that case, its sign depends on whether inflation proceeds along increasing or decreasing $\vphi$, and hence on the choice of the potential and the branch on which inflation occurs.\\

Let us stress that a negative entropic mass squared $m_s^2$ does not by no means imply that the background is unstable. A direct measure of the stability of the latter is provided by the sign of $\msh=m_s^2+4 H^2 \etaperp^2$, which is the mass of the fluctuations orthogonal to the background trajectory in the $k \to 0$ limit. While this quantity becomes negative along $\chi=0$ after the critical time, signaling the instability of this inflationary solution, and hence the geometrical destabilization, the large positive value of $\msh/H^2$ in sidetracked inflation was expected, as the latter corresponds by definition to the stable attractor trajectory in these models. 

In addition, we saw in section \ref{single-field} that one can achieve an effective description of this attractor in terms of one degree of freedom only, in which case the curvature perturbation $\zeta$ is conserved on super-Hubble scales \cite{Wands:2000dp}. Using the expression of $N(\vphi_{{\rm ini}})$ given in table \ref{tab:comparison}, and the $\delta N$-separate universe picture, one then obtains 
$\zeta=N_{,\vphi} Q_{\vphi}
\footnote{Note that in table \ref{tab:comparison}, the number of e-folds of inflation as a function of initial conditions is evaluated on the sidetracked attractor, in particular with $\chi$ determined as a function of $\vphi$. This is different from the quantity $N(\vphi_{{\rm ini}},\chi_{{\rm ini}})$ one should compute in the $\delta N$ formalism, but it is legitimate to do so given the strong attractor solution, and hence the negligible dependence of $N$ on $\chi_{{\rm ini}}$}$, where the right-hand side is evaluated at Hubble crossing such that $k=aH$,  and hence
\beq
{\cal P}_{\zeta_k}= \left( \frac{2 V}{3 (\Mp M m_h)^2} {\cal P}_{Q_{\vphi}} \right)_{k=aH}\,.
\label{zeta-N}
\eeq
Taking into account the fact that
\beq
Q_\sigma=\left\{
\begin{array}{ll}
\left(1+\frac{2 \chi^2}{M^2} \right)^{1/2} Q_{\vphi} \qquad \qquad \qquad \qquad \qquad \qquad \qquad \quad{\rm minimal}\\
\left(1+\frac{2 \chi^2}{M^2} \right)^{1/2}\, Q_{\vphi} +\sqrt{2} \chi/M \left(1+\frac{2 \chi^2}{M^2} \right)^{-1/2} \, Q_{\chi} \qquad \, {\rm hyperbolic}&
\end{array}\right. \,,
\eeq 
Eq.~\eqref{zeta-N} consistently coincides with evaluating ${\cal P}_{\zeta_k}=\left(H/\dot \sigma \right)^2 {\cal P}_{Q_\sigma}$ at Hubble crossing (let us recall Eq.~\eqref{zeta-Qsigma}), when neglecting $Q_{\chi}$ fluctuations in the hyperbolic model. This is indeed a good approximation, as 
\beq
Q_s=\left\{
\begin{array}{ll}
 Q_{\chi} \qquad \qquad \qquad \qquad \quad {\rm minimal}\\
\left(1+\frac{2 \chi^2}{M^2} \right)^{-1/2}\, Q_{\chi}  \qquad\,\,\, {\rm hyperbolic}&
\end{array}\right.\,,
\eeq 
and, as we will see, entropic fluctuations are already negligible compared to adiabatic ones at Hubble crossing. However, note that contrary to standard situations, Eq.~\eqref{zeta-N} is of little practical use without further input, as the non-trivial sub-Hubble dynamics caused by the bending trajectory renders $\left({\cal P}_{Q_{\vphi}}\right)_{k=aH}$ unknown, or more precisely, it can substantially differ from the purely adiabatic result $\left( \frac{H}{2 \pi}\right)^2_{k=aH}$.

Some analytical understanding can however be achieved. In the hyperbolic geometry in particular, we have seen that the entropic mass $m_s^2$ is much larger than the Hubble rate. This type of framework has been extensively studied (see \textit{e.g.} \cite{Tolley:2009fg,Cremonini:2010ua,Achucarro:2010da,Baumann:2011su,Shiu:2011qw,Cespedes:2012hu,Achucarro:2012sm,Avgoustidis:2012yc,Achucarro:2012yr,Gwyn:2012mw,Gong:2013sma,Cespedes:2013rda,Gong:2014rna,Gwyn:2014doa}), and it has been shown that the heavy entropic fluctuation can then be integrated out, resulting in a single-field effective theory for the adiabatic fluctuation, with a non-trivial speed of sound different from unity. More surprisingly at first sight, when the entropic mass is much smaller than the Hubble scale, one can still integrate out the entropic fluctuation in the presence of a large bending, as is relevant in the minimal geometry, obtaining then a single-field effective theory with a modified dispersion relation \cite{Cremonini:2010ua,Baumann:2011su,Gwyn:2012mw,Gwyn:2014doa}. We make use of these tools in the following section, treating each of the two situations in turn.

\subsection{Effective single-field theory for the fluctuations}
\label{EFT}

\begin{figure*}
        \centering
        \begin{subfigure}[b]{0.475\textwidth}
            \centering
            \includegraphics[width=\textwidth]{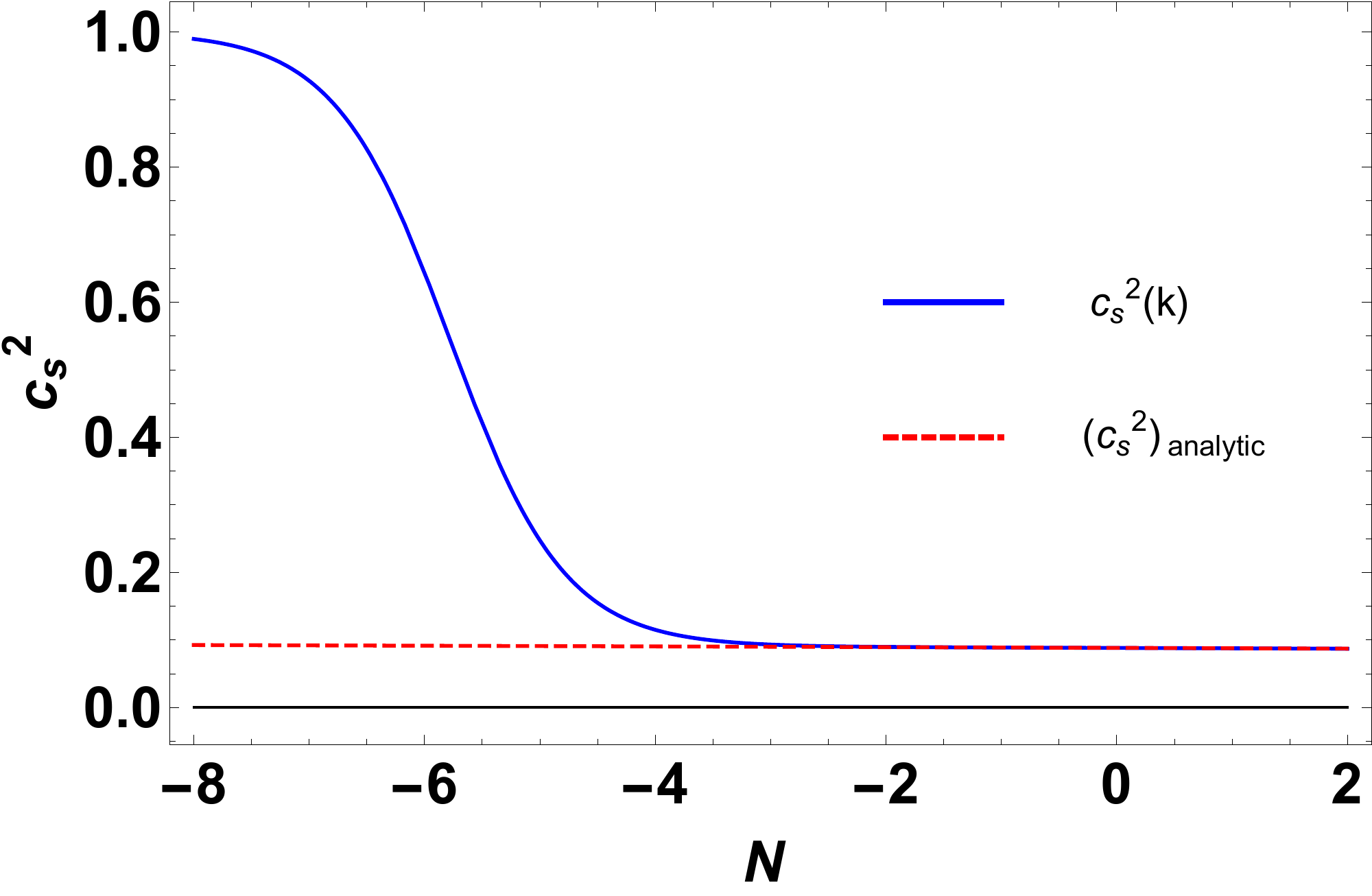}
            \caption{{\small Starobinsky inflation}}    
            \label{fig:cs-SI}
        \end{subfigure}
        \hfill
        \begin{subfigure}[b]{0.475\textwidth}  
            \centering 
            \includegraphics[width=\textwidth]{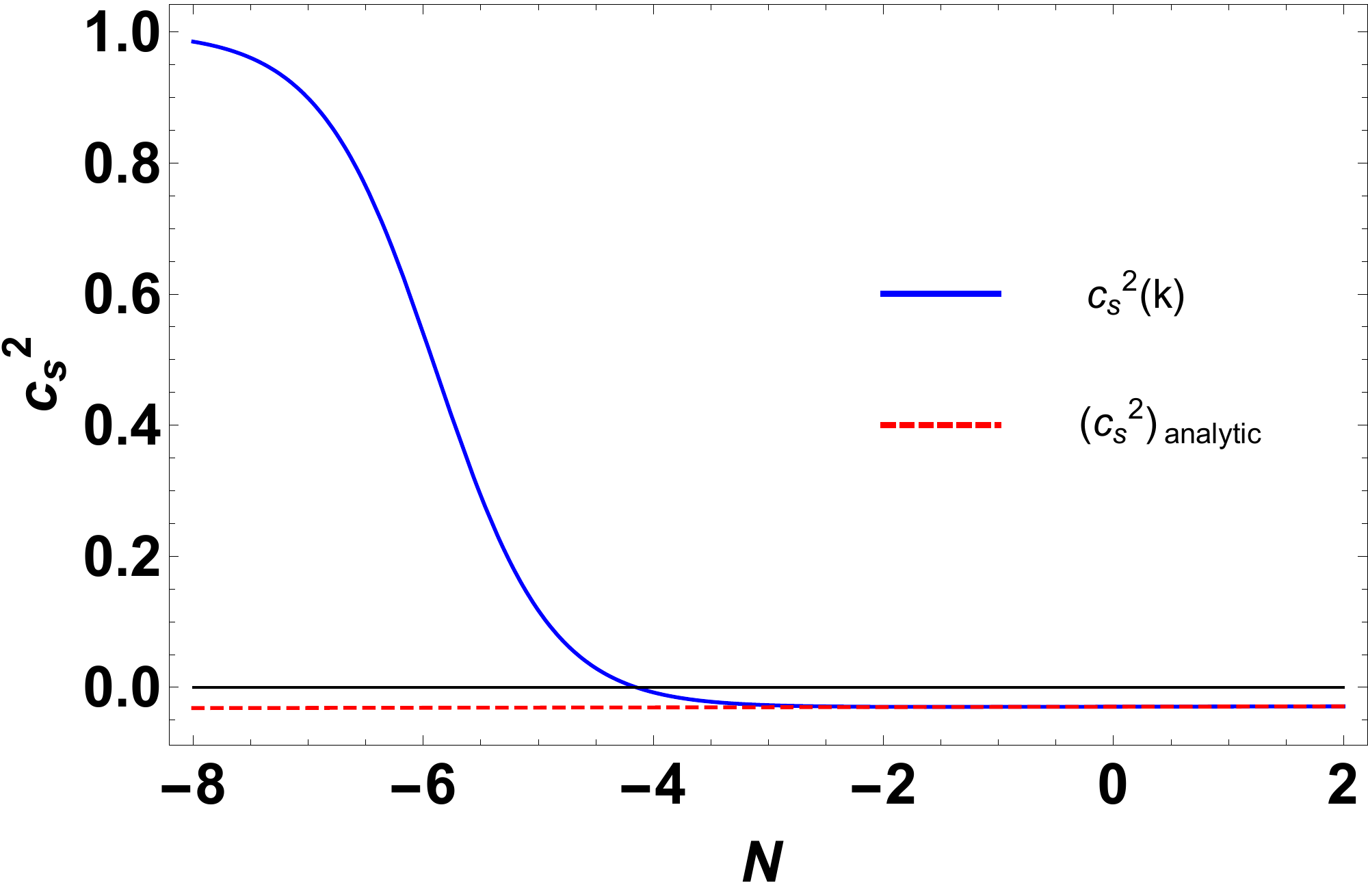}
            \caption{{\small Natural inflation ($f=10$)}}    
            \label{fig:cs-NI}
        \end{subfigure}
        \caption{Fully numerical $c_s^2(k)$ \eqref{cs2k} and analytical result \eqref{cs-analytical} for its `late time' behaviour when $k^2/a^2 \ll m_s^2$, for the potentials of SI (left) and NI, $f=10$ (right). The corresponding scale crosses the Hubble radius 55 e-folds before the end of inflation, at $N=0$ in the plots.}
        \label{fig:cs}
    \end{figure*}
When the entropic mass of the entropic fluctuation $m_s^2$ is large compared to the Hubble scale, as it is relevant in the hyperbolic geometry, one can integrate it out: neglecting the first two terms in its equation of motion \eqref{delta_s_all_scales}, one can express $Q_s$ in terms of the curvature perturbation $\zeta$, plug it back into the second-order action and deduce 
 \begin{eqnarray}
S_{(2) \,({\rm EFT})}&=&\int dt \, d^3 k \, \frac{a^3\, \epsilon\,\Mp^2}{c_s^2(k)} \left[  {\dot {\zeta}_k}^2  + c_s^2(k) k^2  \frac{\zeta_k^2}{a^2} \right]\,,
\label{S2-EFT}
\end{eqnarray}
where
\beq
\frac{1}{c_s^2(k)} \equiv 1+\frac{4 H^2 \etaperp^2}{k^2/a^2+m_s^2}\,.
\label{cs2k}
\eeq
We will discuss below the conditions determining the validity of such an effective field theory (EFT), a subject which has been extensively studied (see \textit{e.g.} \cite{Tolley:2009fg,Cremonini:2010ua,Achucarro:2010da,Baumann:2011su,Shiu:2011qw,Cespedes:2012hu,Achucarro:2012sm,Avgoustidis:2012yc,Achucarro:2012yr,Gwyn:2012mw,Gong:2013sma,Cespedes:2013rda,Gong:2014rna,Gwyn:2014doa}). Note already that in addition to the condition of a heavy entropic field, $m_s^2 \gg H^2$, one should also verify the generalized adiabaticity condition \cite{Cespedes:2012hu}
\beq
| \dot{\eta}_\perp /( m_s \etaperp) | \ll 1\,,
\label{adiabaticity-condition}
\eeq
expressing the fact that the rate of change of the bending should be smaller than the large entropic mass, so as not to excite high-frequency modes that are not captured by the low-energy effective field theory \eqref{S2-EFT}. There is no restriction on the amplitude of the bending however, which can consistently be large, as in sidetracked inflation. From the expression \eqref{etaperp-hyper} of $\etaperp$ (see also the numerical results in Fig.\ \ref{fig:masses}), it is easy to see that the stronger condition $| \dot{\eta}_\perp /( H \etaperp) | \ll 1$ holds in sidetracked inflation in the hyperbolic geometry, so that the condition \eqref{adiabaticity-condition} is safely verified. 

Note that deep on sub-Hubble scales, when $k^2/a^2 \gg (m_s^2,H^2 \etaperp^2)$, the speed of sound equals unity and one recovers the Bunch-Davies behaviour of the full two-field situation, as it should be. One can not integrate out the entropic field in this regime, but the adiabatic and entropic fluctuations behave as uncoupled free fields, and including the  gradient terms in \eqref{cs2k} can be seen as an effective way to treat in a unified manner this regime and the subsequent one, with good results as we will see. As soon as $k^2/a^2$ drops below $m_s^2$, the speed of sound becomes $k$-independent, and approximately reads, using \eqref{etaperp-hyper}-\eqref{ms2-hyper-2}
\beq
\frac{1}{c_s^2}-1 \simeq 4 H^2 \etaperp^2/m_s^2 \simeq 
\left\{
\begin{array}{ll}
 \frac23 \frac{m_h}{H}\, {\rm sign}(V_{,\vphi}) \qquad   {\rm leading-order}\\
\frac{1}{1+\frac{2\chi^2}{M^2} } \frac{4 V_{,\vphi}}{9 \sqrt{2} H^2 M} \qquad \qquad \,\,  {\rm refined}&
\end{array}\right.
\label{cs-k-independent}
\eeq
where, as before, the refined estimate is very accurate (to the level $5 \times 10^{-3}$ in SI for instance), while the leading-order one is less accurate (to the level $8 \times 10^{-2}$ for the same model), but easier to grasp the physics: as $m_h \gg H$, one has $|1/c_s^2-1| \gg 1$, and therefore a low speed of sound determined by the hierarchy between the heavy mass $m_h$ and the Hubble scale, given by
\beq
c_s^2 \simeq \frac{3 H}{2 m_h} \, {\rm sign}(V_{,\vphi})\,.
\label{cs-analytical}
\eeq
We show in Fig.\ \ref{fig:cs} the fully numerical result \eqref{cs2k} for $c_s^2(k)$ and the analytical result \eqref{cs-analytical} for its `late time' behaviour when $k^2/a^2 \ll m_s^2$, for the scale $k_{55}$ that crosses the Hubble radius 55 e-folds before the end of inflation, at $N=0$ in the plot, and for the potentials of SI (left) and NI, $f=10$ (right). Note that $c_s^2$ is moderately small in the first case, but the agreement is nonetheless excellent.

As the reader should have noticed, we treated in a unified manner the situations in which the large entropic mass squared $m_s^2/H^2$ is positive, like in SI, or in which it is negative, like in NI. Although the latter situation is unusual, as it corresponds to a negative speed of sound squared, it does not violate the conditions under which the effective field theory \eqref{S2-EFT} has been derived, and its predictive power is equally applicable here. The physical consequences are however very different and we treat each of them separately, beginning with the more conventional situation.

\subsubsection{Positive speed of sound}
\label{reduced}

\begin{figure*}
        \centering
        \begin{subfigure}[b]{0.475\textwidth}
            \centering
            \includegraphics[width=\textwidth]{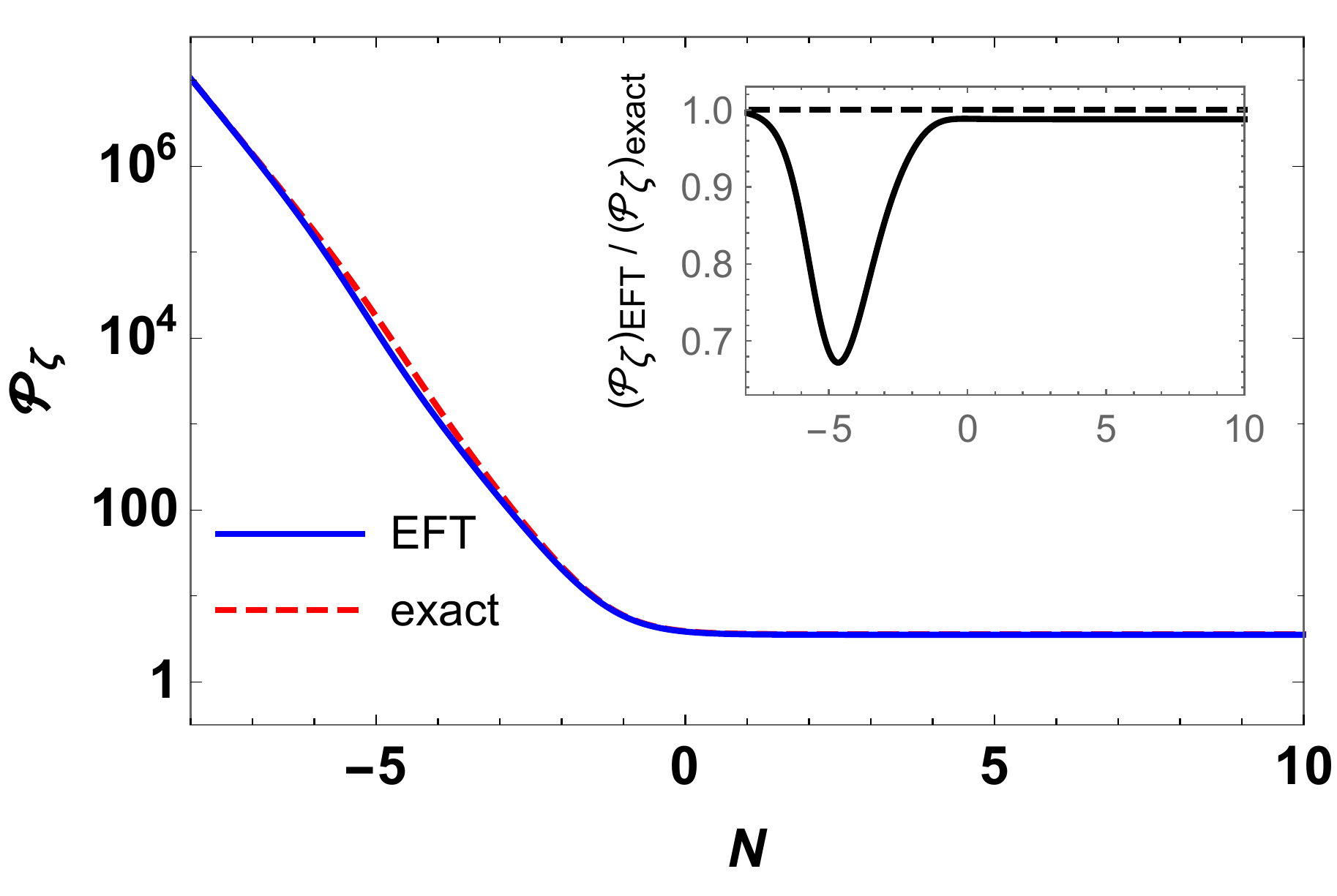}
            \caption{{\small Starobinsky inflation}}    
            \label{fig:EFT-exact-SI-hyper}
        \end{subfigure}
        \hfill
        \begin{subfigure}[b]{0.475\textwidth}  
            \centering 
            \includegraphics[width=\textwidth]{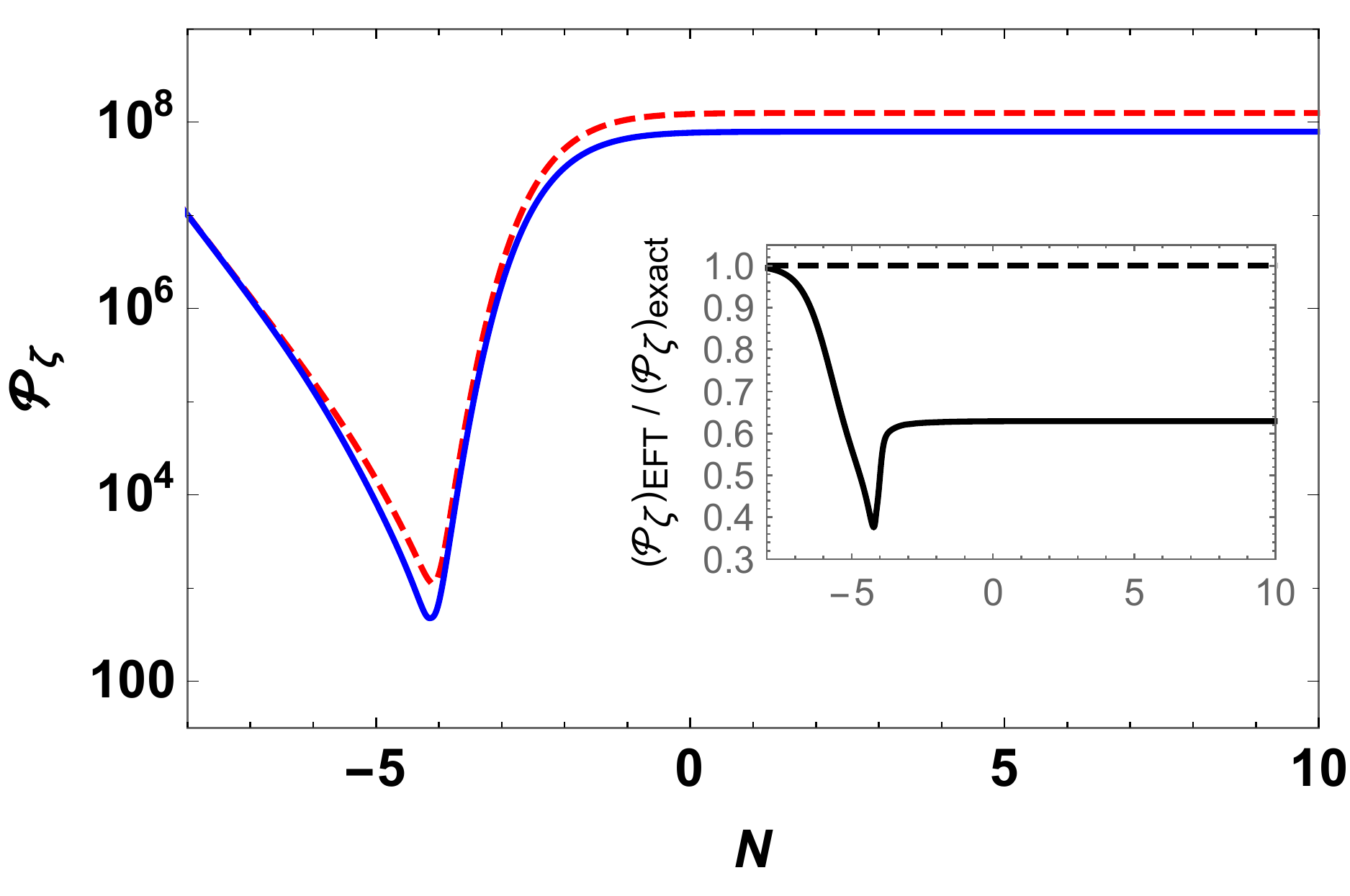}
            \caption{{\small Natural inflation ($f=10$)}}    
            \label{fig:EFT-exact-NI-hyper}
        \end{subfigure}
        \caption{Power spectra of the curvature perturbation as functions of the number of e-folds, computed numerically in the full two-field model (exact, in dashed red), and from the effective field theory \eqref{S2-EFT} using \eqref{evolution-power-spectrum}, for SI (left) and NI with $f=10$ (right) in the hyperbolic geometry. The spectra are evaluated for the scale that crosses the Hubble radius 55 e-folds before the end of inflation, at $N=0$ in the plots, and are normalized by the adiabatic result \eqref{Pad}. The insets show the ratios between the EFT and the exact results.}
        \label{fig:EFT-exact}
    \end{figure*}

\begin{figure*}
        \centering
        \begin{subfigure}[b]{0.475\textwidth}
            \centering
            \includegraphics[width=\textwidth]{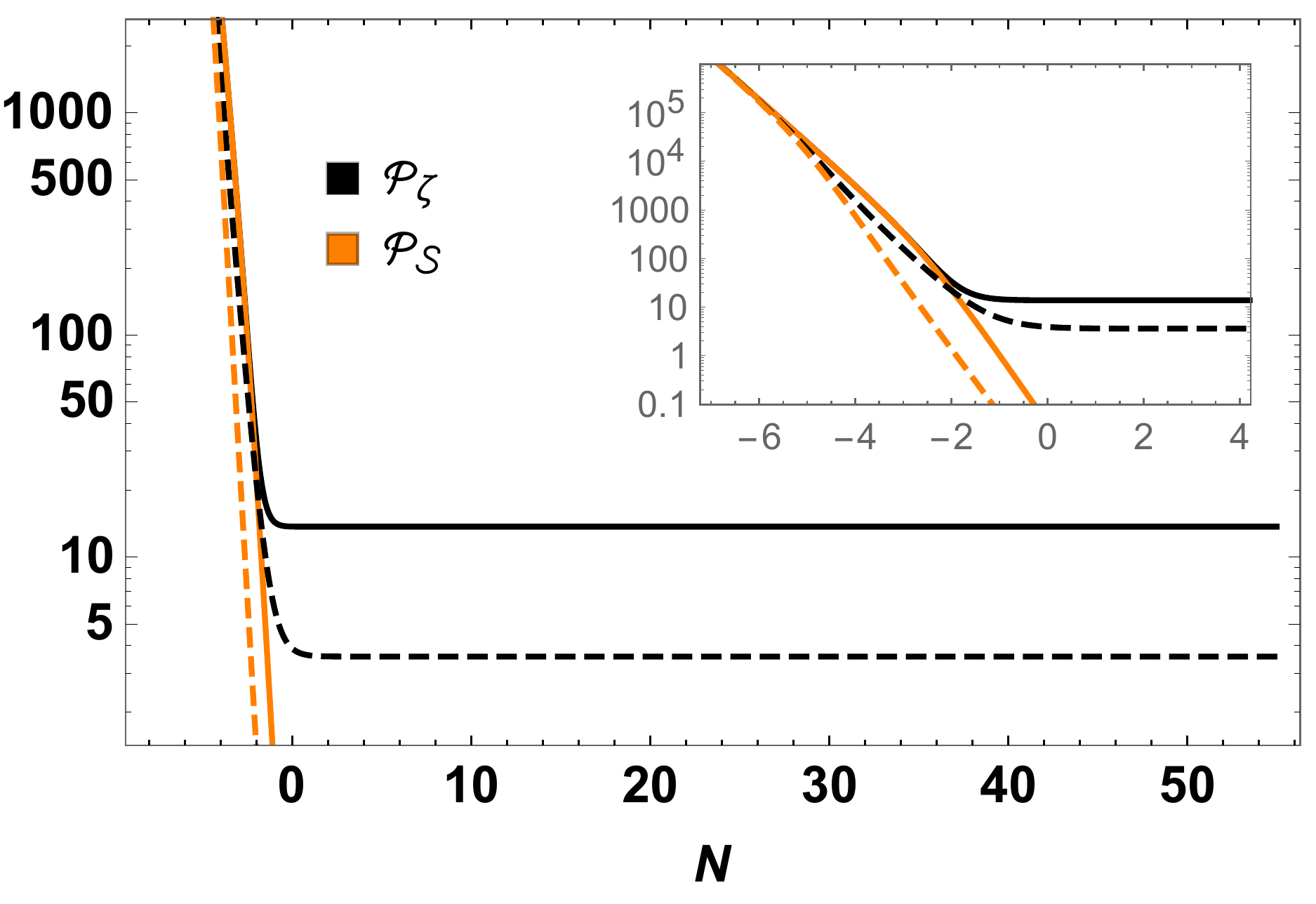}
            \caption{{\small Starobinsky inflation}}    
            \label{fig:powspec SI}
        \end{subfigure}
        \hfill
        \begin{subfigure}[b]{0.475\textwidth}  
            \centering 
            \includegraphics[width=\textwidth]{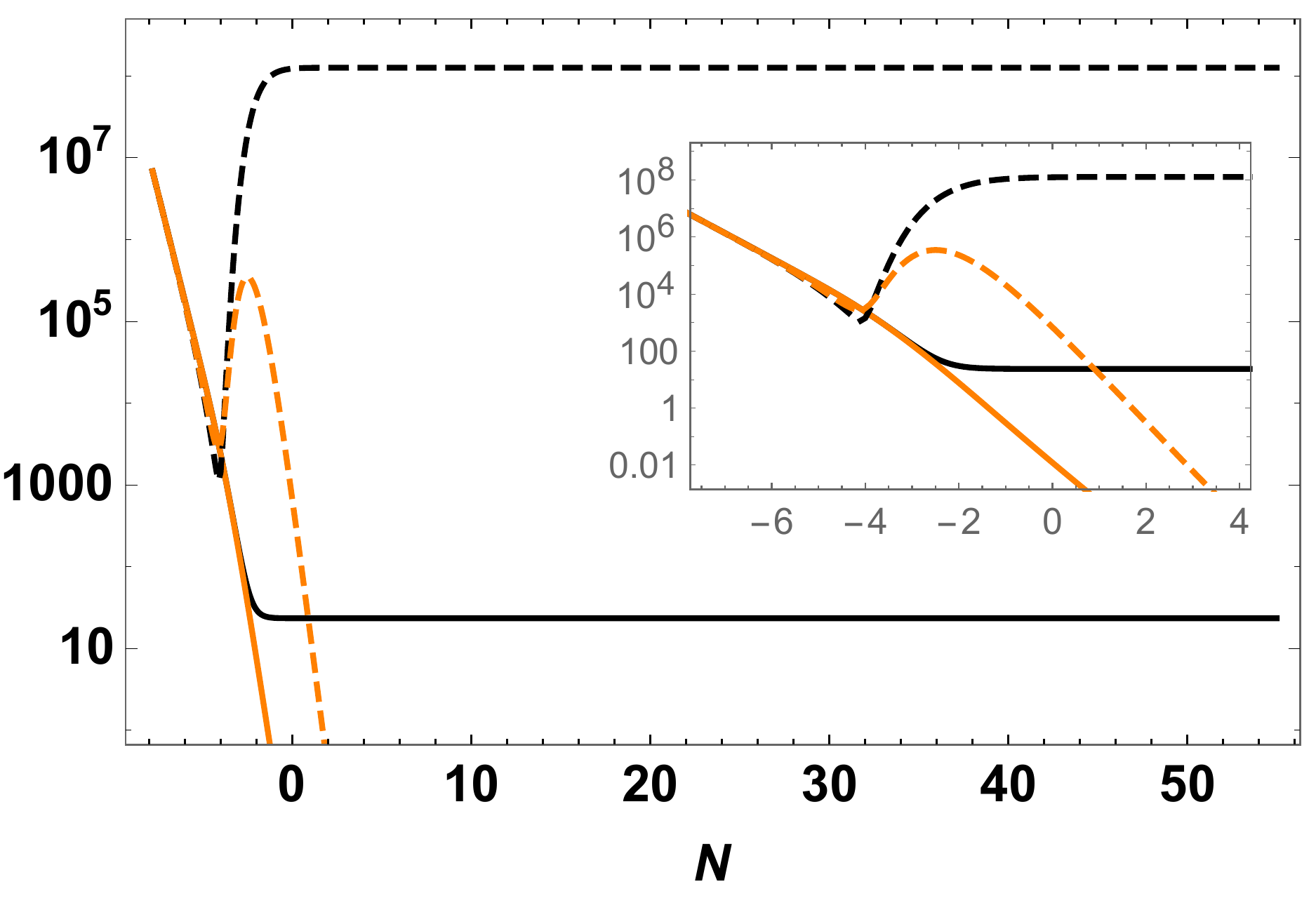}
            \caption{{\small Natural inflation ($f=10$)}}    
            \label{fig:powspec NI}
        \end{subfigure}
        \caption{Adiabatic (${\cal P}_\zeta$) and entropic (${\cal P_S}$) power spectra as functions of the number of e-folds, for both the minimal (solid lines) and hyperbolic (dashed lines) internal metrics. The spectra are evaluated for the scale crosses the Hubble radius 55 e-folds before the end of inflation, at $N=0$ in the plots, and are normalized by the adiabatic result \eqref{Pad}. The insets are details of the same curves around the time of Hubble crossing.}
        \label{fig:powspec}
    \end{figure*}

When the entropic mass $m_s^2$ is positive, $c_s^2(k)$ is always positive, and the action \eqref{S2-EFT} for $k^2/a^2 \ll m_s^2$ describes a standard set-up with a reduced speed of sound $c_s$ given by \eqref{cs-k-independent}-\eqref{cs-analytical}. One then finds that $\zeta_k$ becomes constant soon after sound Hubble crossing such that $k c_s=a H$, with the usual result \cite{Garriga:1999vw}
\beq
{\cal P}_{\zeta_k}\simeq \left(\frac{H^2}{8 \pi^2 \epsilon c_s}\right)_\star\,,
\label{Pzeta-kinflation}
\eeq
where here $\star$ denotes evaluation at $k c_s=a H$ (In the three different situations studied respectively in sections \ref{reduced}, \ref{imaginary-speed-sound}, \ref{modified-dispersion}, the subscript $\star$ indicates an evaluation at different times. This is summarized in table \ref{tab:sidetracked}). In SI, and for the scale $k_{55}$ that crosses the Hubble radius 55 e-folds before the end of inflation, this predicts a value of the enhancement of the curvature power spectrum compared to the adiabatic result ${\cal P}_{\zeta}/{\cal P}_{{\rm ad}} \simeq 3.50$ (respectively $3.56$ for the two-field numerical result). In Fig.\ \ref{fig:EFT-exact-SI-hyper} one can also see the very good agreement between the full two-field numerical result and the numerical result corresponding to the effective theory \eqref{S2-EFT} (see below for the detailed procedure of the computation). Additionally, one can see in Fig.\ \ref{fig:powspec SI} how the entropic power spectrum decreases as $1/a^3$ as soon as $k^2/(a^2 m_s^2)$ drops below one.
Note that all quantities in the various plots of the power spectra in this paper are for the scale $k_{55}$, and are normalized by
\beq
{\cal P}_{{\rm ad}}=\left(H^2/(8 \pi^2 \epsilon) \right)_{k=aH}\,,
\label{Pad}
\eeq
which, as we have explained in section \ref{fluctuations-adiabatic-entropic}, is the prediction for the curvature power spectrum if the effects of the bending were negligible, which we call the adiabatic result. Thus, the deviation from one of the final value observed for ${\cal P}_\zeta$ in these plots is a measure of the non-trivial multifield effects, that occur on sub-Hubble scales in sidetracked inflation. Note also that deep inside the Hubble radius, the Bunch-Davies behaviour implies that all plotted quantities behave as $\simeq k^2/(a^2 H^2)$. 

From the result \eqref{Pzeta-kinflation} for the power spectrum, one deduces the familiar expression of the scalar spectral index $n_s-1 \simeq -2 \epsilon_\star-\eta_\star-s_\star$, where $s \equiv c_s'/c_s$. For the scale $k_{55}$, this gives $n_s \simeq 0.969$, whereas the full two-field numerical result gives $n_s=0.970$, and the adiabatic result would give $n_s = 0.965$; the agreement between the effective field theory and the full result is thus very good. Note eventually that with the expression \eqref{cs-analytical} for $c_s^2$, one obtains $2s \simeq - \epsilon$, and hence the simplified form of the result $n_s-1 \simeq -\frac32 \epsilon_\star-\eta_\star$.

\subsubsection{Imaginary speed of sound}
\label{imaginary-speed-sound}

We now discuss situations in which the effective speed of sound squared is negative (situations in which $m_s^2$ is negative but $c_s^2$ is still positive do not arise in our framework, but results of the previous section would apply in that case). Let us first give a few details about how we calculate the EFT prediction. From the action \eqref{S2-EFT}, one deduces the equation of motion for the complex mode function
\beq
\ddot{\zeta}_k+H(3+\eta-2s) \dot{\zeta}_k+\frac{c_s^2(k)k^2}{a^2} \zeta_k=0\,,
\label{zeta-cs}
\eeq
and the quantization condition, which states that
\beq
\zeta_k \dot{\zeta}^*_k-\dot{\zeta}_k \zeta^*_k=\frac{i c_s^2(k)}{2 \epsilon a^3}
\eeq
holds at all time. Following \cite{Romania:2011ez,Romania:2012tb}, one can then easily deduce a non-linear evolution equation for the power spectrum ${\cal P}_{\zeta_k} \equiv k^3/(2 \pi^2) |\zeta_k|^2$ itself, which reads, in e-fold time:
\beq
{\cal P}_{\zeta_k}^{''}+(3-\epsilon+\eta-2s) {\cal P}_{\zeta_k}^{'}+\frac{2 c_s^2(k)k^2}{a^2 H^2} {\cal P}_{\zeta_k}=\frac{1}{2 {\cal P}_{\zeta_k}} \left( {\cal P}_{\zeta_k}^{'2}+\left(\frac{k^3 c^2_s(k)}{4 \pi^2 \epsilon H a^3}\right)^2\right)\,.
\label{evolution-power-spectrum}
\eeq
One can then easily solve this equation numerically, with initial conditions $ {\cal P}_{\zeta_k}=\frac{1}{8 \pi^2 \epsilon} \frac{k^2}{a^2} \left(1+{\cal O}((aH)^2/k^2) \right)$ deep inside the Hubble radius, where we took into account that $c_s$ equals unity in this regime. Note that in Eqs.~\eqref{zeta-cs}-\eqref{evolution-power-spectrum}, $2 s \equiv c_s^{2'}(k)/c_s^2(k)$. An important subtlety in models with $m_s^2 <0$, and hence in which $c_s^2(k)$ crosses $0$ around $k^2/a^2 \simeq |m_s^2|$, is that $s$, which multiplies ${\cal P}_{\zeta_k}^{'}$ in \eqref{evolution-power-spectrum}, blows up at that time. We then solve Eq.~\eqref{evolution-power-spectrum} in two steps, before this jump, and afterwards, imposing continuity and the regularity condition ${\cal P}_{\zeta_k}^{'}=0$ as an initial condition for the second phase. In this respect, note that the artifact of $c_s^2$ crossing zero comes from our will to have a single EFT that captures both the Bunch-Davies regime $k^2/a^2 \gg |m_s^2|$, and the subsequent period. Contrary to what the action \eqref{S2-EFT} might suggest, there is no ghost or strong coupling problem in the full two-field theory, and $c_s^2(k)$ becoming negative around $k^2/a^2 \simeq |m_s^2|$ simply signals the tachyonic growth of the entropic fluctuation, which in turns feeds the curvature perturbation. Our matching procedure can thus be physically motivated as interpolating between the Bunch-Davies behaviour, in which ${\cal P}_{\zeta_k}$ decreases as $1/a^2$, and the subsequent phase in which ${\cal P}_{\zeta_k}$ grows, hence having ${\cal P}_{\zeta_k}^{'}=0$ at the transition.

We show in Fig.\ \ref{fig:EFT-exact-NI-hyper} the result of this procedure for the model of NI with $f=10$, together with the numerical result of the full two-field theory. The agreement between the exact result and the one derived from our effective field theory treatment is impressive: the two differ only by a factor of $2$ despite the unusually large growth of the power spectrum on sub-Hubble scales, by five orders of magnitude.
In addition, we show in what follows that one can derive an analytical understanding of this large growth, as well as the final result for the power spectrum and its running, building on the matching procedure described above.

Let us consider the action \eqref{S2-EFT} in the regime where $k^2/a^2$ has dropped below $|m_s^2|$, so that $c_s^2$ is $k$-independent and negative. The canonically normalized field $v_k=z \zeta_k$ with $z=a \sqrt{2 \epsilon}/|c_s|$ verifies the standard equation, of $k$-inflationary type, $\frac{{\rm d}^2 v_{k}}{{\rm d} \tau^2}+\left( c_s^2 k^2-\frac{1}{z}\frac{{\rm d}^2 z}{{\rm d} \tau^2}\right) v_{k} =0$.
As we discussed below Eq.~\eqref{xi}, we assume that $\epsilon$ and $c_s$ evolve much less rapidly than the Hubble scale, which is well verified in our setup, so that one approximately obtains, with $a \simeq -1/(H \tau)$:
\beq
\frac{{\rm d}^2 v_{k}}{{\rm d} \tau^2}+\left( c_s^2\,k^2-\frac{2}{\tau^2}\right) v_{k} \simeq 0\,.
\label{eq-v-tau}
\eeq 
For practical purposes, we can take $c_s^2$ to be constant, and the general solution of \eqref{eq-v-tau} is simply obtained from the usual situation, in which $c_s^2>0$, by changing $c_s$ into $i \c$, where we use the notation $\c \equiv \sqrt{|c_s^2|}$ (and similarly for analogous quantities). It reads
\beq
v_k=A_k e^{k \c \tau} \left(1-\frac{1}{k \c \tau} \right)+B_k e^{-k \c \tau} \left(1+\frac{1}{k \c \tau} \right)\,,
\label{two-modes}
\eeq
where the standard oscillatory behaviour is now turned into increasing and decreasing exponential ones, and where $A_k$ and $B_k$ are two constants to determined. As explained above, we determine them by requiring that $\zeta'_k=0$ (implying ${\cal P}^{'}_k=0$) at the matching time such that $k^2=a^2 |m_s^2|$, denoted by a $\m$, and the continuity with the standard Bunch-Davies result $v_k \sim \frac{1}{\sqrt{2k}}e^{-i k \tau}$. This readily gives
\beq
B_k=A_k e^{2 k |c_s| \tau_\m}\,,
\eeq
and then
\beq
A_k=\frac{1}{2 \sqrt{2 k}}e^{-k |c_s| \tau_\m}\,,
\eeq
where we omitted an irrelevant phase factor. The time dependent power spectrum then reads
\beq
{\cal P}_{\zeta_k}(\tau) =\frac{H^2}{32 \pi^2 \epsilon} \left[e^{x-x_\m}(x-1)+e^{-(x-x_\m)}(x+1) \right]^2\,,
\label{result-power-tau-NI-hyper}
\eeq
where $x=k |c_s| \tau \simeq -k|c_s|/(aH)$ is negative and grows with time, from $x_\m$ such that $ x^2_\m \gg 1$, towards zero on super-Hubble scales. Eq.~\eqref{result-power-tau-NI-hyper}, which rightly reproduces the time-dependence of the power spectrum seen in Fig.\ \ref{fig:EFT-exact-NI-hyper}, shows that the two modes are equally important at the transition time, although it is dominated very rapidly by the exponentially growing mode. With $|x_\m|\simeq \left(|c_s | |m_s|/H\right)_\m$, this gives the final result for the power spectrum
\beq
{\cal P}_{\zeta_k} =\left(\frac{H^2}{8 \pi^2 \epsilon}\, \frac14\, e^{2 |c_s |\frac{|m_s|}{H}}\right)_\m=\left(\frac{H^2}{8 \pi^2 \epsilon}\, \frac14\, e^{  |\frac{m_s^2}{H^2 \etaperp}|} \right)_\m\,,
\label{result-power-NI-hyper}
\eeq
where we should consistently take for the values of the slowly-varying quantities the ones at the matching time $\m$, \textit{i.e} at entropic mass crossing such that $k^2=a^2 |m_s^2|$. Let us stress that $|c_s|_\m$ here is evaluated using the $k$-independent limit \eqref{cs-k-independent} and not \eqref{cs2k}, which would give a vanishing value. Note also that the result \eqref{result-power-NI-hyper} is expressed in terms of general quantities and holds beyond our particular framework of sidetracked inflation. 

Now specifying the general result \eqref{result-power-NI-hyper} to this setup, and using Eqs.~\eqref{etaperp-hyper}-\eqref{ms2-hyper-1}-\eqref{cs-analytical}, one finds that the exponential enhancement simply reads $e^{6 \sqrt{2} \chi_\star/M}$, so that the scalar spectral index reads
\beq
n_s-1=-2 \epsilon_\m-\eta_\m+6 \sqrt{2}\chi^{'}_\m/M\,.
\label{ns-instability}
\eeq 
For NI with $f=10$, these results predict a value of the enhancement of the power spectrum compared to the adiabatic result ${\cal P}_{\zeta}/{\cal P}_{{\rm ad}} \simeq 4.4 \times 10^8$ (respectively $1.26 \times 10^8$ for the two-field numerical result), as well as $n_s=0.973$ (respectively $0.974$). Given the highly non-trivial and very large growth of the power spectrum, and the degree of arbitrariness in our matching procedure, we find this order one agreement very good for the first result, and rather remarkable for $n_s$. In addition, although we concentrated here on our representative example, we will comment in section \ref{numerical-results} how our analytical formulae enable one to reproduce and understand the full numerical results for a variety of models and parameters.

Before closing this section, let us note that a similar sub-Hubble growth of the curvature perturbation induced by a transient tachyonic instability has already been observed in the literature in reference \cite{Cremonini:2010ua}, although it has not been studied in detail. More recently, a two-field model with hyperbolic geometry and that features the same type of behaviour has also been studied \cite{Brown:2017osf,Mizuno:2017idt}. The authors there used a full two-field description, but we note that our description in terms of an effective single-field theory with an imaginary speed of sound seems equally applicable there.

\subsubsection{Modified dispersion relation}
\label{modified-dispersion}
\begin{figure*}
        \centering
        \begin{subfigure}[b]{0.475\textwidth}
            \centering
            \includegraphics[width=\textwidth]{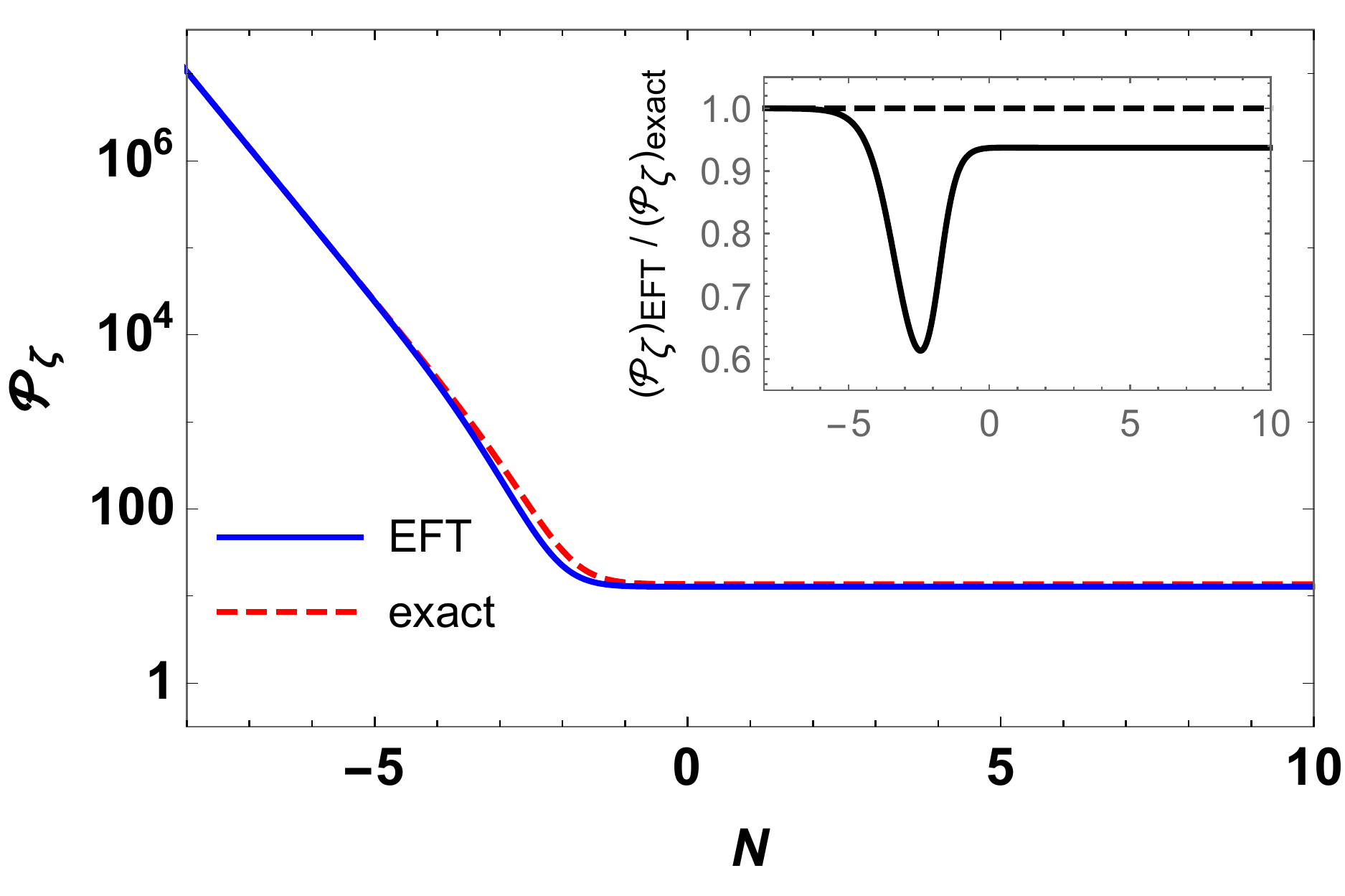}
            \caption{{\small Starobinsky inflation}}    
            \label{fig:EFT-exact-SI-minimal}
        \end{subfigure}
        \hfill
        \begin{subfigure}[b]{0.475\textwidth}  
            \centering 
            \includegraphics[width=\textwidth]{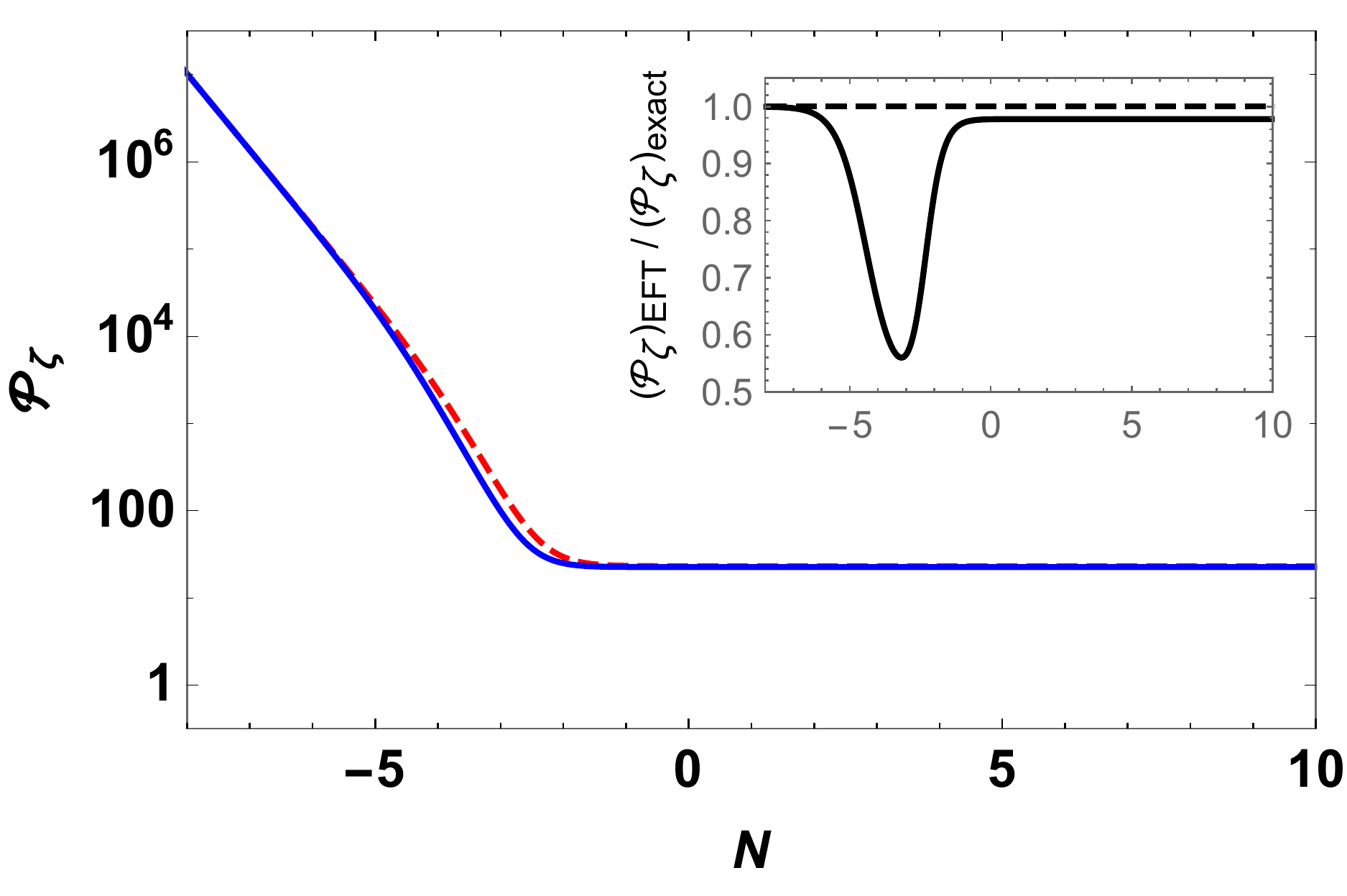}
            \caption{{\small Natural inflation ($f=10$)}}    
            \label{fig:EFT-exact-NI-minimal}
        \end{subfigure}
        \caption{Power spectra of the curvature perturbation as functions of the number of e-folds, computed numerically in the full two-field model (exact, in dashed red), and from the effective field theory \eqref{S2-EFT} using \eqref{evolution-power-spectrum}, for SI (left) and NI with $f=10$ (right) in the minimal geometry. The spectra are evaluated for the scale that crosses the Hubble radius 55 e-folds before the end of inflation, at $N=0$ in the plots, and are normalized by the adiabatic result \eqref{Pad}. The insets show the ratios between the EFT and the exact results.}
        \label{fig:EFT-exact-minimal}
    \end{figure*}
We now discuss the single-field effective field theory behind sidetracked inflation in the minimal geometry. As we have seen, we have $|m_s^2|/H^2 \ll 1$ in that case, and one would usually not expect to be able to integrate out a light field around Hubble crossing. However, this picture can be modified when the background trajectory does not follow a geodesic, as this introduces the new mass scale $H^2 \etaperp^2$. When it is much larger than the Hubble scale, like in sidetracked inflation, a non-trivial dynamics is arising on sub-Hubble scales, and it is then legitimate to integrate out the entropic field. This situation has been studied in references \cite{Cremonini:2010ua,Baumann:2011su,Gwyn:2012mw,Gwyn:2014doa}, to which we refer the reader for more details. The resulting effective action for the curvature perturbation is still formally given by \eqref{S2-EFT}, but the relevant energy scale of applicability and phenomenology are markedly different from what we discussed previously.

 On sub-Hubble scales, one can now neglect $m_s^2$ with respect to $k^2/a^2$ in the expression of the effective speed of sound \eqref{cs2k}. And while the dynamics is naturally of Bunch-Davies type deep on sub-Hubble scales, with $c_s^2(k)Ê\simeq 1$ for $k^2/a^2 \gg H^2 \etaperp^2$, one obtains $c_s^2(k) \simeq  k^2/(4 a^2 H^2 \etaperp^2)$ in the relevant intermediate regime $4 H^2 \etaperp^2 \gg k^2/a^2 \gg m_s^2$. The dynamics of the cosmological fluctuations is hence characterized by a non-linear dispersion relation $\omega(k) \propto k^2$, similarly to what arises in ghost inflation \cite{ArkaniHamed:2003uz}. It is distinct however, contrary to what the familiar form of the evolution equation \eqref{zeta-cs} might suggest. In that case, indeed, the speed of sound is not slowly evolving compared to the scale factor, and the friction term $-2 s H \dot{\zeta}_k$ is not a small correction to the Hubble friction. With $s \simeq-1$ and with $\eta \ll1$, Eq.~\eqref{zeta-cs} indeed reads
\beq
\ddot{\zeta}_k+5H \dot{\zeta}_k+\frac{1}{4 H^2 \etaperp^2}\frac{k^4}{a^4} \zeta_k \simeq0\,,
\label{zeta-equation-modified-dispersion}
\eeq
which displays both a quadratic dispersion relation and an unusual friction term. Upon quantization and the choice of the Bunch-Davies vacuum, the relevant solution reads \cite{Baumann:2011su} 
\beq
\zeta_k=\frac{H}{k^{3/2}} \sqrt{\frac{\pi}{2 \epsilon}} \etaperp^{1/4} y^{5/2} H_{5/4}^{(1)}(y^2)
\label{zeta-nonlinear-dispersion}
\eeq
where $y \equiv -k \tau/(2 \sqrt{\etaperp})$, and all slowly evolving parameters have taken to be constants here to obtain an analytical solution. With $y^{5/2}H_{5/4}^{(1)}(y^2)  \underset{y \sim 0}{\sim} -i \frac{2^{5/4}}{\pi} \Gamma(\frac54)$, one finds that the curvature perturbation becomes constant soon after $y \sim 1$, with an almost scale-invariant power spectrum
\beq
{\cal P}_{\zeta_k} = \frac{ \sqrt{2} \Gamma(5/4)^2}{\pi^3} \left(\frac{H^2}{\epsilon}\sqrt{\etaperp}\right)_\star\,
\label{result-power-minimal}
\eeq
(note that the enhancement of the power spectrum by $\sqrt{\etaperp}$ in this kind of setup was first given in \cite{Cremonini:2010ua}).
Determining at which time $\tau_\star$ exactly should slowly evolving parameters be evaluated exceeds the accuracy of the calculation here, but it is natural to choose it such that $y_\star=1$, at the transition between the two asymptotic regimes of the solution \eqref{zeta-nonlinear-dispersion}. From \eqref{result-power-minimal}, one then finds
\beq
n_s-1=-2 \epsilon_\star-\eta_\star+\frac12 \left( \frac{\etaperp'}{\etaperp} \right)_\star\,,
\label{ns-dispersion}
\eeq
where the new last term is small, as we explained below Eq.~\eqref{adiabaticity-condition}. With the explicit expression \eqref{etaperp-minimal} of $\etaperp$, one finds $ \frac{\etaperp'}{\etaperp}  \simeq \epsilon+\frac{\chi'}{\chi}\left( 1+\frac{2\chi^2}{M^2} \right)^{-1}$, where it can easily be checked that the second term is negligible compared to the first, simplifying the general result \eqref{ns-dispersion} to 
\beq
n_s-1=-\frac32 \epsilon_\star-\eta_\star.
\label{ns-dispersion-simple}
\eeq

In Fig.\ \ref{fig:EFT-exact-minimal}, we show the curvature perturbation power spectrum computed numerically in the full two-field model, and from the effective field theory \eqref{S2-EFT} using \eqref{evolution-power-spectrum}, for SI (left) and NI with $f=10$ (right) in the minimal geometry. The two results are in very good agreement, as well as with the analytical solution \eqref{zeta-nonlinear-dispersion}. In SI, the latter predict a value of the enhancement of the power spectrum compared to the adiabatic result ${\cal P}_{\zeta}/{\cal P}_{{\rm ad}} \simeq 12.9$ (respectively $13.6$ for the two-field numerical result), as well as $n_s=0.969$ (respectively $0.968$). In NI, one predicts ${\cal P}_{\zeta}/{\cal P}_{{\rm ad}} \simeq 22.5$ (respectively $23.0$ for the two-field numerical result), as well as $n_s=0.969$ (respectively $0.969$).

\subsubsection{Summary}    

Here, we simply collect and present in a unified manner in table \ref{tab:sidetracked} the characteristic features and analytical results for the observables $n_s$ and $r$, for the three different types of sidetracked inflationary scenarios that we encountered. We give expressions for the observables that are expressed in terms of general quantities, applicable to other multifield scenarios with the same characteristics, as well as specific results that take into account the particular background of sidetracked inflation. Note that the factor $\left(\frac{H_{k=aH}}{H_\star} \right)^2$ appearing in the expression of $r$ comes from the different time around which the curvature power spectrum and the tensor one become constant.

\setlength{\tabcolsep}{6pt}
\renewcommand{\arraystretch}{1.75}
\begin{table}[h]
\centering

\begin{tabular*}{0.7\linewidth}{@{\extracolsep{\fill}}ccc}
\hline
 & Minimal geometry \\
\hline\hline
Entropic mass & $|m_s^2| \ll H^2$ \\
\hline
Single-field EFT& Quadratic dispersion relation   \\
\hline
Relevant time $\star$ & $k/a=2 H \sqrt{|\etaperp|}$  \\
 \hline
$r$& $  \frac{\sqrt{2} \pi}{(\Gamma(5/4))^2} \frac{\epsilon_\star}{\sqrt{\eta_{\perp\star}}}  \left(\frac{H_{k=aH}}{H_\star} \right)^2  $   \\
\hline
$n_s-1$&$-2 \epsilon_\star-\eta_\star+\frac12 \left( \frac{\etaperp'}{\etaperp} \right)_\star \simeq -\frac32 \epsilon_\star-\eta_\star$ \\
\hline
& \\
\end{tabular*}

\begin{tabular*}{1\linewidth}{@{\extracolsep{\fill}}ccc}
\hline
\multirow{2}{*}{} & Hyperbolic geometry & Hyperbolic geometry \\
& and $V_{,\vphi}>0$ & and $V_{,\vphi}<0$ \\
\hline\hline
Entropic mass &  $ m_s^2 \gg H^2, \, m_s^2 > 0$ & $ |m_s^2 | \gg H^2, \, m_s^2 < 0$   \\
\hline
\multirow{2}{*}{Single-field EFT} & Reduced speed of sound & Imaginary speed of sound \\
& $0<c_s^2 \ll1$ & $c_s^2<0$ \\
\hline
Relevant time $\star$ &  $k/a=H/c_s $ & $k/a=|m_s|$ \\
\hline
\multirow{2}{*}{$r$} & \multirow{2}{*}{$16 \,\epsilon_\star c_{s \star} \left(\frac{H_{k=aH}}{H_\star} \right)^2$} & $64  \,\epsilon_\star e^{- |\frac{m_s^2}{H^2 \etaperp}|} \left(\frac{H_{k=aH}}{H_\star} \right)^2$  \\
& & $\simeq
64\,\epsilon_\star e^{-6 \sqrt{2} \chi_\star/M} \left(\frac{H_{k=aH}}{H_\star} \right)^2$ \\
\hline
\multirow{2}{*}{$n_s-1$} & \multirow{2}{*}{$-2 \epsilon_\star-\eta_\star-s_\star \simeq -\frac32 \epsilon_\star-\eta_\star$} & $-2 \epsilon_\m-\eta_\m+ |\frac{m_s^2}{H^2 \etaperp}|^{'}$ \\
& & $\simeq-2 \epsilon_\m-\eta_\m+6 \sqrt{2}\chi^{'}_\m/M$ \\
\hline
\end{tabular*}

\caption{Comparison between the three different sidetracked inflationary scenarios.}
\label{tab:sidetracked}
\end{table}

\subsection{Numerical results for all potentials}
\label{numerical-results}

In this section, we give the results for the observables $r$ and $n_s$ for the four type of potentials that we have studied, and the various lists of parameters indicated in table \ref{tab:models}, both for the minimal and the hyperbolic geometry. It is interesting to compare our results with the values of $r$ and $n_s$ of single-field inflation, that is the results in the absence of any geometrical destabilization. The difference between the two outcomes is therefore a measure of the overall observable consequences of the instability and the second sidetracked phase. To better quantify how the predictions are affected by the non-trivial multifield effects, we further do another comparison by displaying the power spectrum parameters calculated on the sidetracked trajectory using the adiabatic description, \textit{i.e.} by completely neglecting entropic perturbations. As the evolution of the scale factor is close to the de-Sitter one, in the sense made precise in section \ref{fluctuations-adiabatic-entropic}, the adiabatic power spectrum is given to a very good approximation by Eq.~\eqref{Pad}, which hence gives
\beq \label{eq:adiabatic result}
r_{{\rm ad}}=16\,\epsilon\,,\qquad n_{s \, {\rm ad}}=1-2\epsilon-\eta\,,
\eeq
where all quantities are evaluated at Hubble crossing such that $k=aH$.

We present the results in the $(n_s,r)$ plane in figs.\ \ref{fig:r and ns minimal} and \ref{fig:r and ns hyperbolic}, while precise values are also listed in the appendix \ref{sec:app1}. The different markers used in the plots relate to the three descriptions above: the results labeled ``exact'' correspond to the numerical results in the full sidetracked inflation set-up; the ones labeled ``without GD'' mean those obtained along $\chi=0$, \textit{i.e.} when the geometrical destabilization is overlooked; and the description called ``adiabatic'' is the one where we use Eq.\ \eqref{eq:adiabatic result} to compute the results of the sidetracked phase.

All the results that we obtained are in very good agreement with the predictions of the single-field effective theories that we have derived in section \ref{EFT}, and although it will be tedious to make a detailed account of all the 36 models, we will comment on how the EFT results, summarized in table \ref{tab:sidetracked}, enable one to explain the different behaviours and parameters' dependences that we observe.

\subsubsection{Minimal geometry}
\label{results-minimal}

\begin{figure*}
        \centering
        \begin{subfigure}[b]{0.475\textwidth}
            \centering
            \includegraphics[width=\textwidth]{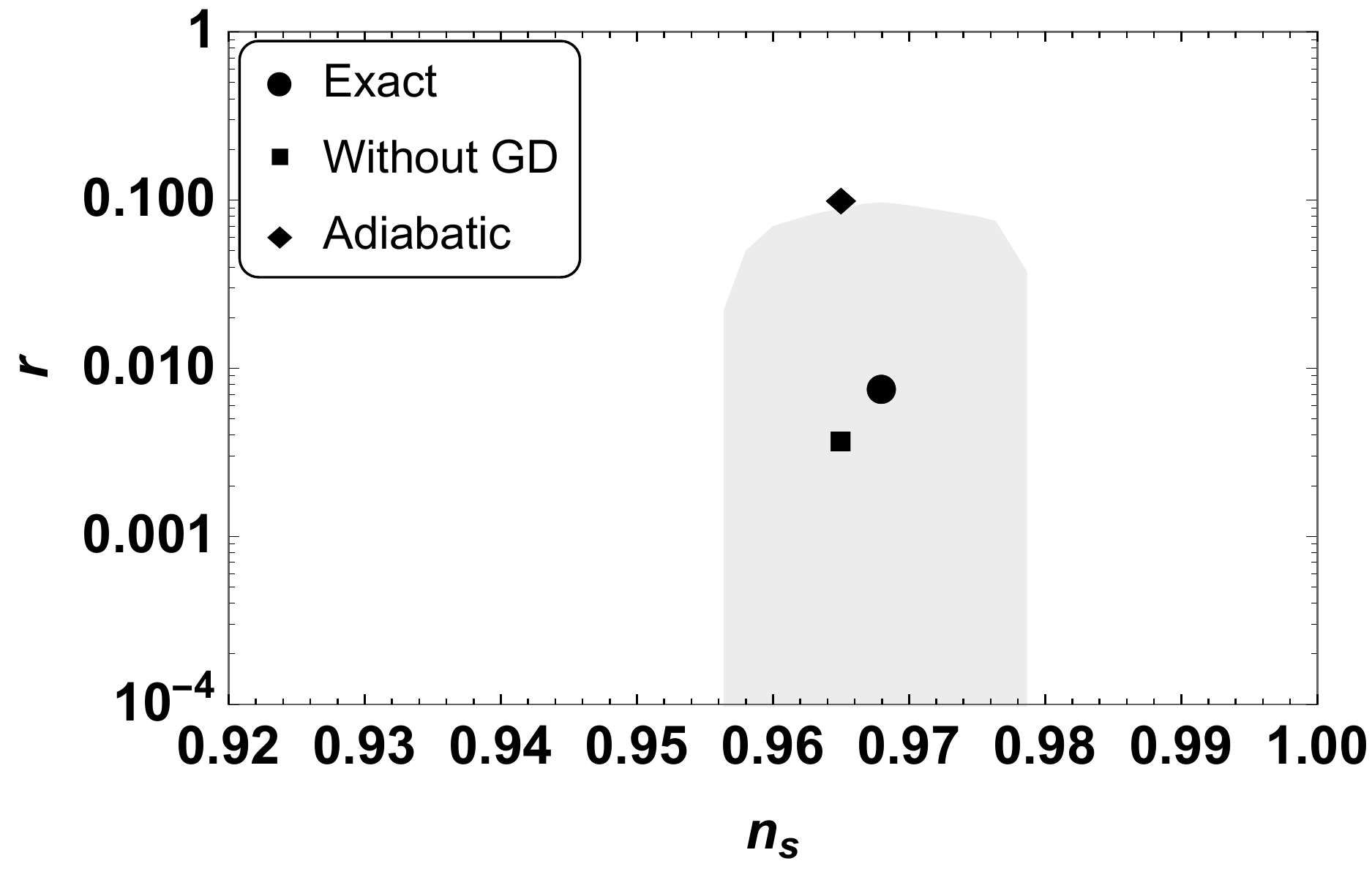}
            \caption{{\small Starobinsky inflation}}    
            \label{fig:r and ns minimal SI}
        \end{subfigure}
        \hfill
        \begin{subfigure}[b]{0.475\textwidth}  
            \centering 
            \includegraphics[width=\textwidth]{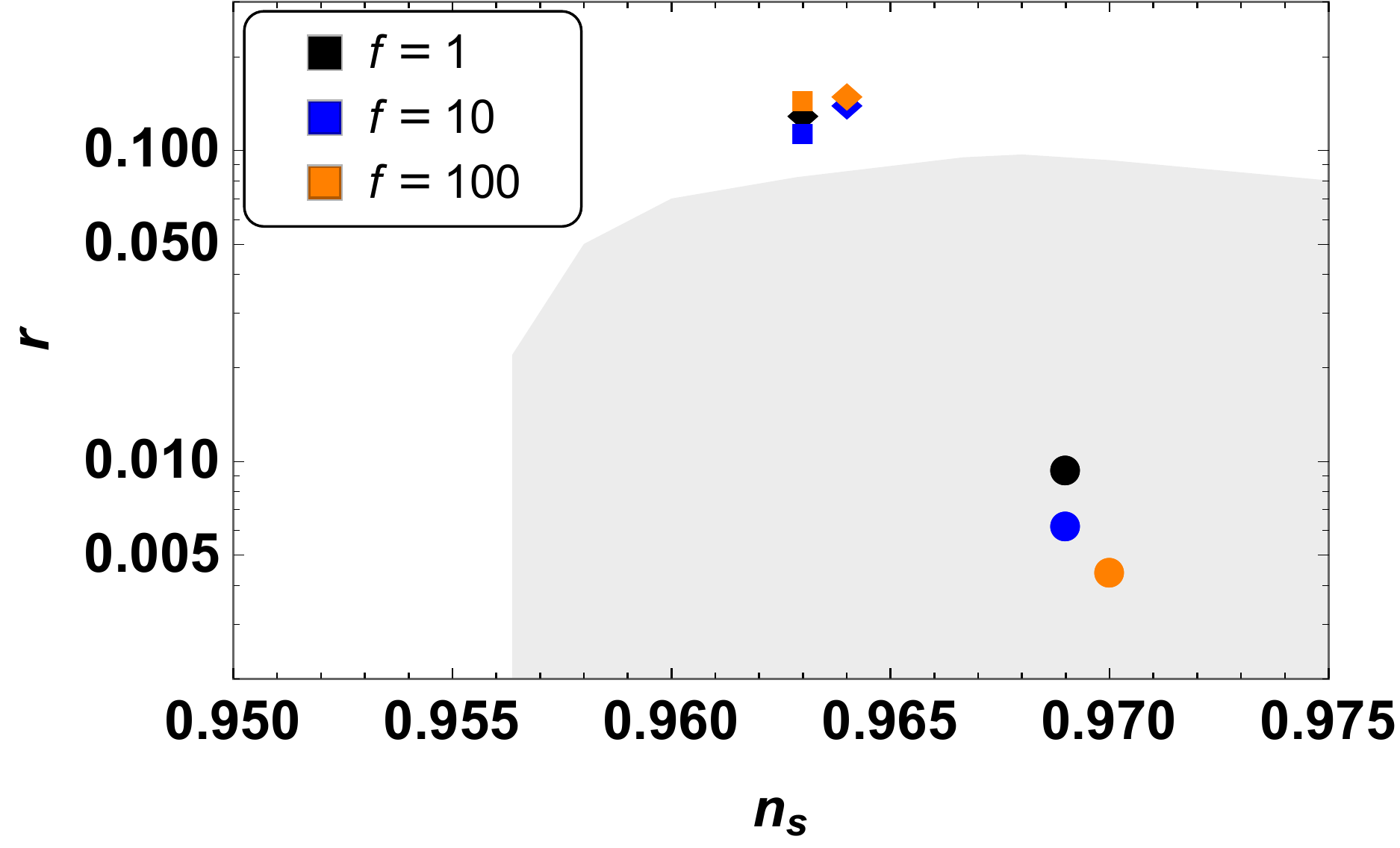}
            \caption{{\small Natural inflation}}    
            \label{fig:r and ns minimal NI}
        \end{subfigure}
        \vskip\baselineskip
        \begin{subfigure}[b]{0.475\textwidth}   
            \centering 
            \includegraphics[width=\textwidth]{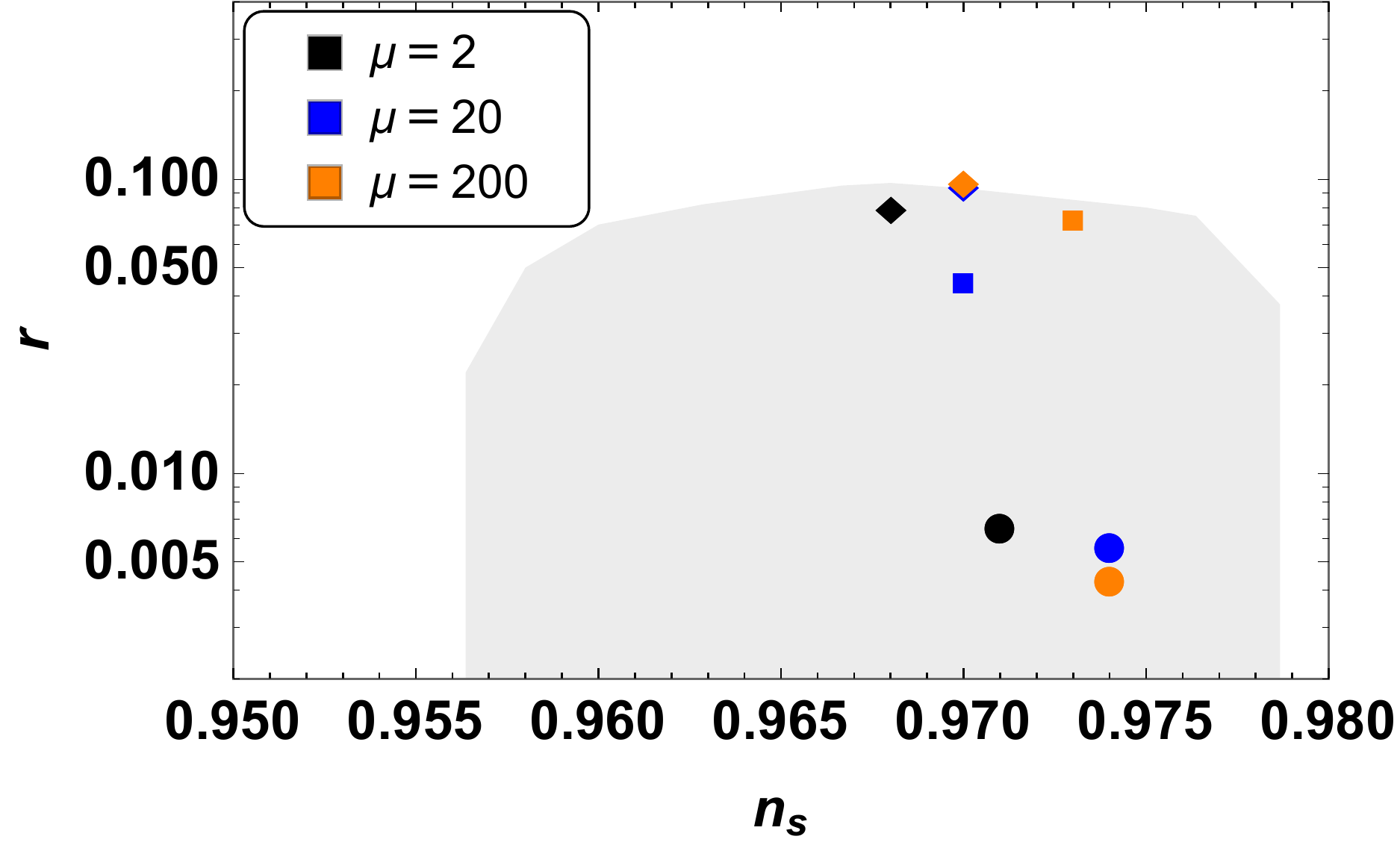}
            \caption{{\small Quadratic small field inflation}}    
            \label{fig:r and ns minimal SFI}
        \end{subfigure}
        \quad
        \begin{subfigure}[b]{0.475\textwidth}   
            \centering 
            \includegraphics[width=\textwidth]{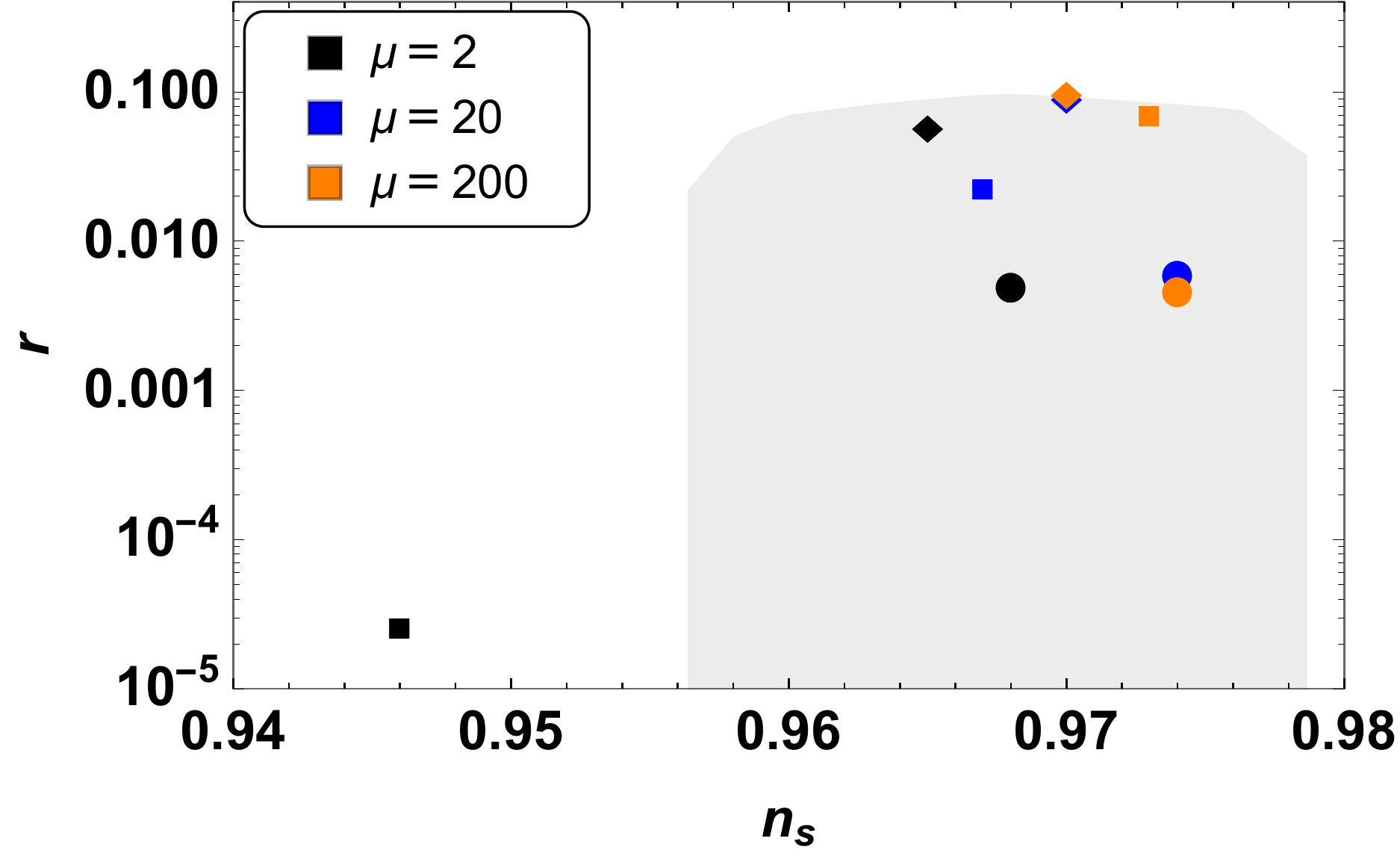}
            \caption{{\small Quartic small field inflation}}    
            \label{fig:r and ns minimal IP}
        \end{subfigure}
        \caption{Results for the power spectrum parameters $n_s$ and $r$ in the minimal field space geometry defined by the metric \eqref{eq:minimalmetric}. The shaded region represents approximately the experimental bounds of Planck 2015 \cite{Ade:2015lrj}. As indicated in the legends, different colors label different models, while the marker shapes correspond to the three descriptions we consider, as explained in the main text.} 
        \label{fig:r and ns minimal}
    \end{figure*}

We show in fig.\ \ref{fig:r and ns minimal} the results in the minimal geometry \eqref{eq:minimalmetric} for the ten models under study, at the fixed curvature scale $M=10^{-3} \Mp$. One of the first thing to notice is that the exact tensor-to-scalar ratio $r$ is always smaller than its adiabatic counterpart \eqref{eq:adiabatic result}. This is well understood using the analytical result for $r$ in table \ref{tab:sidetracked}, which gives
\beq
\frac{r}{r_{{\rm ad}}} \simeq 0.34 \frac{\epsilon_\star}{\epsilon_{k=aH}}  \left(\frac{H_{k=aH}}{H_\star} \right)^2 \frac{1}{\sqrt{|\eta_{\perp \star}|}}\,.
\label{r/rad-minimal}
\eeq
Because $\epsilon$ grows in time in these models, the time $\star$ is earlier than the one of Hubble crossing, and more importantly because of the large bending, all the factors in \eqref{r/rad-minimal} are indeed smaller than unity. Note that, since the adiabatic result can be greater than without geometrical destabilization (and is often so), the exact $r$ can also be bigger, like in SI and $\mathrm{SFI}_4$.
Concerning the scalar spectral index, one can observe that in all models, $n_s> _{s \, {\rm ad}}$. This can also be easily understood, as the result of two effects: first, as $\epsilon$ and $\eta$ are increasing functions of time in these models, one has 
$\left(-2 \epsilon-\eta\right)_{k=aH} < \left(-2 \epsilon-\eta\right)_{\star}$. Moreover, compared to the adiabatic result, $n_s$ in Eq.~\eqref{ns-dispersion-simple} has the additional positive contribution $+\frac12 \left( \frac{\etaperp'}{\etaperp} \right)_\star \simeq \frac12 \epsilon_\star$.

One can also observe that for a given model, the bigger the scale $f$ or $\mu$ in its potential, and the larger the decrease of $r$ compared to the adiabatic result. One should be careful in the comparison, because the various trajectories are different then. One can nonetheless explain this trend using our analytical formulae. For this, note that the suppression in Eq.~\eqref{r/rad-minimal} is dominated by the large bending, with $\etaperp$ given in \eqref{etaperp-minimal} which depends on $m_h/H$, and $\chi/M$. The bigger the scale $f$ or $\mu$ in its potential, and the flatter it is. As the potential gets flatter, the duration of the sidetracked phase increases. And as we used the same initial condition $m_h/H_c=10$ at the critical time preceding the sidetracked phase, this gives a larger $m_h/H$ when evaluated $\simeq 55$ e-folds before Hubble crossing. Using \eqref{chi-phi-2}, this effect, combined with a flatter potential, generates smaller values of $\chi/M$. This effect is however numerically milder than the growth of $m_h/H$. This explains why the bending is larger for flatter potentials, and hence why the suppression of $r$ by multifield effects is more important.

Let us also briefly comment on the dependence on the observables on the curvature scale $M$, with results listed in table \ref{T:Table_4}. While the dependence of $n_s$ on $M$ is mild, we observe for all models (except SI) that a smaller $M$ comes with a smaller $r$. Again, one should be careful in comparing different models, but this result can be understood intuitively: as the field space curvature increases, one expects the sidetracked phase to display more bending, and hence more multifield effects. Indeed, one can check more quantitatively that both $\chi/M$ and $\etaperp$ increase as $M$ decrease, hence the smaller tensor-to-scalar ratio.

\subsubsection{Hyperbolic geometry}

\begin{figure*}
        \centering
        \begin{subfigure}[b]{0.475\textwidth}
            \centering
            \includegraphics[width=\textwidth]{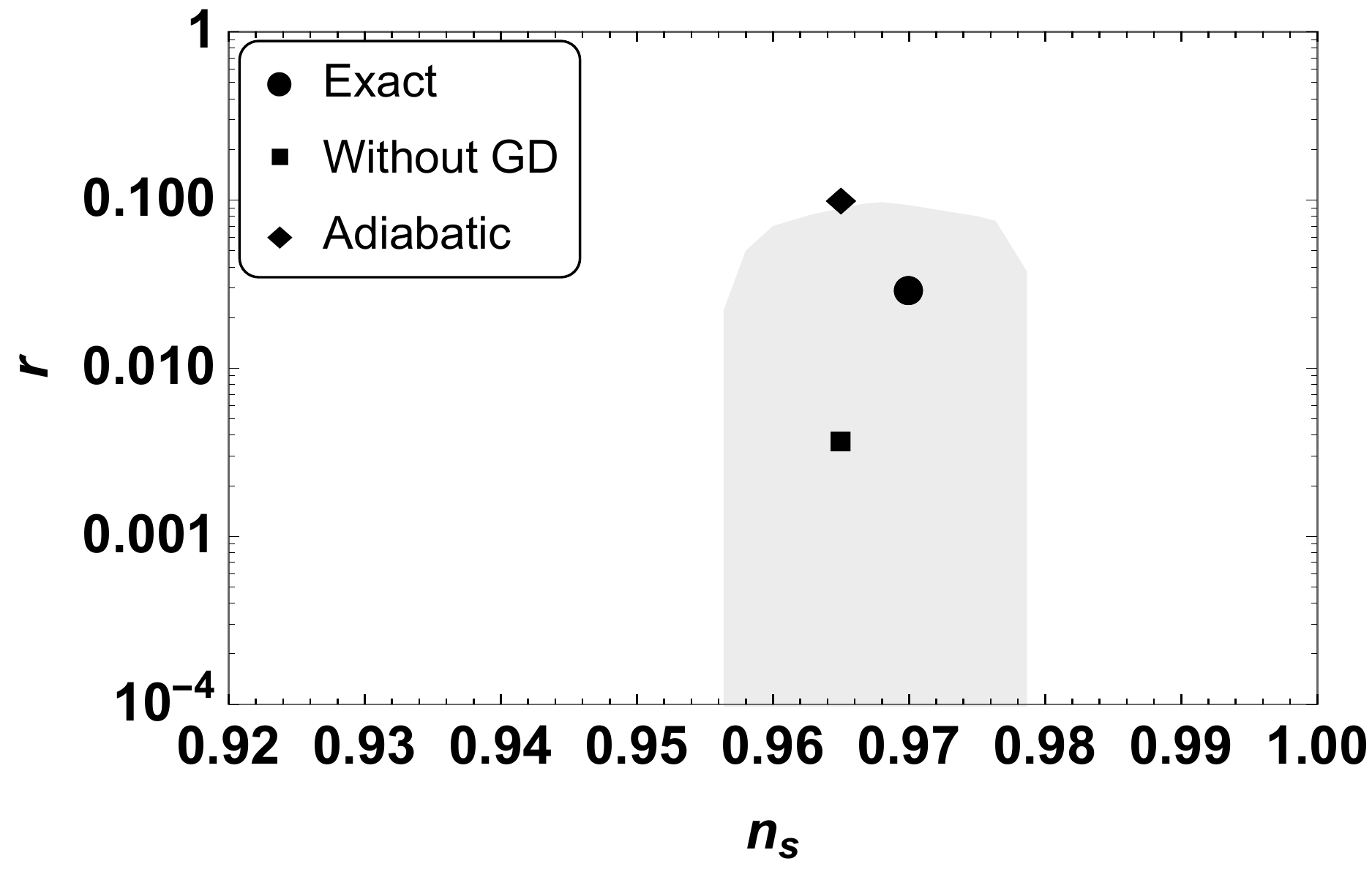}
            \caption{{\small Starobinsky inflation}}    
            \label{fig:r and ns hyperbolic SI}
        \end{subfigure}
        \hfill
        \begin{subfigure}[b]{0.475\textwidth}  
            \centering 
            \includegraphics[width=\textwidth]{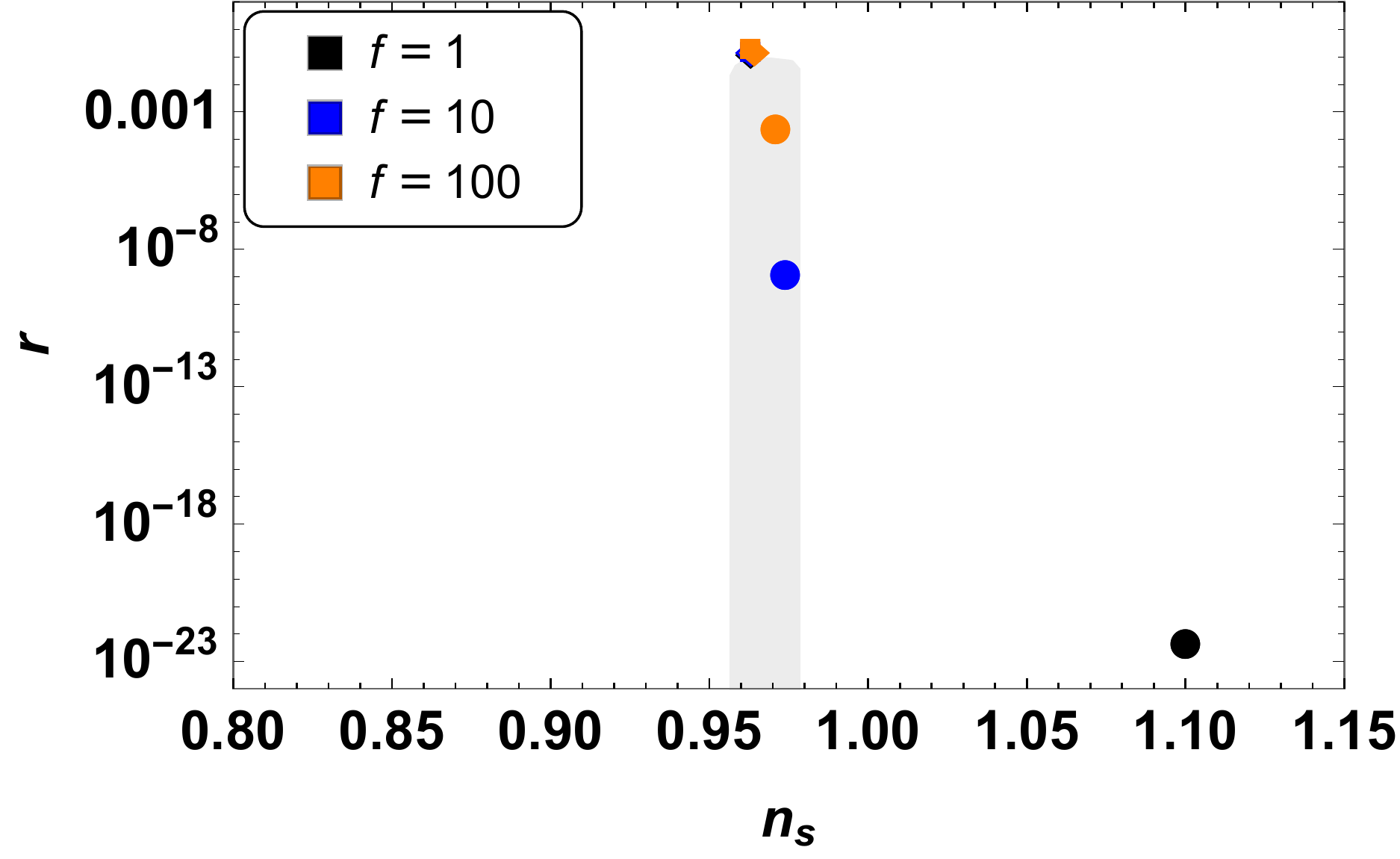}
            \caption{{\small Natural inflation}}    
            \label{fig:r and ns hyperbolic NI}
        \end{subfigure}
        \vskip\baselineskip
        \begin{subfigure}[b]{0.475\textwidth}   
            \centering 
            \includegraphics[width=\textwidth]{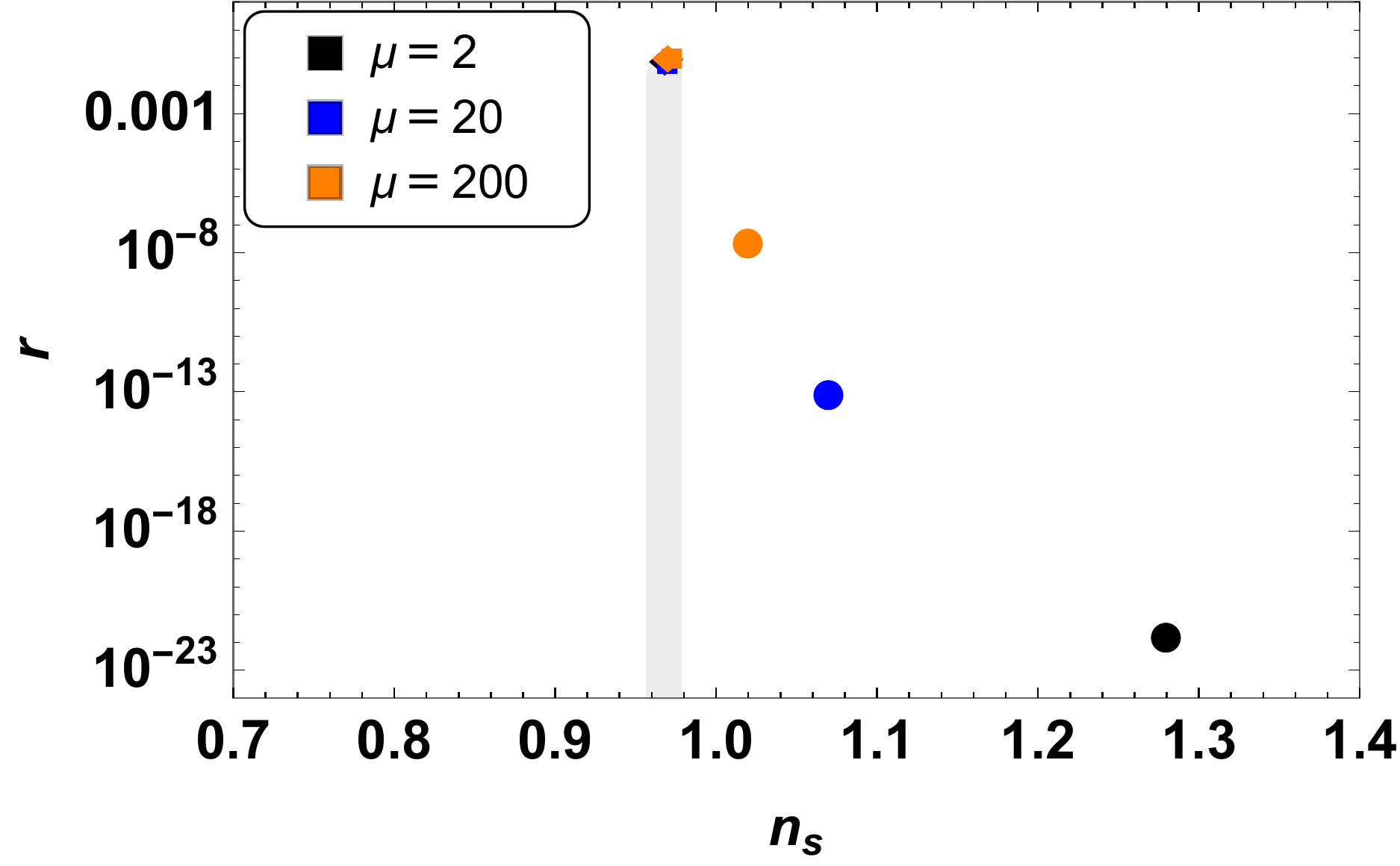}
            \caption{{\small Quadratic small field inflation}}    
            \label{fig:r and ns hyperbolic SFI}
        \end{subfigure}
        \quad
        \begin{subfigure}[b]{0.475\textwidth}   
            \centering 
            \includegraphics[width=\textwidth]{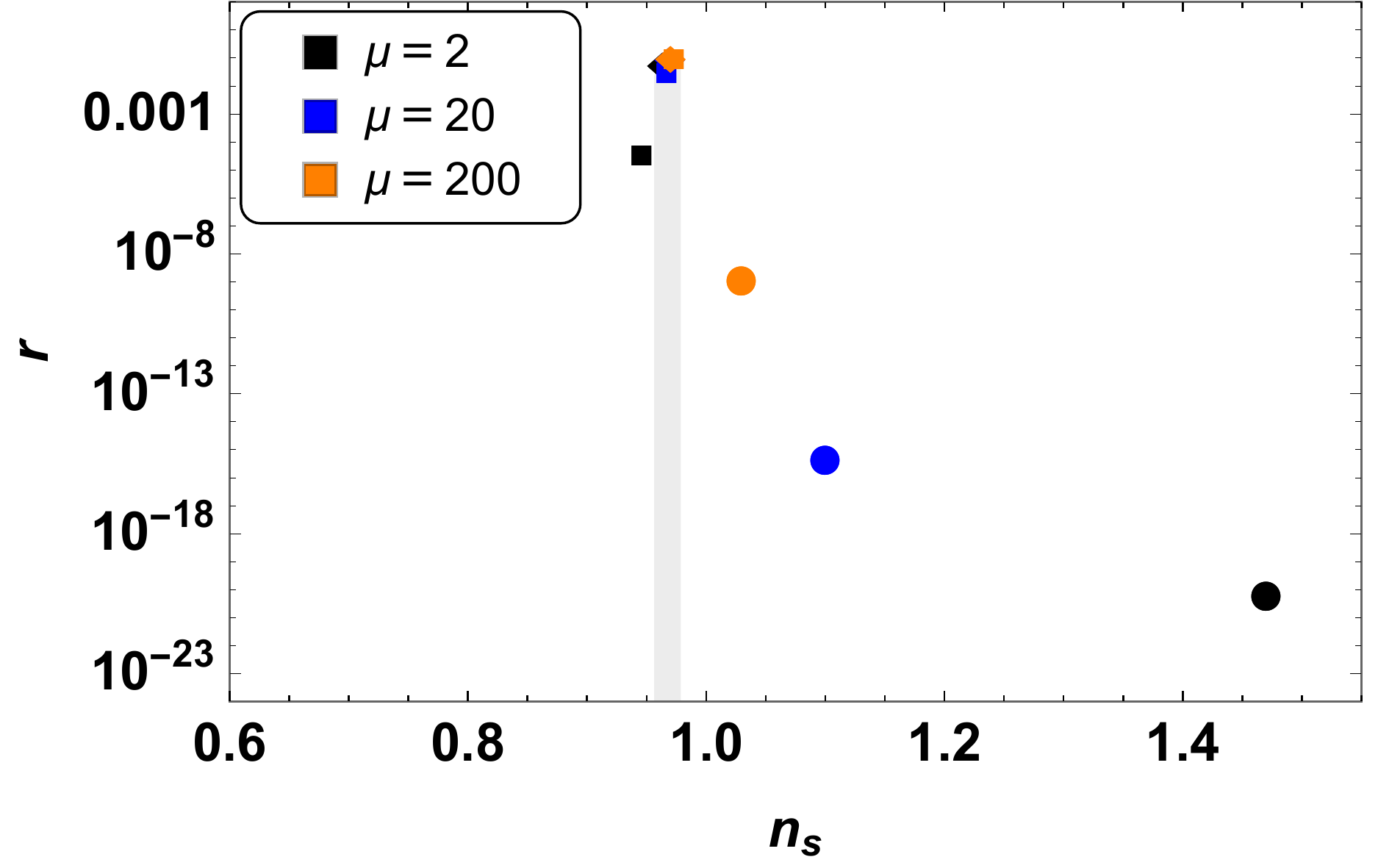}
            \caption{{\small Quartic small field inflation}}    
            \label{fig:r and ns hyperbolic SFI4}
        \end{subfigure}
        \caption{Results for the power spectrum parameters $n_s$ and $r$ in the hyperbolic field space geometry defined by the metric \eqref{eq:hyperbolicmetric}. The shaded region represents approximately the experimental bounds of Planck 2015 \cite{Ade:2015lrj}. As indicated in the legends, different colors label different models, while the marker shapes correspond to the three descriptions we consider, as explained in the main text. Note the logarithmic scale used to represent the very large spread of values of $r$. 
        } 
        \label{fig:r and ns hyperbolic}
    \end{figure*}

Continuing with the hyperbolic field space geometry \eqref{eq:hyperbolicmetric}, we present in fig.\ \ref{fig:r and ns hyperbolic}  the results of $r$ and $n_s$ for the ten models under study, at the fixed curvature scale $M=10^{-3} \Mp$.
The most striking fact lies in the very small values of $r$ obtained in all models except SI, with a decrease with respect to the adiabatic result by several orders of magnitude. This is in perfect agreement with the understanding gained in section \ref{EFT}: like NI, $\mathrm{SFI}_2$ and $\mathrm{SFI}_4$ have a negative slope, hence they fall in the category studied in \ref{imaginary-speed-sound} of models with a transient tachyonic instability, and that can be described by an effective single-fied theory with an imaginary speed of sound around Hubble crossing. This leads to a very large enhancement of the power spectrum compared to the adiabatic result, and hence a decreased $r$, following
\beq
\frac{r}{r_{{\rm ad}}} \simeq 4 \frac{\epsilon_\star}{\epsilon_{k=aH}}  \left(\frac{H_{k=aH}}{H_\star} \right)^2e^{-6 \sqrt{2} \chi_\star/M}\,.
\label{r/rad-instability}
\eeq
Like in the minimal case, all the factors on the right hand are smaller than unity, but with $\chi_\star={\cal O}(M)$, the effect is largely dominated by the exponential factor. 

Another observation is that $n_s$ in these models can deviate rather strongly from scale invariance, notably with a blue spectrum in $\mathrm{SFI}_2$ and $\mathrm{SFI}_4$. This can be understood using our estimate \eqref{ns-instability}: $n_s-1=-2 \epsilon_\m-\eta_\m+6 \sqrt{2}\chi^{'}_\m/M$. With the approximate expressions \eqref{velocity-chi}-\eqref{velocity-phi-1}, one can find indeed
\beq
\frac{\chi'}{M}\simeq \frac14 \frac{M}{\chi} \left[\frac{\Mp^2}{2} \left(\frac{V_{,\vphi}}{V} \right)^2 -\Mp^2 \frac{V_{,\vphi \vphi}}{V}\right]\,,
\label{chi'M}
\eeq
where all the terms are positive for the concave potentials of $\mathrm{SFI}_2$ and $\mathrm{SFI}_4$ (and the net result is also positive for NI),
so that the last contribution to $n_s-1$ is positive.

Using these formulae, one can also understand the dependence of the observables on the parameter ($f$ and $\mu$) controlling the steepness of the potential. We have seen in section \ref{results-minimal} that the smaller this scale, the bigger the value of $\chi_\star/M$ (remember that the background in the minimal and the hyperbolic geometry are the same to a good approximation). As $r$ depends exponentially on $\chi_\star/M$, this well explains the huge decrease of $r$ as this scale gets lower. As $\mu$ say, decreases, two competing effects arise for $\chi'/M$ in Eq.~\eqref{chi'M}: $M/\chi$ decreases, but $\left(V_{,\vphi}/V \right)^2$ and $V_{,\vphi \vphi}/V$ decrease. The latter effect, as $1/\mu^2$, is however more important than the decrease of $M/\chi$, which roughly scales as $\mu^{1/2}$. As a result, $\chi'_\star/M$ increases when lowering $\mu$ or $f$, and so does its large positive contribution to $n_s-1$, in plain agreement with the results visible in fig.\ \ref{fig:r and ns hyperbolic}.
Eventually, one can see in table \ref{T:Table_4} that for all models with negative slope, $r$ decreases exponentially as $M$ decreases. We have indeed indicated that $\chi/M$ increases as $M$ decrease, so this result is well understood as a result of the exponential dependence of $r \propto e^{-6 \sqrt{2} \chi_\star/M}$.


\section{Primordial non-Gaussianities} \label{sec:bispectrum}

In the preceding section, we have seen that the curvature power spectrum generated in sidetracked inflation can be understood by an effective single-field description of the fluctuations with, depending on the type of scenarios, an imaginary speed of sound, a reduced speed of sound, or a modified dispersion relation. It is well understood in the framework of the effective field theory of fluctuations that the two latter situations come along with large primordial non-Gaussianities of the curvature perturbation \cite{Cheung:2007st}. Hence it is a natural question to investigate the non-Gaussian signal generated in sidetracked inflation (see \textit{e.g.} \cite{Wands:2010af,Chen:2010xka,Wang:2013eqj,Renaux-Petel:2015bja} for reviews about primordial non-Gaussianities). For this purpose, we make a preliminary analysis by numerically calculating the bispectrum for the various models under study. It is a non-trivial task, both theoretically and numerically, to calculate the bispectrum from generic nonlinear sigma models of inflation with curved field spaces, and it is only recently that the powerful transport approach has been numerically implemented to calculate the bispectrum in this framework, with the codes \texttt{PyTransport 2.0} \cite{Ronayne:2017qzn} and  \texttt{CppTransport} \cite{Butchers:2018hds} (see \cite{Chen:2006xjb,Chen:2008wn,Funakoshi:2012ms,Hazra:2012yn,Horner:2013sea,Horner:2015gza} for previous works on numerical computation of the bispectrum in single-field models). In this work, we made use of the former package. We stress that with this approach, one can efficiently compute from first principles the amplitude of the bispectrum for all types of triangle configurations, taking into account all physical effects at tree-level in the `in-in' formalism. We thus have access, not only to overall amplitude of the bispectrum, but also to its full shape and scale dependencies.

 \subsection{Non-Gaussian shapes and correlators}
 \label{shapes}
 
We are interested in the three-point correlation function of the curvature perturbation: 
 \beq
 \langle \zeta_{\boldsymbol{k}_1} \zeta_{\boldsymbol{k}_2} \zeta_{\boldsymbol{k}_3} \rangle \equiv (2\pi)^3 \delta(\sum_i \boldsymbol{k}_i) B_{\zeta}(k_1,k_2,k_3)\,,
 \label{Bispectrum}
 \eeq
 where the factor $\delta(\sum_i \boldsymbol{k}_i)$, coming from statistical homogeneity, implies that the wavevectors form a triangle in Fourier space. Additionally, statistical isotropy entails that the orientation of this triangle irrelevant, so that only its shape and overall scale matter, hence the dependence on the three wavenumbers $k_i=|\boldsymbol{k}_i|$ only. 
 
We will plot the reduced bispectrum $f_{nl}$, defined as
\begin{equation}
\frac{6}{5}f_{nl}\left(k_1,k_2,k_3\right) \equiv \frac{B_{\zeta}(k_1,k_2,k_3)}{P_{\zeta}(k_1) P_{\zeta}(k_2)+P_{\zeta}(k_1)P_{\zeta}(k_3)+P_{\zeta}(k_2)P_{\zeta}(k_3)},
\label{fNL}
\end{equation}
where, by putting a constraint on the overall scale, $k_s=k_1+k_2+k_3$, we can parameterize any shape by $\alpha$ and $\beta$ such that\footnote{When studying the shape dependence of the bispectrum, we always use $k_s=3\, k_{55}$.},
\begin{equation}
\begin{split}
k_1=&\frac{k_s}{4}(1+\alpha+\beta)\\
k_2=&\frac{k_s}{4}(1-\alpha+\beta)\\
k_3=&\frac{k_s}{2}(1-\beta)\,,
\end{split}
\end{equation}
with the allowed values of $(\alpha,\beta)$ falling inside a triangle in the $\alpha,\beta$ plane with vertices $(-1,0)$, $(1,0)$ and $(0,1)$ (see figs.~\ref{shapeNI} and \ref{shapeNIH} for explicit representations). In the literature, it is also customary to use the shape function $S$ such that
\beq
B_{\zeta} \equiv (2 \pi)^4 \frac{S(k_1,k_2,k_3)}{(k_1 k_2 k_3)^2} A_s^2\,,
\eeq
where $A_s \simeq 2.4 \times 10^{-9}$ denotes the amplitude of the dimensionless curvature power spectrum ${\cal P}_\zeta$ at the pivot scale $k=0.05\,{\rm Mpc}^{-1}$. Note that in the approximation of an exactly scale-invariant power spectrum, $f_{nl}(k_i)$ and $S(k_i)$ are related by 
\beq
f_{nl}\left(k_1,k_2,k_3\right) \approx \frac{10}{3} \frac{S(k_1,k_2,k_3)}{k_1^2/(k_2 k_3)+2\, {\rm perms}}\,.
\eeq

As we expect non-Gaussian shapes typical of derivative interactions, we compare the shape dependence of sidetracked inflation with the equilateral shape \cite{Creminelli:2005hu}:
\begin{equation}
\label{templateEQ}
S^{\eq}=\frac{9}{10}f^{\eq}_{nl}\left[-\left(\frac{k_1^2}{k_2k_3}+2\mathrm{perm.}\right)+\left(\frac{k_1}{k_2}+5\mathrm{perm.}\right)-2\right] \,,
\end{equation}
and the orthogonal shape \cite{Senatore:2009gt}:
\begin{equation}
\label{templateORT}
S^{\orth}=\frac{27}{10}f^{\orth}_{nl}\left[-\left(\frac{k^2_1}{k_2k_3}+2\mathrm{perm.}\right)+\left(\frac{k_1}{k_2}+5\mathrm{perm.}\right)-\frac{8}{3}\right]\,.
\end{equation}
For this, we compute the shape overlap of the various models with the templates given in Eqs.~\eqref{templateEQ}-\eqref{templateORT}, by calculating the correlation factor between the shape of our given model $S$ and the equilateral or orthogonal shape templates $S'$. To do so we make use of the shape correlator \cite{Fergusson:2008ra}
\begin{equation}
\bar{\mathcal{C}}(S,S')=\frac{F(S,S')}{\sqrt{F(S,S)F(S',S')}},
\end{equation}
where $F(S,S')$ is the inner product of $S$ and $S'$ with a weight function $1/k_s$.
In order to compare our results with observational constraints we use the `fudge factor' \cite{Babich:2004gb} defined as $f(S)=F(S,S')/F(S',S')$, which corrects for the misalignment between our shape and the corresponding template. The values that we thus obtain are to be compared with the observational constraints from the Planck mission \cite{Ade:2015ava}:
\begin{eqnarray}
f_{nl}^\eq&=&-4\pm 43 \quad (68 \% {\rm CL}, T+E) \\
f_{nl}^\orth&=&-26\pm 21  \quad (68 \% {\rm CL}, T+E)\,.
\label{constraints-fNL}
\end{eqnarray}
Eventually, let us make explicit the link between the parameterization of the shape in terms of the wavevectors $k_i$ and the one in terms of $(\alpha,\beta)$, for several configurations of interest (see figs.~\ref{shapeNI} and \ref{shapeNIH} for explicit $(\alpha,\beta)$ representations). The equilateral configuration ($k_1= k_2= k_3$) lies at the center of the the triangle $(\alpha,\beta)=(0,1/3)$. Its corners at $(\alpha,\beta)=(-1,0),(1,0),(0,1)$ represent the squeezed limits, in which one of the momenta is much smaller than the two others (say $k_3 \ll k_1 \simeq k_2$). Eventually, flattened configurations (with $k_2+k_3 \simeq k_1$ for instance) correspond to the edges of the triangle, with squashed configurations, such that $k_3=k_2=k_1/2$ and cyclic permutations, at $(\alpha,\beta)=(0,0),(-1/2,1/2),(1/2,1/2)$.

\subsection{Numerical results}
\label{numerical-results-bispectrum}

We now discuss our numerical results for the bispectrum, which can all be found in a tabulated form in the appendix \ref{sec:app1}. In particular, for the minimal geometry (respectively the hyperbolic one), we list in table \ref{T:Table_1} (respectively \ref{T:Table_2}) our numerical results for $f_{nl}$ in the equilateral configuration for the pivot scale $k_{55}$, for the 10 models under study, and at the fixed curvature scale $M=10^{-3} \Mp$. In table \ref{T:Table_3} (respectively \ref{T:Table_4}), similar results are shown when varying the curvature scale. Eventually, in table \ref{T:Table_5} (respectively \ref{T:Table_6}) we give for all the models the correlations between the shape of the bispectrum and the equilateral and orthogonal templates, as well as the corresponding amplitudes $f_{nl}^{\eq}$ and $f_{nl}^{\orth}$.

\subsubsection{Minimal geometry}

\begin{figure}[h]
\centering
\includegraphics[width=0.6\linewidth]{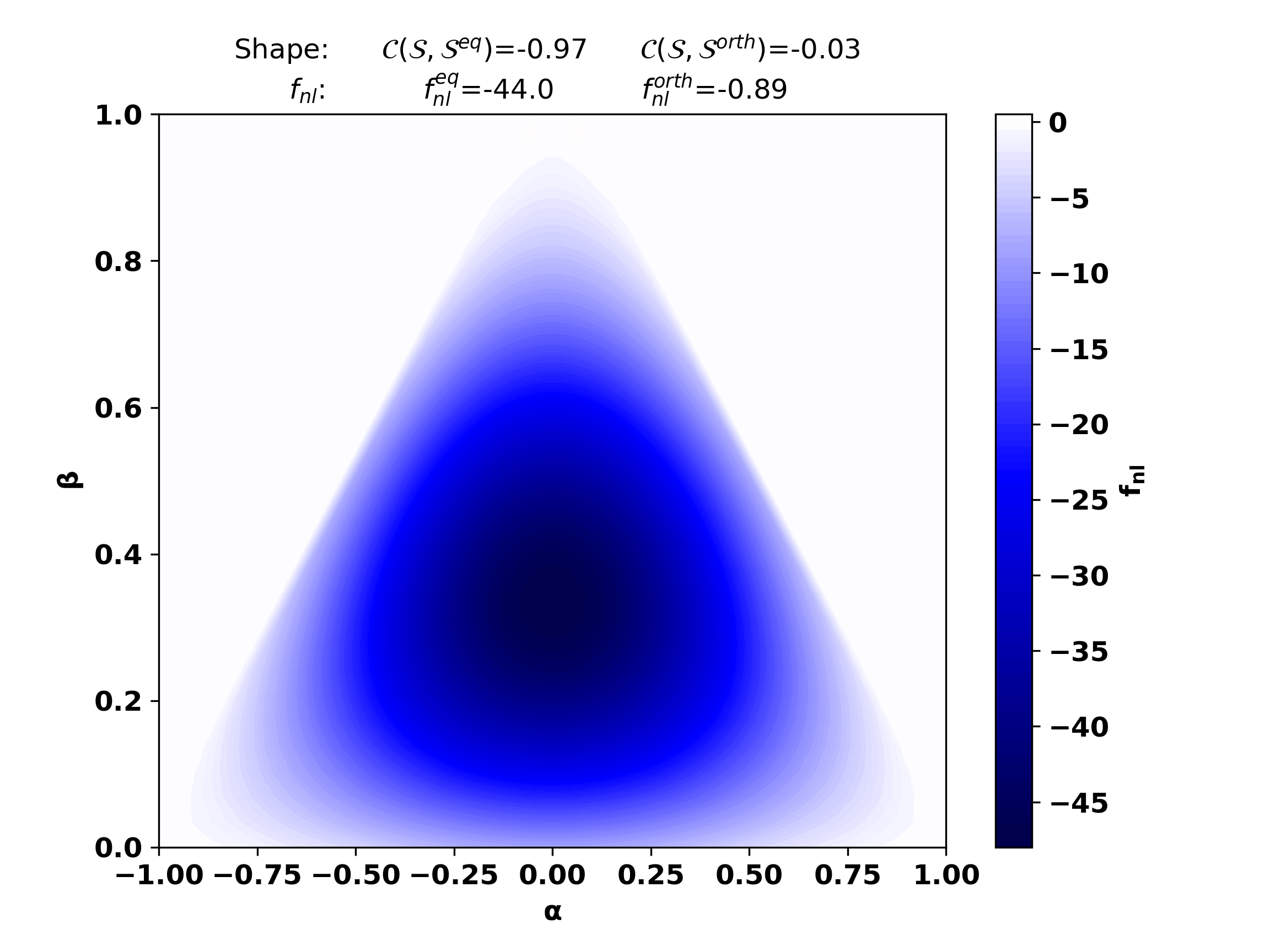}
\caption{Shape dependence $f_{nl}$($\alpha,\beta$) generated for NI with $f=10$ and $M=10^{-3}$ in the minimal geometry. We used $k_s=3\,k_{55}$. The shape has a very large correlation with the equilateral template and a very small one with the orthogonal template.}
\label{shapeNI}
\end{figure}
The first thing to notice is that a large non-Gaussian signal is generated in the minimal geometry for all type of potentials, parameters in the potential and curvature scales, as indicated by the consistently large (negative) values of $f_{nl}$ in the equilateral limit that we find, typically of a few tens, ranging from $-7$ to $-98$. 
The second striking result is that the non-Gaussian shape has a very large (anti) correlation with the equilateral template --- we find an overlap of $-0.97$ for all models --- and a negligible correlation with the orthogonal one --- with an overlap always less than $0.03$ (see table \ref{T:Table_5}). The fact that the shape is almost indistinguishable from the equilateral one is illustrated in fig.~\ref{shapeNI} for the representative example of NI with $f=10$.
More quantitatively, the fact that the shape is faithfully represented by the equilateral template implies a very low value of $f_{nl}^\orth \lesssim 1$, and a value of $f_{nl}^\eq$ almost identical to the reduced bispectrum \eqref{fNL} in the equilateral limit (only lowered by few percents).
Eventually, we observe a very clear correlation between the curvature scale and the parameter controlling the steepness of the potential on the one hand, and the value of $f_{nl}^\eq$ on the other hand: the latter grows as $M$ decreases, or the steepness parameter $f$ or $\mu$ increases, in the same way as the bending parameter $\etaperp$ does. More quantitatively, we find that all the results are in very good agreement with the simple behaviour
\beq
f_{nl}^\eq \simeq \eta_{\perp \star}\,,
\eeq
up to an order one coefficient. Similar results for the shape and the amplitude of the bispectrum have been found in related contexts in references \cite{Baumann:2011su,Gwyn:2012mw,Gwyn:2014doa} by using the effective field theory of fluctuations. There, however, only the quadratic action for the entropic field was taken into account in the unitary gauge. The fact that our full numerical results agree with this picture hints at the fact that the interactions taken into account there are dominant, and it would be interesting to study this further. Note also that although one obtains large negative values for $f_{nl}^\eq$, they lie within the Planck constraints \eqref{constraints-fNL} for all the models we have studied, with the only exception of NI with $f=100$ and $M=10^{-3}$.
\begin{figure}
\centering
\includegraphics[width=0.6\linewidth]{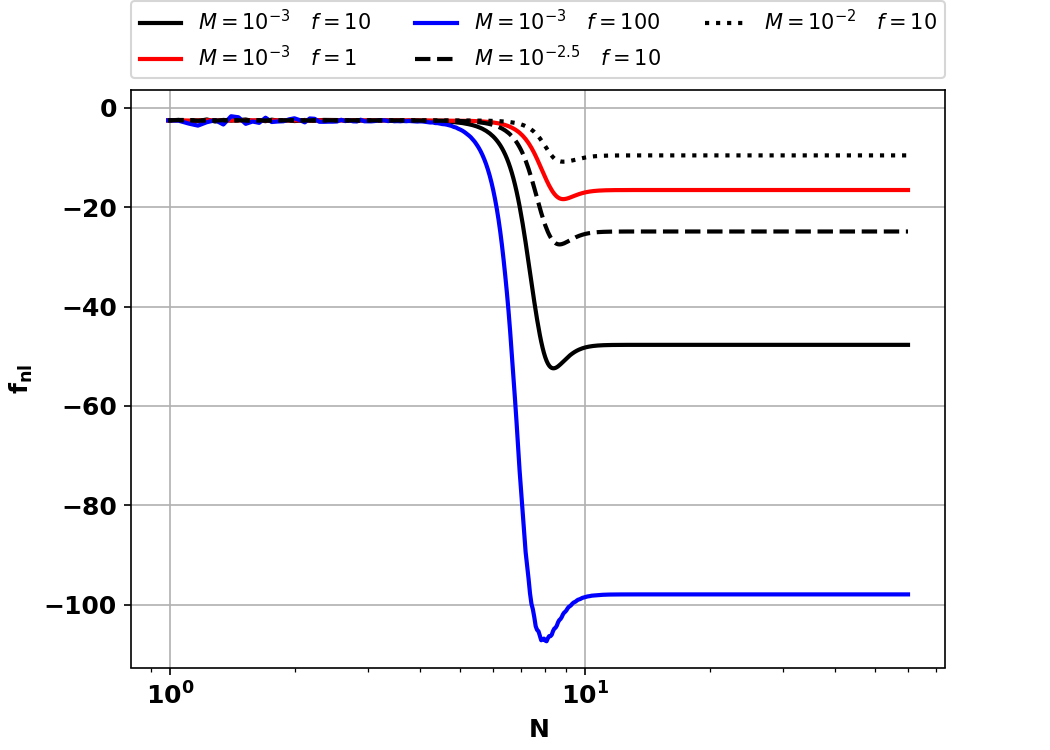}
\caption{Time evolution of the reduced bispectrum \eqref{fNL}, evaluated on the equilateral triangle for the pivot scale $k_{55}$, for NI in the minimal geometry, and different parameters for the curvature scale $M$ and steepness parameter $f$. Note that that we use of a logarithmic scale for the number of e-folds, and that Hubble crossing arises at $N=10$ e-folds.}
\label{evoNI}
\end{figure}

Eventually, we display in figure~\ref{evoNI} the time
 evolution of the reduced bispectrum in the equilateral configuration for the scale $k_{55}$, for the representative model of NI with the five different combinations of parameters that we studied. We can see that the bispectrum starts to differ from the Bunch-Davies regime a few e-folds before Hubble crossing (arising at $N=10$ e-folds in the plot), and one can check that this arises when $2 H \sqrt{|\etaperp|}$ becomes non-negligible compared to $k/a$, in agreement with the identification in section \ref{modified-dispersion} of this relevant timescale for the physics of the fluctuations. After a rapid growth, the bispectrum then stabilizes at its final value soon after Hubble crossing.

\subsubsection{Hyperbolic geometry}

As far as the hyperbolic geometry is concerned, based on the understanding of the linear fluctuations in section \ref{EFT}, one can expect two qualitatively different results, respectively for models with $V_{,\vphi} >0$, which are characterized by a reduced speed of sound, and for the ones with $V_{,\vphi}<0$, that feature a transient tachyonic instability induced by an effective imaginary speed of sound. In our models, only Starobinsky inflation belong to the first class. Unfortunately, we have not been able to reliably compute the bispectrum numerically for this model. As the effective field theory of fluctuations indicate though \cite{Cheung:2007st}, a reduced sound speed implies the appearance of boosted cubic interactions leading to an equilateral-type contribution to the bispectrum of amplitude $f_{nl}^\eq \sim 1/c_s^2$. Additional cubic interactions can however be present, and we leave for future work a more in-depth study of this setup, both numerically and analytically. \\

Concentrating on the other class of models, with NI, $\mathrm{SFI}_2$ and $\mathrm{SFI}_4$, we find again, for all models with curvature scale $M=10^{-3}$, a large negative reduced bispectrum of a few tens in the equilateral limit, with values ranging from $-16$ to $-57$. The same qualitatively holds true when varying the curvature scale, but the value of $f_{nl}$ can sometimes be reduced to $\simeq 1$, as we find for NI ($f=10$) and $\mathrm{SFI}_2$ $(\mu=20)$ with $M=10^{-2}$. 
However, the striking difference compared to the minimal geometry concerns the shape of the bispectrum: although the values of the two correlations depend on the precise model, we find across all of them a small overlap with the equilateral template (typically $\simeq -0.1$, ranging from $-0.01$ to $-0.47$), and a very significant one with the orthogonal shape (typically $\simeq -0.78$, ranging from $-0.66$ to $-0.80$). Its is rather unusual to generate orthogonal non-Gaussianities. Let us recall indeed that the orthogonal shape has been designed in the context of the effective field theory of inflation by carefully choosing a linear combinations of otherwise equilateral-type shapes \cite{Senatore:2009gt}, so as to fully cover the space of possible shapes in the simplest singe-field extensions of slow-roll inflation. As a result, in the past it is only for rather fine-tuned parameters that this shape of the bispectrum has been shown to be generated in explicit models (see \textit{e.g.} \cite{RenauxPetel:2011dv,RenauxPetel:2011uk,Green:2013rd}). 

\begin{figure*}
        \centering
        \begin{subfigure}[b]{0.475\textwidth}
            \centering
            \includegraphics[width=\textwidth]{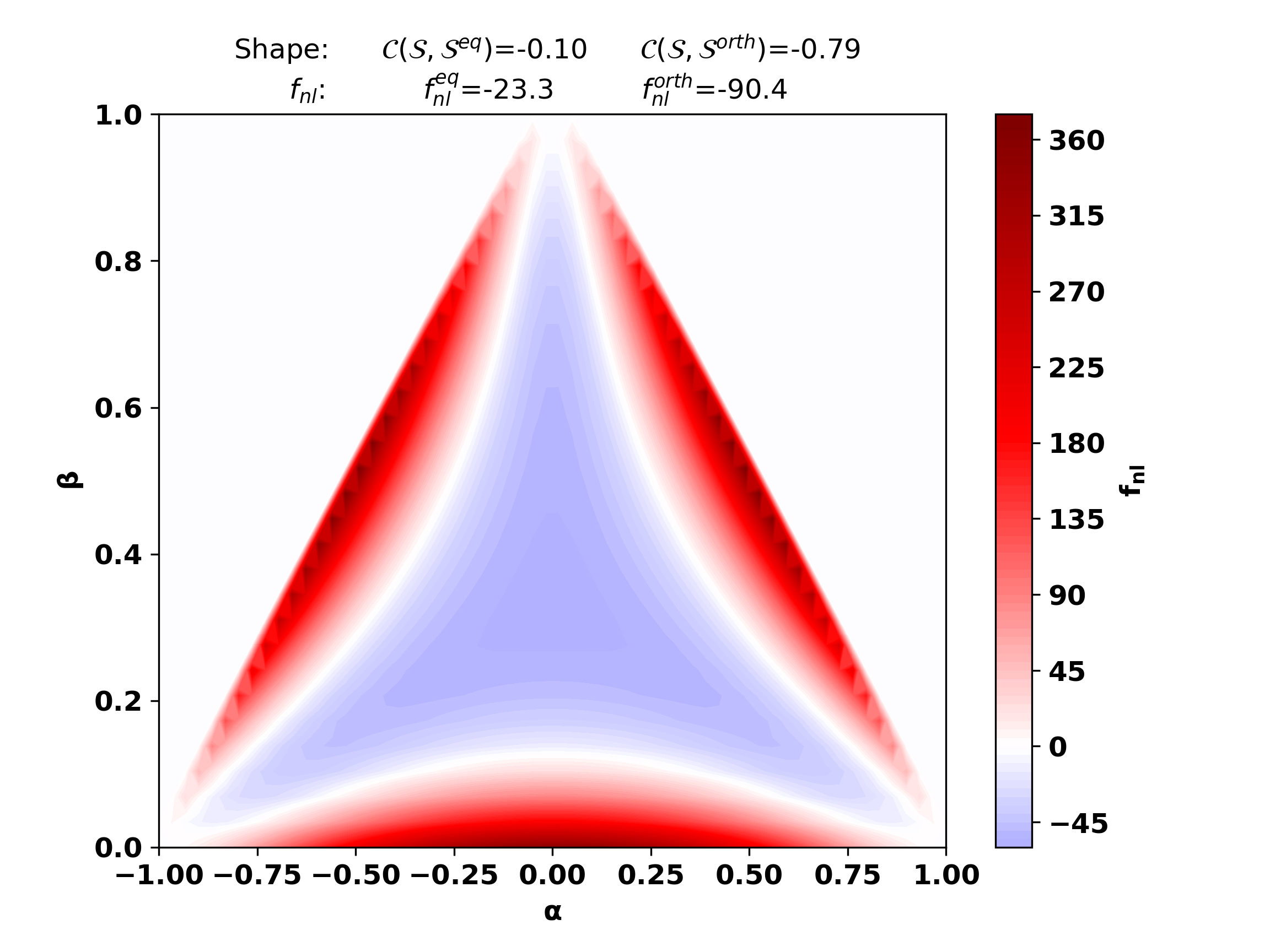}
            \caption{{\small Natural inflation ($f=10$)}}    
            \label{fig:shapeNIH-f10}
        \end{subfigure}
        \hfill
        \begin{subfigure}[b]{0.475\textwidth}  
            \centering 
            \includegraphics[width=\textwidth]{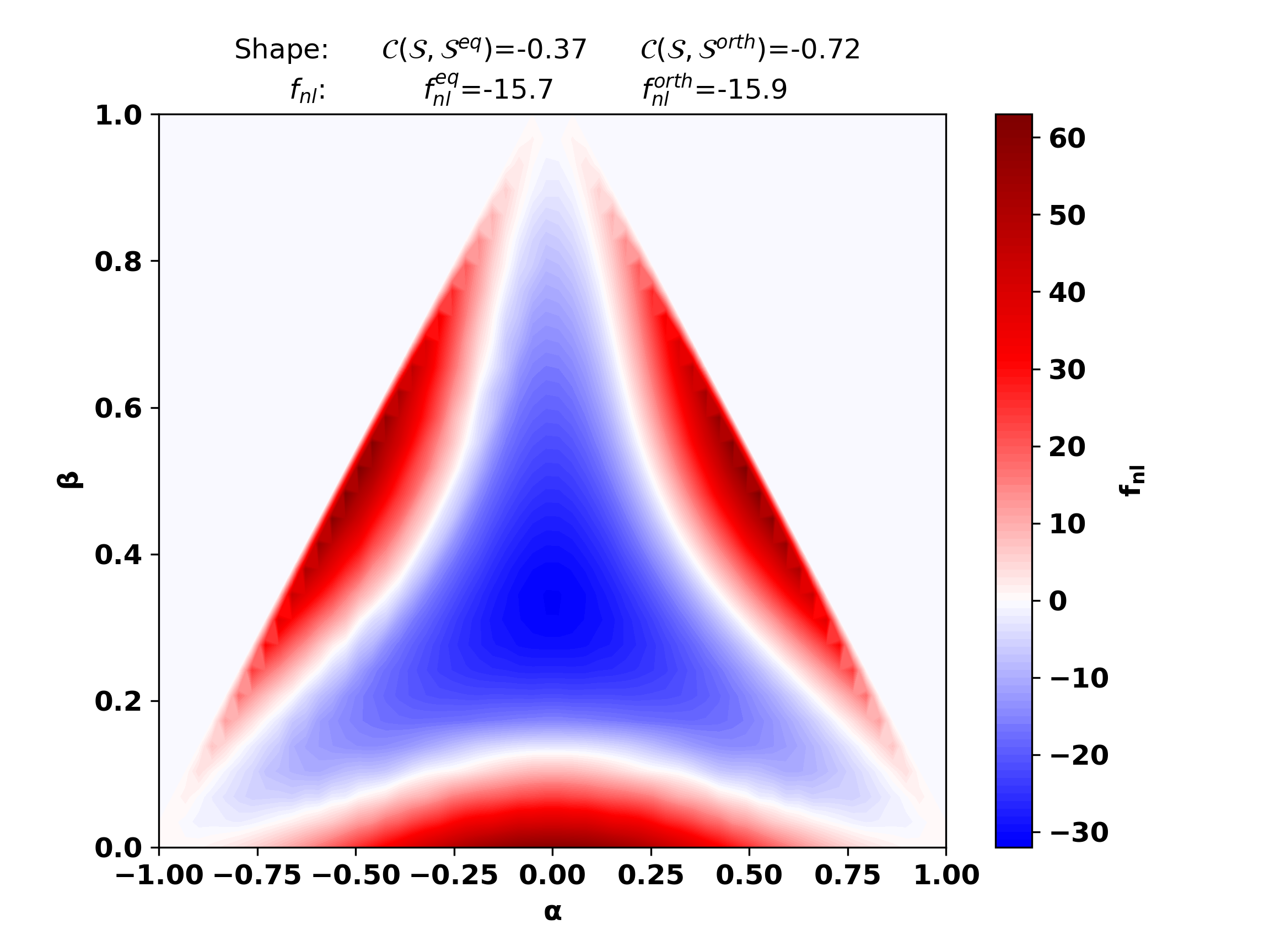}
            \caption{{\small Natural inflation ($f=100$)}}    
            \label{fig:shapeNIH-f100}
        \end{subfigure}
        \caption{Shape dependence $f_{nl}$($\alpha,\beta$) generated for NI with $f=10$ (left) and $f=100$ (right), in the hyperbolic geometry for $M=10^{-3}$. We used $k_s=3\,k_{55}$. We see for both of them a modest anti-correlation with the equilateral shape template and  a large anti-correlation with the orthogonal shape template.}
        \label{shapeNIH}
    \end{figure*}

We show in fig.~\ref{shapeNIH} the shapes of the bispectra obtained for NI with $M=10^{-3}$, for the two parameters $f=10$ and $f=100$, as they are representative of the other models. 
The main difference with the shape obtained in the minimal geometry in fig.~\ref{shapeNI} is readily apparent: the bispectrum is still negative in the equilateral limit, but it changes sign and becomes positive for squashed, and more generally, for flattened triangles, a distinctive feature of the orthogonal shape (with negative $f_{nl}^\orth$). In both cases, the amplitude of the signal is even more important in the squashed configuration than in the equilateral limit. The difference between the amplitudes in the two configurations is comparatively less pronounced for $f=100$ however, which explains why the correlation with the equilateral shape is larger in that case than for $f=10$.

\begin{figure}
\centering
\includegraphics[width=0.6\linewidth]{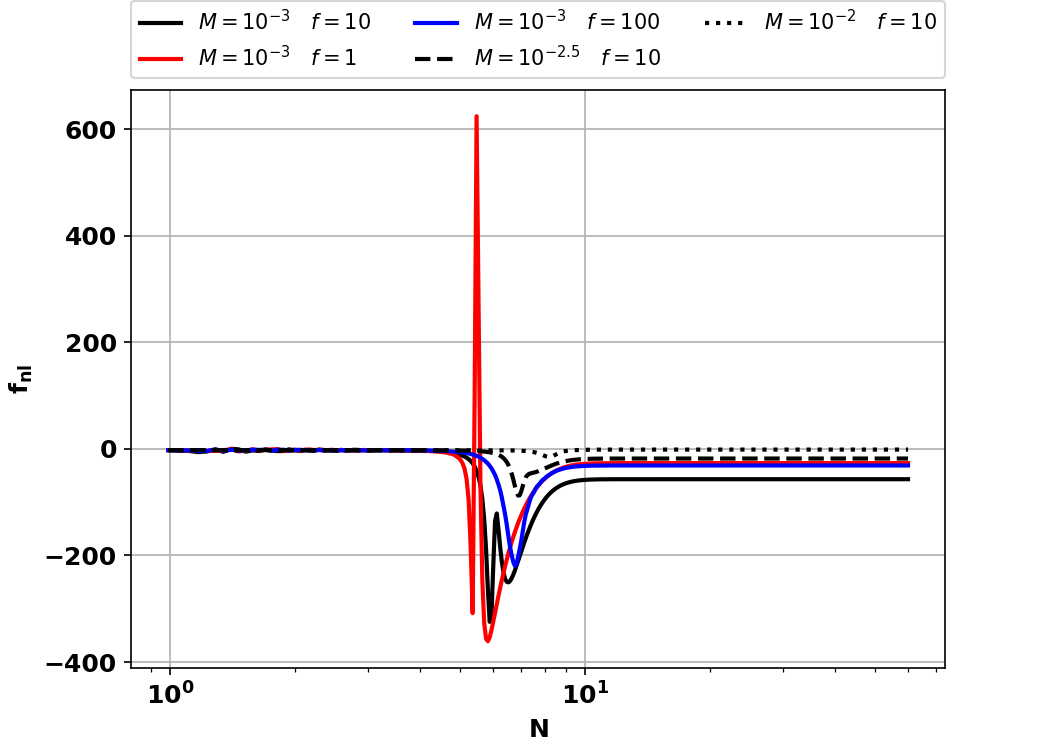}
\caption{Time evolution of the reduced bispectrum \eqref{fNL}, evaluated on the equilateral triangle for the pivot scale $k_{55}$, for NI in the hyperbolic geometry, and different parameters for the curvature scale $M$ and steepness parameter $f$. Note that that we use of a logarithmic scale for the number of e-folds, and that Hubble crossing arises at $N=10$ e-folds.}
\label{evohyp}
\end{figure}

Similarly to the minimal geometry, we show in fig.~\ref{evohyp} the time evolution of the reduced bispectrum in the equilateral configuration for the scale $k_{55}$, for the representative model of NI with the five different combinations of parameters that we studied. We can check that the bispectrum starts to differ from the Bunch-Davies regime when $|m_s^2|$ becomes non-negligible compared to $k^2/a^2$, and that the bispectrum has become constant already at Hubble crossing. The time-dependence however is more complicated than in the minimal geometry, especially for $M=10^{-3}$ with an oscillatory behaviour of the bispectrum for $f=10$ and $f=1$, and in that case, with even a brief spike of $f_{nl}$ to large positive values before returning to negative values. 

As for the parameters' dependence is concerned, one globally observe the same trends as in the minimal geometry: decreasing the curvature scale, or the steepness parameter of the potential, comes with an increase of the bispectrum in the equilateral configuration. The latter monotonous behaviour is broken for NI though, which generates $f_{nl}$ in this limit smaller for $f=100$ than for $f=10$. More importantly, as the shape differs strongly from the equilateral one, the reduced bispectrum in the equilateral configuration, although instructive, does not faithfully represent the overall amplitude of the bispectrum. This can be easily seen in fig.~\ref{shapeNIH} where, as we have noticed, the bispectrum is larger in the squashed configuration than in the equilateral one. A more robust measure of the non-Gaussianities is provided by the amplitude of the orthogonal signal $f_{nl}^\orth$. The latter is always found to be roughly minus a quarter of the reduced bispectrum in the squashed configuration, and contrary to the signal in the equilateral limit, the global trend is that $f_{nl}^\orth$ decreases as the steepness parameter increases (see table \ref{T:Table_6}). This is at least what we observe for NI and $\mathrm{SFI}_2$, with $\mathrm{SFI}_4$ breaking this monotonous behaviour in that case. Thus there is no obvious universal relationship that we can observe between values of background quantities and $f_{nl}^\orth$, although for NI and $\mathrm{SFI}_2$, one has the rough behaviour $f_{nl}^\orth \sim {\cal O}\left(6 \sqrt{2} \chi_\star/M \right)$. It would of course be interesting to understand analytically the appearance of the orthogonal shape as well as its amplitude, which we leave for future work.\footnote{It has later been shown in reference \cite{Garcia-Saenz:2018vqf} how an orthogonal-type shape originates on general grounds from an imaginary speed of sound.} 

Note eventually that the values one obtains for $f_{nl}^\orth$ often lie beyond the Planck observational bound \eqref{constraints-fNL}, contrary to the equilateral bispectrum generated in the minimal geometry. Moreover, when $f_{nl}^\orth$ is within the observational bound, it is the spectral index that is often too blue compared to the observations, leaving only, within the models we have studied, NI with $f=100$ and $M=10^{-3}$ as a viable model (and marginally NI with $f=10$ and $M=10^{-2.5}$).

\subsubsection{Squeezed limit and scale dependence of the bispectrum}

\begin{figure}
\centering
\includegraphics[width=1.0\linewidth]{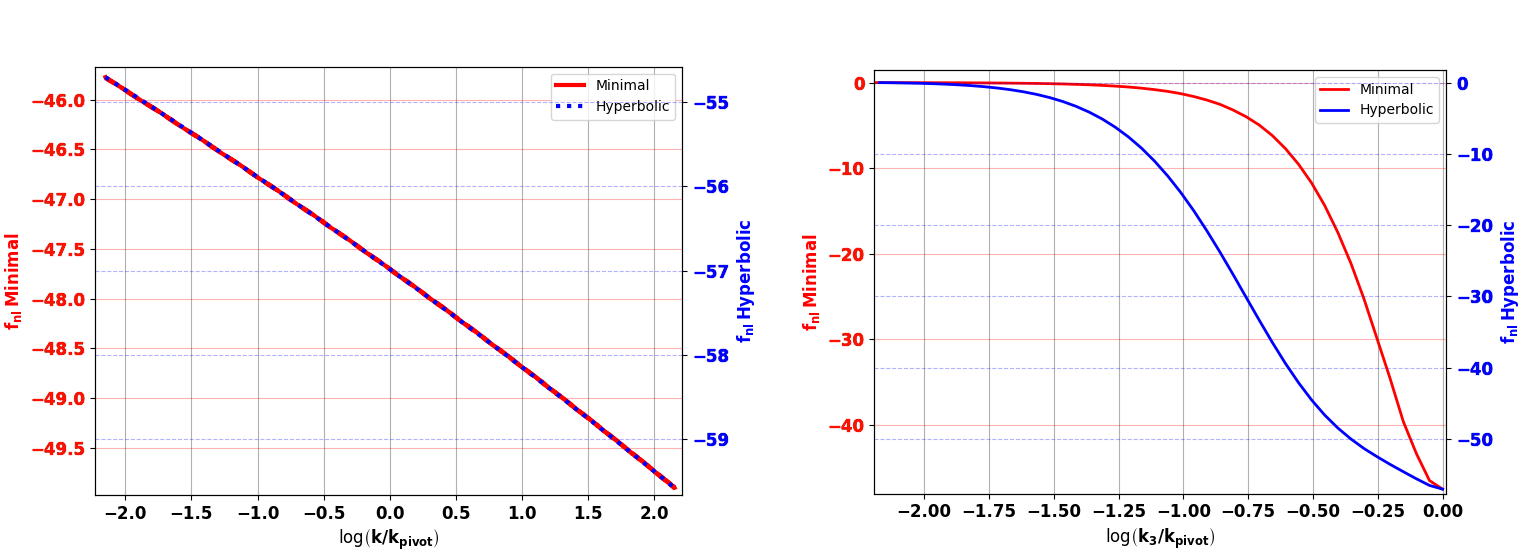}
\caption{Left: scale dependence of the reduced equilateral bispectrum over a range of scales exiting the Hubble radius between $60$ and $50$ e-folds before the end of inflation, for NI with $f=10$ and $M=10^{-3}$, in the minimal geometry (red) and the hyperbolic one (blue). Right: for the same models and color coding, reduced bispectrum $f_{nl}(k_1,k_2,k_3)$ for $k_1=k_2=k_{{\rm pivot}}=k_{55}$ and $k_3\to 0$, showing how the single-clock consistency relation is verified in the squeezed limit. Note each time the two different scales for the vertical axes.}
\label{evoNIbi}
\end{figure}

Eventually, as the reader might have noticed from figs.~\ref{shapeNI} and \ref{shapeNIH}, all the bispectra that we have computed have a small amplitude in the squeezed limit. This is expected theoretically, as we have seen that one can derive an effective single-field theory for the fluctuations, so that all models should verify the single-clock consistency relation \cite{Maldacena:2002vr,Creminelli:2004yq}. This relation is indeed satisfied, as one shows in fig.~\ref{evoNIbi} (right) for the two models of NI with $f=10$ and $M=10^{-3}$, in the minimal and in the hyperbolic geometry. There, we plot $f_{nl}(k_1,k_2,k_3)$, keeping two modes the same ($k_1=k_2=k_{{\rm pivot}}=k_{55})$, and letting $k_3\to 0$, i.e. approaching the squeezed limit, finding that the single-clock consistency relation $f_{nl}^{{\rm squeezed}}=\frac{5}{12} (1-n_s)(k_{{\rm pivot}})$ is well verified as soon as $k_3 \ll k_{{\rm pivot}}$.

Eventually, while we concentrated our efforts on studying the shape of the bispectrum generated in sidetracked inflation, it is useful to comment on its scale dependence. For this, we show in fig.~\ref{evoNIbi} (left) the reduced bispectrum in the equilateral configuration over a range of scales exiting the Hubble radius between $60$ and $50$ e-folds before the end of inflation, for the same models of NI with $f=10$ and $M=10^{-3}$ in the two geometries. We have chosen to overlap the two cases for better comparison (note therefore the two different scales for the vertical axis). It is clear they both have very small scale dependencies with $d \log |f_{nl}|/d\log(k)\sim 0.02$.


\section{Comparisons with cosmological attractors} \label{sec:alpha}

Cosmological attractors, such as $\alpha$-attractor \cite{Kallosh:2013yoa} and $\xi$-attractor models \cite{Kallosh:2013tua}, have the common characteristic of exhibiting a pole in the kinetic term of the inflaton \cite{Galante:2014ifa,Broy:2015qna}, and this is precisely the point near which slow-roll inflation takes place. The universal predictions of these models can then be seen to be a consequence of the exponential stretching of the potential as a result of eliminating the coordinate singularity of the internal space metric by a redefinition of the fields. In the context of multi-field inflation such a pole can arise naturally if the field space manifold is negatively curved, and the hyperbolic plane is particularly well motivated because of its origin in supergravity \cite{Kallosh:2014rga}.

In this section we would like to address the question of whether the geometrical destabilization, and sidetracked inflation in particular, could play a relevant role in cosmological attractors. A recent work \cite{Achucarro:2017ing} has established that the predictions for the power spectrum parameters in $\alpha$-attractors remain universal even when the multi-field dynamics is important (see also the recent work \cite{Carrilho:2018ffi} in which similar conclusions are reached in multifield $\xi$ attractors). Sidetracked inflation has some interesting similarities with the axion-dilaton model of \cite{Achucarro:2017ing}---in both scenarios the second field evolves away from the minimum of the potential trough---and yet the predictions are markedly different. We have seen that the results of sidetracked inflation are typically highly sensitive to multi-field effects, while the opposite appears to happen in the $\alpha$-attractor set-up.

One reason for this discrepancy can be easily understood if we express our hyperbolic metric model, eq.\ \eqref{eq:hyperbolicmetric}, in Poincar\'e disk coordinates. It is simpler to first go to half-plane coordinates $(X,Y)$ with
\beq \label{eq:fields to halfplane}
\varphi=\frac{M}{\sqrt{2}}\,\log Y\,,\qquad \chi=\frac{M}{\sqrt{2}}\,\frac{X}{Y}\,,
\eeq
so that the field space metric takes the form
\beq
ds_{\rm fs}^2=2M^2\,\frac{dX^2+dY^2}{4Y^2}\qquad (-\infty<X<\infty\,,\,0<Y)\,.
\eeq
The transformation to disk coordinates $(\rho,\theta)$ is given by the standard formulae
\beq \label{eq:halfplane to disk}
X=\frac{2\rho\sin\theta}{1+\rho^2-2\rho\cos\theta}\,,\qquad Y=\frac{1-\rho^2}{1+\rho^2-2\rho\cos\theta}\,,
\eeq
and
\beq
ds_{\rm fs}^2=2M^2\,\frac{d\rho^2+\rho^2d\theta^2}{(1-\rho^2)^2}\qquad (0\leq\rho<1\,,\,0\leq\theta<2\pi)\,.
\eeq

\begin{figure*}
        \centering
            \includegraphics[width=0.4\textwidth]{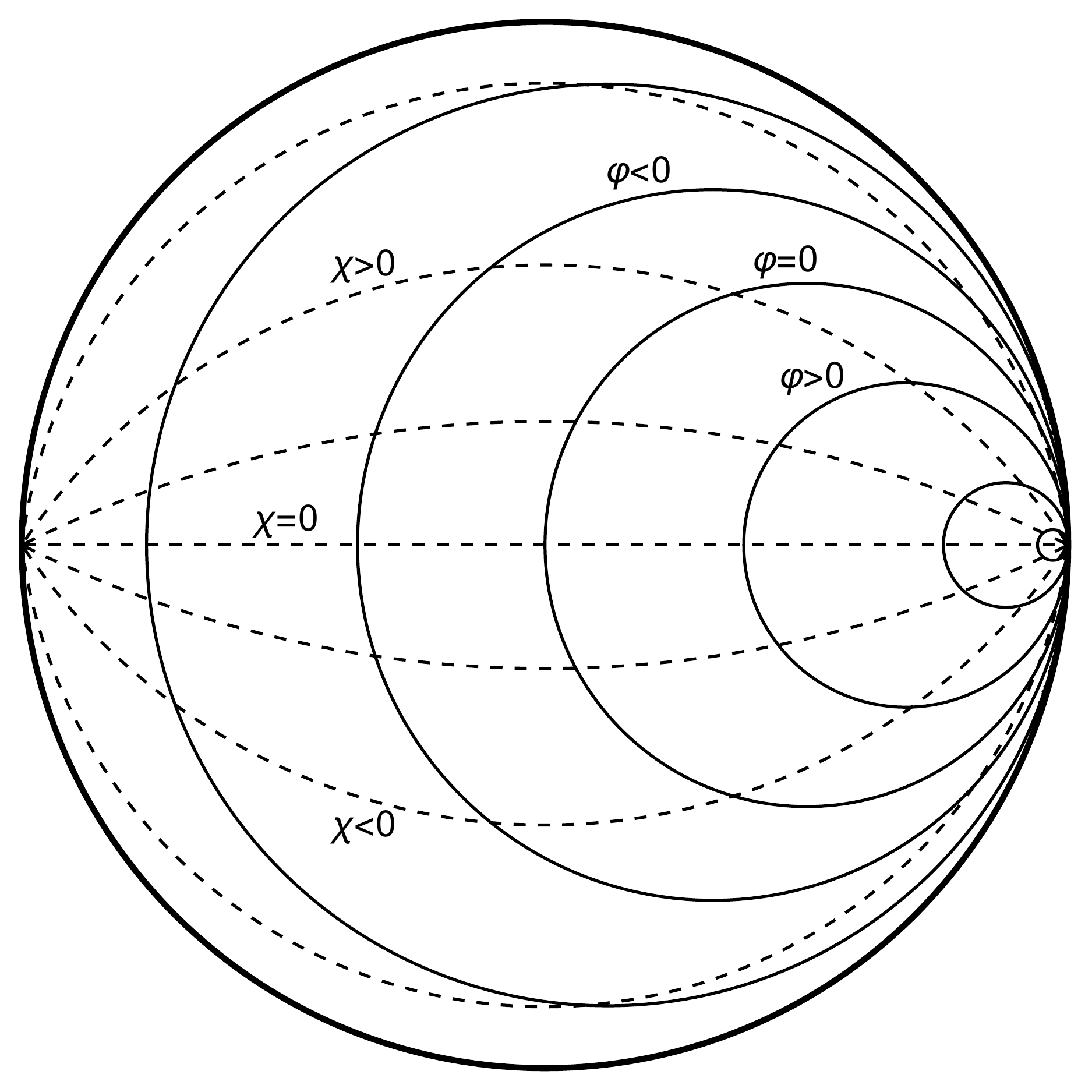}
        \caption{Constant $\varphi$ (solid curves) and $\chi$ (dashed curves) contours in the Poincar\'e disk.}
        \label{fig:contours}
    \end{figure*}

The contours of constant $\varphi$ and $\chi$ in the Poincar\'e disk are shown in fig.\ \ref{fig:contours}. The $\varphi$ contours are given by circles of radius $(1+e^{\sqrt{2}\,\varphi/M})^{-1}$ and centered at the Cartesian point $(x_{\vphi},0)$ with $x_{\varphi}=(1+e^{-\sqrt{2}\,\varphi/M})^{-1}$. The $\chi$ contours are also part of circles of radius $\sqrt{1+M^2/2\chi^2}$ with center at the Cartesian point $(0,y_{\chi})$, with $y_{\chi}=-M/(\sqrt{2}\,\chi)$. Since during most of the sidetracked inflationary phase, $\varphi/M \gg 1$ for the models we have investigated, we observe that inflation is spent very near the rightmost corner of the disk, i.e.\ near the edge of the Poincar\'e disk at $\rho=1$.

The single-field inflationary path $\chi=0$, which is potentially unstable because of the geometrical destabilization, corresponds to the lines $\theta=0,\pi$, on which
\beq
\rho=\tanh \left(\pm\, \vphi/\sqrt{2} M \right)\,,
\eeq
where the $+$ sign (respectively $-$) correspond to $\vphi>0$, $\theta=0$ (respectively $\vphi<0$, $\theta=\pi$).
This is the familiar relation that gives rise to the stretching of the potential in $\alpha$-attractors. The origin of this property can be traced to the presence of a pole at $\rho=1$ in the kinetic term of the inflaton before canonical normalization, and the universality of the predictions of cosmological attractors stems from the fact that inflationary parameters depend, to leading order in the number of e-folds, only on the characteristics of the pole \cite{Galante:2014ifa}.

However, a crucial assumption behind this is that the potential must be regular at $\rho=1$. This is hardly restrictive when only the inflaton potential is concerned, since in $\alpha$-attractors the potential is usually analytic at the pole's location. But things get more intricate in the two-field scenario and the choice of parametrization becomes important. Indeed, from \eqref{eq:fields to halfplane} and \eqref{eq:halfplane to disk} we see that
\beq
\chi=\frac{M}{\sqrt{2}}\,\frac{2\rho\sin\theta}{1-\rho^2}\,,
\eeq
and so any pedestrian polynomial function of the heavy field $\chi$ will have a pole at $\rho=1$ when expressed in disk coordinates. In particular the simple mass term $ m_h^2\,\chi^2/2$ that we have used in our models has a pole of order two, just like the kinetic term.

The conclusion is that the existence of a pole shared by both the kinetic and potential terms in a hyperbolic nonlinear sigma model action can invalidate the universal predictions of cosmological attractors. Our results of sections \ref{sec:powerspectrum} and \ref{sec:bispectrum} are clear evidence that this is the case, as we have seen that the predictions of the sidetracked inflation scenario are quite sensitive to the form of the potential and the scales involved. We therefore expect that our findings may be useful to better understand the constrains that the potentials must satisfy for instance in $\alpha$-attractors, but also more generally.\footnote{We also remark that in the models we have considered the potential is clearly not a monotonic function of $\rho$, which was one of the assumptions made in \cite{Achucarro:2017ing}.} Indeed, although above we focused on the hyperbolic manifold, similar conclusions regarding the presence of coordinate singularities are likely to apply for more general field space metrics.


\section{Conclusion} \label{sec:conclusion}

During inflation, heavy scalar fields can be destabilized from their vacuum state as a result of an instability triggered by negatively curved field spaces \cite{Renaux-Petel:2015mga}. The fate of this geometrical destabilization of inflation is yet uncertain, as its understanding requires going beyond perturbation theory. After having studied the consequences of one possible outcome of it, namely a premature end of inflation \cite{Renaux-Petel:2017dia}, we have studied in this paper the phenomenological consequences of the second possibility: the fact that the system can be driven into a second inflationary phase where the heavy field is displaced from the minimum of its potential. We named this scenario sidetracked inflation in view of the way the inflationary trajectory deviates from its initial effectively single-field path as a result of the geometrical destabilization

Contrary to the premature end of inflation, the sidetracked phase a priori depends on details of the completion of the model away from the original inflationary valley, such that the precise form of the potential and the field space metric. For this reason, we have studied and compared different type of models. First, we made use of two field space manifolds: the first one has been used in reference \cite{Renaux-Petel:2015mga} to exhibit the geometrical destabilization of inflation, and we refer to it as the minimal geometry. The second one is the hyperbolic geometry with constant negative curvature, although in an unusual system of coordinates. We have also chosen parameters such that the geometrical destabilization is triggered at the same location along the initial single-field path. In addition, for each of these geometries, we varied its curvature scale and we studied four different shapes of potentials for the inflaton, also varying their parameters. However, we kept a simple quadratic potential for the heavy field for simplicity. We solved the dynamics of these various models numerically, for the background, the linear fluctuations, and the primordial non-Gaussianities, notably making use of the transport approach and the package \texttt{PyTransport 2.0} \cite{Ronayne:2017qzn}.

We observed that the background dynamics of the sidetracked phase is very similar for the two types of geometries, which we have understood analytically. In each model, the heavy field is adiabatically following the inflaton, stabilized by the competition between the force originating from the field space geometry and the one from its potential, like in the gelaton model \cite{Tolley:2009fg}. This field in turns modifies the dynamics of the inflaton, slowing it down by giving it more inertia. The background dynamics can hence be described by an effective single-field model with unusual properties. In particular it allows inflation on potentials that would otherwise be too steep for standard slow-roll inflation, with the weaker requirement that the potential be flat with respect, not to the Planck scale, but to the curvature scale of the field space manifold.

Equipped with our analytical understanding of the background, we were able to explain the very different behaviours exhibited by the cosmological fluctuations in the two geometries, despite the very similar background dynamics. In both cases, the deviation of sidetracked inflation's trajectory from a geodesic, and hence the coupling between the adiabatic and entropic fluctuations, is very large. 
However, in the minimal geometry, the mass of the entropic fluctuation is small compared to the Hubble scale, whereas it is large in the hyperbolic geometry. Moreover, in this case, the entropic mass squared can be positive or negative, depending on the slope of the inflaton potential being positive or negative.
Building on previous studies, we showed that the resulting dynamics of the curvature perturbation can be described by an effective single-field theory, in which the entropic fluctuations have been integrated out, albeit with very different properties  in each case: a modified dispersion relation when the entropic field is light in the minima geometry, and a reduced speed of sound or an imaginary one, depending on the potential, in the hyperbolic geometry. In each of these three qualitatively different scenarios, we gave analytical formulae for the tensor-to-scalar-ratio $r$ and the spectral index $n_s$, showing how they enable to reproduce and understand the full numerical results and their parameter dependencies.

Scenarios described by a single-field effective theory with an imaginary speed of sound around the time of Hubble crossing arise in the presence of a large and tachyonic mass of the entropic fluctuation. While this generates a transient tachyonic growth of the fluctuations, we stress that these situations do not present any fundamental pathology. In particular, they can be perfectly compatible with a stable background, as it is here, in the presence of a trajectory deviating strongly from a field space geodesic that render the \textit{super-Hubble} entropic mass squared, the true indicator of the stability or not of the background, indeed positive.

We also made a preliminary numerical study of the bispectrum generated in sidetracked inflation, finding for almost all models large non-Gaussianities, but characterized by different shapes: equilateral in the minimal geometry, and orthogonal in scenarios featuring an imaginary speed of sound in the hyperbolic geometry.\footnote{The link between an imaginary speed of sound and orthogonal and flattened non-Gaussianities has later been made in reference \cite{Garcia-Saenz:2018vqf}.} While models of the first type are typically in agreement with observational bounds on non-Gaussianities, for the second type we find large negative values of $f_{nl}^\orth$ that often exceed the Planck constraints, together with marked deviations from scale invariance, although some models are observationally viable.

Lastly we made a brief comparison between sidetracked inflation and the two-field cosmological attractor models that arise in supergravity implementations of inflation. We pointed out that beyond the single-field case the choice of field parametrization becomes subtle---a simple quadratic potential for our heavy field was seen to exhibit a pole at the edge of the Poincar\'e disk. We argued that such singularity evades the assumptions that lead to the universality of the predictions of cosmological attractors, and indeed our results were seen to depend strongly on the details of the potential. It would hence be interesting to gain further insight into the role of the pole structure of the potential beyond the simple case we have considered as well as for more general field space manifolds.

In the future it would be interesting to study if a fully non-linear single-field effective description of sidetracked inflation can be made, that would enable to unify the effective single-field dynamics that we derived separately for the background and the fluctuations. Such a description is known in related frameworks like the gelaton model, giving rise to effective $k$-inflationary models. However, preliminary investigations indicate that the two setup differ, as one can anticipate by noting that sidetracked inflation can exhibit modified dispersion relations that are not present in $k$-inflationary models. 

We should note as well that the nonlinear sigma models we have examined are phenomenological and don't have an obvious theoretical justification, since our goal for now has been to gain insight into the physics of sidetracked inflation rather than to provide a top-down motivation for it. The hyperbolic plane metric is perhaps the most interesting case study because of its maximal isometries and its relation to $\alpha$-attractor models, but the class of potentials we considered is of course restricted. Similarly, our focus on the minimal geometry was motivated by its simplicity. It cannot be regarded as a consistent truncation of an  effective field theory for the reason that the heavy scalar field probes values of order $M$ during the sidetracked inflationary phase, and this is precisely the scale at which the putative effective theory is expected to break down (although as usual the true cutoff may be actually even lower). We see the interesting phenomenological consequences of sidetracked inflationary scenarios as motivations to consider more realistic setups in the future. Eventually, we have seen that several important features of sidetracked inflation are tied to the heavy mass of the additional scalar field and it would be interesting to see how observables may be modified when this field has an intermediate mass of order the Hubble scale. We 
hope to come back to these questions in future works.\\

\noindent
{\it Note added:} While this paper was in its writing stage, two preprints appeared that have links with our work. Reference \cite{Linde:2018hmx} considered generalized $\alpha$-attractor models in which the multifield potential can be singular at the boundary of the Poincar\' e disk, like in our study. Reference \cite{Christodoulidis:2018qdw} similarly studied the multifield dynamics of inflationary models with an hyperbolic geometry. Although the authors did not study this aspect, they also encountered models having negative entropic mass squared, while positive super-Hubble entropic mass squared, and it is likely that our reasoning and effective field theory treatments can also apply in that case.

\begin{acknowledgments}

We would like to thank Yuichiro Tada for collaboration at the early stage of this work. We are grateful to Perseas Christodoulidis, Shuntaro Mizuno, Enrico Pajer, Diederik Roest, Luca Santoni, David Stefanyszyn, Krzysztof Turzy\'nski and Yvette Welling for useful discussions. S.GS is supported by the European Research Council under the European CommunityÕs Seventh Framework Programme (FP7/2007-2013 Grant Agreement no. 307934, NIRG project). S.RP is supported by the European Research Council (ERC) under the European UnionÕs Horizon 2020 research and innovation programme (grant agreement No 758792, GEODESI project). J.W.R. is supported by a studentship jointly funded by Queen Mary University of London and by the Frederick Perren Fund of the University of London.

\end{acknowledgments}


\appendix

\section{Tabulated results} \label{sec:app1}

\begin{table}[h!]
\centering
\begin{tabular}{|l|l|l|l|l|l|l|}
\hline
Acronym   & Parameter        & $r$    & $n_s$ & $f_{nl}$ & $r_{{\rm ad}}$& $n_{s\, {\rm ad}}$\\ \hline
$SI$    & -                & 0.0075 & 0.968 & -16.31 &0.10& 0.965 \\ \hline
        & $f=1$            & 0.0095 & 0.968 & -16.54 & 0.13& 0.963 \\ \cline{2-7}
$NI$    & $f=10$           & 0.0063 & 0.969 & -47.71 & 0.14& 0.964 \\ \cline{2-7} 
        & $f=100$          & 0.0046 & 0.970 & -97.95 &0.15& 0.964\\ \hline
        & $\mu=2$          & 0.0065 & 0.971 & -12.85  &0.079&0.968 \\ \cline{2-7}
$SFI_2$ & $\mu=20$         & 0.0056 & 0.973 & -25.32 &0.094&0.970 \\ \cline{2-7}
        & $\mu=200$        & 0.0043 & 0.974 & -46.63  & 0.097&0.970\\ \hline
        & $\mu=2$          & 0.0049 & 0.967 & -11.40  & 0.057&0.965\\ \cline{2-7} 
$SFI_4$ & $\mu=20$         & 0.0059 & 0.973 & -20.63 &0.090&0.970  \\ \cline{2-7}
        & $\mu=200$        & 0.0046 & 0.974 & -39.14  &0.096&0.970 \\ \hline
\end{tabular}
\caption{Results of sidetracked inflation in the minimal geometry with $M=10^{-3} \Mp$. $f_{nl}$ is the reduced bispectrum \eqref{fNL} evaluated on the equilateral triangle for the pivot scale $k_{55}$.}
\label{T:Table_1}
\end{table}

\begin{table}[h]
\centering
\begin{tabular}{|l|l|l|l|l|l|l|}
\hline
Acronym   & Parameter        & $r$    & $n_s$ & $f_{nl}$ & $r_{{\rm ad}}$& $n_{s\, {\rm ad}}$\\ \hline
$SI$    & -                & 0.029 & 0.969 & *  & 0.10& 0.965 \\ \hline
        & $f=1$            & $4.5\times 10^{-23}$ & 1.10 & -26.77 &0.13& 0.963  \\ \cline{2-7}
$NI$    & $f=10$           & $1.2\times 10^{-9}$ & 0.974 & -56.98  &0.15& 0.963 \\ \cline{2-7}
        & $f=100$          & $5.9\times 10^{-4}$ & 0.970 & -31.24  &0.15& 0.964 \\ \hline
        & $\mu=2$          & $1.5\times 10^{-22}$ & 1.282 & -19.00 &0.079&0.968  \\ \cline{2-7} 
$SFI_2$ & $\mu=20$         & $8.0\times 10^{-14}$ & 1.066 & -31.33  &0.095&0.970  \\\cline{2-7} 
        & $\mu=200$        & $2.0\times 10^{-8}$ & 1.019 & -51.50 & 0.097&0.970  \\  \hline 
        & $\mu=2$          & $5.4\times 10^{-21}$ & 1.476 & --15.92& 0.055&0.963  \\ \cline{2-7} 
$SFI_4$ & $\mu=20$         & $4.0\times 10^{-16}$ & 1.105 & -25.21 & 0.091&0.969 \\ \cline{2-7} 
        & $\mu=200$        & $1.1\times 10^{-9}$ & 1.030 & -46.69 & 0.097&0.970 \\ \hline
\end{tabular}
\caption{Results of sidetracked inflation in the hyperbolic geometry with $M=10^{-3} \Mp$. $f_{nl}$ is the reduced bispectrum \eqref{fNL} evaluated on the equilateral triangle for the pivot scale $k_{55}$.  * for SI indicates that we have not been able to reliably compute the bispectrum, as we explain in the main text.}
\label{T:Table_2}
\end{table}

\begin{table}[h]
\centering
\begin{tabular}{|l|l|l|l|l|l|}
\hline
Acronym   & Parameter        & $r$    & $n_s$ & $f_{nl}$ \\ \hline
    & $M=10^{-2}$               & NA & NA & NA   \\ \cline{2-5} 
$SI$    & $M=10^{-2.5}$            & 0.0057 & 0.965 & -11.31   \\ \cline{2-5}  
    & $M=10^{-3}$              & 0.0075 & 0.968 & -16.31  \\ \hline
        & $M=10^{-2}$            & 0.0131 & 0.969 & -9.56   \\ \cline{2-5} 
$NI$ $f=10$    & $M=10^{-2.5}$     & 0.0085 & 0.969 & -24.89   \\ \cline{2-5} 
        & $M=10^{-3}$           & 0.0063 & 0.969 & -47.71  \\ \hline
        & $M=10^{-2}$          & 0.0088 & 0.975 & -7.62   \\ \cline{2-5} 
$SFI_2$ $\mu=20$ & $M=10^{-2.5}$         & 0.0067 & 0.973 & -16.52   \\ \cline{2-5} 
        & $M=10^{-3}$         & 0.0056 & 0.973 & -25.32  \\ \hline
        & $M=10^{-2}$          & 0.0070 & 0.976 & -6.53  \\ \cline{2-5} 
$SFI_4$ $\mu=20$ & $M=10^{-2.5}$         & 0.0064 & 0.972 & -13.99   \\ \cline{2-5}  
        & $M=10^{-3}$         & 0.0059 & 0.973 & -20.63   \\ \hline
\end{tabular}
\caption{Results of sidetracked inflation in the minimal geometry, here with varying $M$. $f_{nl}$ is the reduced bispectrum \eqref{fNL} evaluated on the equilateral triangle for the pivot scale $k_{55}$. NA indicates that the sidetracked phase lasts less than 55 e-folds.}
\label{T:Table_3}
\end{table}

\begin{table}[h]
\centering
\begin{tabular}{|l|l|l|l|l|l|}
\hline
Acronym   & Parameter        & $r$    & $n_s$ & $f_{nl}$ \\ \hline
    & $M=10^{-2}$               & NA & NA & NA  \\ \cline{2-5} 
$SI$    & $M=10^{-2.5}$              & NA & NA & NA  \\  \cline{2-5} 
    & $M=10^{-3}$               & 0.029 & 0.969 & *    \\ \hline
        & $M=10^{-2}$            & 0.0024 & 0.991 & -1.40   \\ \cline{2-5} 
$NI$ $f=10$    & $M=10^{-2.5}$      & $6.875\times 10^{-6}$ & 0.977 & -18.19    \\ \cline{2-5} 
        & $M=10^{-3}$         & $1.2\times 10^{-9}$ & 0.974 & -56.98   \\ \hline
        & $M=10^{-2}$          & 0.0020 & 1.038 & -1.03    \\ \cline{2-5} 
$SFI_2$ $\mu=20$ & $M=10^{-2.5}$        & $4.3\times 10^{-7}$ & 1.027 & -15.78  \\ \cline{2-5} 
        & $M=10^{-3}$         & $8.0\times 10^{-14}$ & 1.066 & -31.33   \\ \hline
        & $M=10^{-2}$          & NA & NA & NA   \\ \cline{2-5} 
$SFI_4$ $\mu=20$ & $M=10^{-2.5}$         & $9.0\times 10^{-8}$ & 1.054 & -14.74   \\ \cline{2-5} 
        & $M=10^{-3}$      & $4.0\times 10^{-16}$ & 1.105 & -25.21  \\ \hline
\end{tabular}
\caption{Results of sidetracked inflation in the hyperbolic geometry, here with varying $M$. $f_{nl}$ is the reduced bispectrum \eqref{fNL} evaluated on the equilateral triangle for the pivot scale $k_{55}$. NA indicates that the sidetracked phase lasts less than 55 e-folds.  * for SI with $M=10^{-3}$ indicates that we have not been able to reliably compute the bispectrum, as we explain in the main text.}
\label{T:Table_4}
\end{table}

\begin{table}[h]
\centering
\begin{tabular}{|l|l|l|l|l|l|l|}
\hline
Acronym   & Parameter&M        & $\mathcal{C}(S,S^{\eq})$    & $\mathcal{C}(S,S^{\orth})$&$ f_{nl}^{\eq} $&$ f_{nl}^{\orth} $\\ \hline
    & &$10^{-2}$               & NA & NA & NA  & NA   \\ \cline{3-7} 
$SI$    & &$10^{-2.5}$           &-0.97&-0.03& -10.4&-0.2  \\  \cline{3-7} 
   & & $10^{-3}$             &-0.97&-0.03& -15.0&-0.3\\ \hline

	&  $f=1$ & $10^{-3}$           &-0.97&-0.03& -15.2&-0.3  \\ \cline{2-7} 
        &  $f=10$ & $10^{-2}$           &-0.97&-0.01 &-8.99&-0.07  \\ \cline{2-7} 
$NI$    &  $f=10$ & $10^{-2.5}$     &-0.97&-0.03& -23.0&-0.43  \\ \cline{2-7} 
        &  $f=10$ & $10^{-3}$          &-0.97&-0.03& -44.0&-0.89 \\ \cline{2-7} 
	&  $f=100$ & $10^{-3}$          &-0.97&-0.02&-91.1&-1.30 \\ \hline

	&  $\mu=2$ & $10^{-3}$           &-0.97&-0.03& -11.8&-0.23  \\ \cline{2-7} 
        &  $\mu=20$ & $10^{-2}$           &-0.97&-0.01 &-7.19&-0.05  \\ \cline{2-7} 
$SFI_2$    &  $\mu=20$ & $10^{-2.5}$     &-0.97&-0.03& -15.2&-0.28  \\ \cline{2-7} 
        &  $\mu=20$ & $10^{-3}$          &-0.97&-0.03& -23.3&-0.46 \\ \cline{2-7} 
	&  $\mu=200$ & $10^{-3}$          &-0.97&-0.03& -43.0&-0.84 \\ \hline

	&  $\mu=2$ & $10^{-3}$           &-0.97&-0.03& -10.5&-0.20  \\ \cline{2-7} 
        &  $\mu=20$ & $10^{-2}$           &-0.97&-0.01 &-6.21&-0.01  \\ \cline{2-7} 
$SFI_4$    &  $\mu=20$ & $10^{-2.5}$     &-0.97&-0.03& -12.9&-0.24 \\ \cline{2-7} 
        &  $\mu=20$ & $10^{-3}$          &-0.97&-0.03& -19.0&-0.37 \\ \cline{2-7} 
	&  $\mu=200$ & $10^{-3}$          &-0.97&-0.03& -36.1&-0.71 \\ \hline
\end{tabular}
\caption{Results for the bispectrum generated in sidetracked inflation in the minimal geometry, indicating the correlation of the shape with the equilateral and orthogonal templates, as well as the corresponding amplitudes. NA indicates that the sidetracked phase lasts less than 55 e-folds.}
\label{T:Table_5}
\end{table}

\begin{table}[h]
\centering
\begin{tabular}{|l|l|l|l|l|l|l|}
\hline
Acronym   & Parameter&M        & $\mathcal{C}(S,S^{\eq})$    & $\mathcal{C}(S,S^{\orth})$&$ f_{nl}^{\eq} $&$ f_{nl}^{\orth} $\\ \hline
    & &$10^{-2}$               & NA & NA & NA  & NA   \\ \cline{3-7} 
$SI$    & &$10^{-2.5}$           &NA&NA&NA&NA \\  \cline{3-7} 
   & & $10^{-3}$             &*&*&*&*\\ \hline

	&  $f=1$ & $10^{-3}$           &-0.03&-0.77& -13.9&-176.0  \\ \cline{2-7} 
        &  $f=10$ & $10^{-2}$           &-0.31&-0.71 &-0.66&-0.78  \\ \cline{2-7} 
$NI$    &  $f=10$ & $10^{-2.5}$     &-0.22&-0.77& -7.73&-13.6  \\ \cline{2-7} 
        &  $f=10$ & $10^{-3}$          &-0.10&-0.79& -23.3&-90.4 \\ \cline{2-7} 
	&  $f=100$ & $10^{-3}$          &-0.37&-0.72&-15.7&-15.9 \\ \hline

	&  $\mu=2$ & $10^{-3}$           &-0.01&-0.78& -3.98&-136.0  \\ \cline{2-7} 
        &  $\mu=20$ & $10^{-2}$           &-0.47&-0.66 &-0.62&-0.46  \\ \cline{2-7} 
$SFI_2$    &  $\mu=20$ & $10^{-2.5}$     &-0.18&-0.78& -6.67&-14.5  \\ \cline{2-7} 
        &  $\mu=20$ & $10^{-3}$          &-0.02&-0.79& -6.57&-105.0 \\ \cline{2-7} 
	&  $\mu=200$ & $10^{-3}$          &-0.15&-0.78& -22.9&-59.6 \\ \hline

	&  $\mu=2$ & $10^{-3}$           &-0.01&-0.79& -1.49&-106.0  \\ \cline{2-7} 
        &  $\mu=20$ & $10^{-2}$           &NA&NA &NA&NA  \\ \cline{2-7} 
$SFI_4$    &  $\mu=20$ & $10^{-2.5}$     &-0.15&-0.79& -6.04&-15.7 \\ \cline{2-7} 
        &  $\mu=20$ & $10^{-3}$          &-0.01&-0.80& -0.69&-123.0 \\ \cline{2-7} 
	&  $\mu=200$ & $10^{-3}$          &-0.11&-0.79& -19.5&-71.8 \\ \hline
\end{tabular}
\caption{Results for the bispectrum generated in sidetracked inflation in the hyperbolic geometry, indicating the correlation of the shape with the equilateral and orthogonal templates, as well as the corresponding amplitudes. NA indicates that the sidetracked phase lasts less than 55 e-folds. * for SI with $M=10^{-3}$ indicates that we have not been able to reliably compute the bispectrum, as we explain in the main text.}
\label{T:Table_6}
\end{table}


\bibliographystyle{apsrev4-1}
\bibliography{Biblio-2018}

\end{document}